\documentstyle[12pt]{article}

\catcode`\@=11
\def\marginnote#1{}
\newcount\hour
\newcount\minute
\newtoks\amorpm
\hour=\time\divide\hour by60
\minute=\time{\multiply\hour by60 \global\advance\minute
by-\hour}\edef\standardtime{{\ifnum\hour<12
\global\amorpm={am}%
        \else\global\amorpm={pm}\advance\hour by-12 \fi
        \ifnum\hour=0 \hour=12 \fi
        \number\hour:\ifnum\minute<10
0\fi\number\minute\the\amorpm}}
\edef\militarytime{\number\hour:\ifnum\minute<10
0\fi\number\minute}

\def\draftlabel#1{{\@bsphack\if@filesw {\let\thepage\relax
   \xdef\@gtempa{\write\@auxout{\string
      \newlabel{#1}{{\@currentlabel}{\thepage}}}}}\@gtempa
   \if@nobreak \ifvmode\nobreak\fi\fi\fi\@esphack}
        \gdef\@eqnlabel{#1}}
\def\@eqnlabel{}
\def\@vacuum{}
\def\draftmarginnote#1{\marginpar{\raggedright\scriptsize\tt#1}}
\def\draft{\oddsidemargin -.5truein
        \def\@oddfoot{\sl preliminary draft \hfil
        \rm\thepage\hfil\sl\today\quad\militarytime}
        \let\@evenfoot\@oddfoot \overfullrule 3pt
        \let\label=\draftlabel
        \let\marginnote=\draftmarginnote

\def\@eqnnum{(\theequation)\rlap{\kern\marginparsep\tt\@eqnlabel}%
\global\let\@eqnlabel\@vacuum}  }

\def\numberbysection{\@addtoreset{equation}{section}
        \def\theequation{\thesection.\arabic{equation}}}

\def\underline#1{\relax\ifmmode\@@underline#1\else
 $\@@underline{\hbox{#1}}$\relax\fi}

\catcode`@=12
\relax

\numberbysection

\advance \topmargin by -\headheight
\advance \topmargin by -\headsep

\evensidemargin \oddsidemargin
\begin{document}

\topmargin -.6in

\def\rh{{\hat \rho}}
\def\alie{{\hat{\cal G}}}
\def\sect#1{\section{#1}}

\def\rf#1{(\ref{#1})}
\def\lab#1{\label{#1}}
\def\nonu{\nonumber}
\def\br{\begin{eqnarray}}
\def\er{\end{eqnarray}}
\def\be{\begin{equation}}
\def\ee{\end{equation}}
\def\eq{\!\!\!\! &=& \!\!\!\! }
\def\foot#1{\footnotemark\footnotetext{#1}}
\def\lb{\lbrack}
\def\rb{\rbrack}
\def\llangle{\left\langle}
\def\rrangle{\right\rangle}
\def\blangle{\Bigl\langle}
\def\brangle{\Bigr\rangle}
\def\llbrack{\left\lbrack}
\def\rrbrack{\right\rbrack}
\def\lcurl{\left\{}
\def\rcurl{\right\}}
\def\({\left(}
\def\){\right)}
\newcommand{\nit}{\noindent}
\newcommand{\ct}[1]{\cite{#1}}
\newcommand{\bi}[1]{\bibitem{#1}}
\def\lskip{\vskip\baselineskip\vskip-\parskip\noindent}
\relax

\def\tr{\mathop{\rm tr}}
\def\Tr{\mathop{\rm Tr}}
\def\v{\vert}
\def\bv{\bigm\vert}
\def\Bgv{\;\Bigg\vert}
\def\bgv{\bigg\vert}
\newcommand\partder[2]{{{\partial {#1}}\over{\partial {#2}}}}
\newcommand\funcder[2]{{{\delta {#1}}\over{\delta {#2}}}}
\newcommand\Bil[2]{\Bigl\langle {#1} \Bigg\vert {#2} \Bigr\rangle}  
\newcommand\bil[2]{\left\langle {#1} \bigg\vert {#2} \right\rangle} 
\newcommand\me[2]{\left\langle {#1}\bv {#2} \right\rangle} 
\newcommand\sbr[2]{\left\lbrack\,{#1}\, ,\,{#2}\,\right\rbrack}
\newcommand\pbr[2]{\{\,{#1}\, ,\,{#2}\,\}}
\newcommand\pbbr[2]{\lcurl\,{#1}\, ,\,{#2}\,\rcurl}
%
\def\a{\alpha}
\def\at{{\tilde A}^R}
\def\atc{{\tilde {\cal A}}^R}
\def\atcm#1{{\tilde {\cal A}}^{(R,#1)}}
\def\b{\beta}
\def\btil{{\tilde b}}
\def\dc{{\cal D}}
\def\d{\delta}
\def\D{\Delta}
\def\eps{\epsilon}
\def\vareps{\varepsilon}
\def\fptil{{\tilde F}^{+}}
\def\fmtil{{\tilde F}^{-}}
\def\gh{{\hat g}}
\def\g{\gamma}
\def\G{\Gamma}
\def\grad{\nabla}
\def\h{{1\over 2}}
\def\l{\lambda}
\def\L{\Lambda}
\def\m{\mu}
\def\n{\nu}
\def\o{\over}
\def\om{\omega}
\def\O{\Omega}
\def\p{\phi}
\def\P{\Phi}
\def\pa{\partial}
\def\pr{\prime}
\def\pt{{\tilde \Phi}}
\def\qs{Q_{\bf s}}
\def\ra{\rightarrow}
\def\s{\sigma}
\def\S{\Sigma}
\def\t{\tau}
\def\th{\theta}
\def\Th{\Theta}
\def\tpp{\Theta_{+}}
\def\tmm{\Theta_{-}}
\def\tpg{\Theta_{+}^{>}}
\def\tms{\Theta_{-}^{<}}
\def\tp0{\Theta_{+}^{(0)}}
\def\tm0{\Theta_{-}^{(0)}}
\def\ti{\tilde}
\def\wti{\widetilde}
\def\jc{J^C}
\def\bj{{\bar J}}
\def\sj{{\jmath}}
\def\bsj{{\bar \jmath}}
\def\bp{{\bar \p}}
\def\vp{\varphi}
\def\vt{{\tilde \varphi}}
\def\faa{Fa\'a di Bruno~}
\def\ca{{\cal A}}
\def\cb{{\cal B}}
\def\ce{{\cal E}}
\def\cg{{\cal G}}
\def\cgh{{\hat {\cal G}}}
\def\ch{{\cal H}}
\def\chh{{\hat {\cal H}}}
\def\cl{{\cal L}}
\def\cm{{\cal M}}
\def\cn{{\cal N}}
\def\ns{N_{{\bf s}}}
\newcommand\sumi[1]{\sum_{#1}^{\infty}}   
\newcommand\fourmat[4]{\left(\begin{array}{cc}  
{#1} & {#2} \\ {#3} & {#4} \end{array} \right)}

%
\def\lie{{\cal G}}
\def\kmlie{{\hat{\cal G}}}
\def\dlie{{\cal G}^{\ast}}
\def\elie{{\widetilde \lie}}
\def\edlie{{\elie}^{\ast}}
\def\hlie{{\cal H}}
\def\flie{{\cal F}}
\def\wlie{{\widetilde \lie}}
\def\f#1#2#3 {f^{#1#2}_{#3}}
\def\winf{{\sf w_\infty}}
\def\win1{{\sf w_{1+\infty}}}
\def\hwinf{{\sf {\hat w}_{\infty}}}
\def\Winf{{\sf W_\infty}}
\def\Win1{{\sf W_{1+\infty}}}
\def\hWinf{{\sf {\hat W}_{\infty}}}
\def\Rm#1#2{r(\vec{#1},\vec{#2})}          
\def\OR#1{{\cal O}(R_{#1})}           
\def\ORti{{\cal O}({\widetilde R})}           
\def\AdR#1{Ad_{R_{#1}}}              
\def\dAdR#1{Ad_{R_{#1}^{\ast}}}      
\def\adR#1{ad_{R_{#1}^{\ast}}}       
\def\KP{${\rm \, KP\,}$}                 
\def\KPl{${\rm \,KP}_{\ell}\,$}         
\def\KPo{${\rm \,KP}_{\ell = 0}\,$}         
\def\mKPa{${\rm \,KP}_{\ell = 1}\,$}    
\def\mKPb{${\rm \,KP}_{\ell = 2}\,$}    
%
\def\rlx{\relax\leavevmode}
\def\inbar{\vrule height1.5ex width.4pt depth0pt}
\def\IZ{\rlx\hbox{\sf Z\kern-.4em Z}}
\def\IR{\rlx\hbox{\rm I\kern-.18em R}}
\def\IC{\rlx\hbox{\,$\inbar\kern-.3em{\rm C}$}}
\def\IN{\rlx\hbox{\rm I\kern-.18em N}}
\def\IO{\rlx\hbox{\,$\inbar\kern-.3em{\rm O}$}}
\def\IP{\rlx\hbox{\rm I\kern-.18em P}}
\def\IQ{\rlx\hbox{\,$\inbar\kern-.3em{\rm Q}$}}
\def\IF{\rlx\hbox{\rm I\kern-.18em F}}
\def\IG{\rlx\hbox{\,$\inbar\kern-.3em{\rm G}$}}
\def\IH{\rlx\hbox{\rm I\kern-.18em H}}
\def\II{\rlx\hbox{\rm I\kern-.18em I}}
\def\IK{\rlx\hbox{\rm I\kern-.18em K}}
\def\IL{\rlx\hbox{\rm I\kern-.18em L}}
\def\one{\hbox{{1}\kern-.25em\hbox{l}}}
\def\0#1{\relax\ifmmode\mathaccent"7017{#1}%
B        \else\accent23#1\relax\fi}
\def\omz{\0 \omega}
%
\def\ltimes{\mathrel{\vrule height1ex}\joinrel\mathrel\times}
\def\rtimes{\mathrel\times\joinrel\mathrel{\vrule height1ex}}
%
\def\mark{\noindent{\bf Remark.}\quad}
\def\prop{\noindent{\bf Proposition.}\quad}
\def\theor{\noindent{\bf Theorem.}\quad}
\def\name{\noindent{\bf Definition.}\quad}
\def\exam{\noindent{\bf Example.}\quad}
\def\proof{\noindent{\bf Proof.}\quad}
%
%
\def\PRL#1#2#3{{\sl Phys. Rev. Lett.} {\bf#1} (#2) #3}
\def\NPB#1#2#3{{\sl Nucl. Phys.} {\bf B#1} (#2) #3}
\def\NPBFS#1#2#3#4{{\sl Nucl. Phys.} {\bf B#2} [FS#1] (#3) #4}
\def\CMP#1#2#3{{\sl Commun. Math. Phys.} {\bf #1} (#2) #3}
\def\PRD#1#2#3{{\sl Phys. Rev.} {\bf D#1} (#2) #3}
\def\PRv#1#2#3{{\sl Phys. Rev.} {\bf #1} (#2) #3}
\def\PLA#1#2#3{{\sl Phys. Lett.} {\bf #1A} (#2) #3}
\def\PLB#1#2#3{{\sl Phys. Lett.} {\bf #1B} (#2) #3}
\def\JMP#1#2#3{{\sl J. Math. Phys.} {\bf #1} (#2) #3}
\def\PTP#1#2#3{{\sl Prog. Theor. Phys.} {\bf #1} (#2) #3}
\def\SPTP#1#2#3{{\sl Suppl. Prog. Theor. Phys.} {\bf #1} (#2) #3}
\def\AoP#1#2#3{{\sl Ann. of Phys.} {\bf #1} (#2) #3}
\def\PNAS#1#2#3{{\sl Proc. Natl. Acad. Sci. USA} {\bf #1} (#2) #3}
\def\RMP#1#2#3{{\sl Rev. Mod. Phys.} {\bf #1} (#2) #3}
\def\PR#1#2#3{{\sl Phys. Reports} {\bf #1} (#2) #3}
\def\AoM#1#2#3{{\sl Ann. of Math.} {\bf #1} (#2) #3}
\def\UMN#1#2#3{{\sl Usp. Mat. Nauk} {\bf #1} (#2) #3}
\def\FAP#1#2#3{{\sl Funkt. Anal. Prilozheniya} {\bf #1} (#2) #3}
\def\FAaIA#1#2#3{{\sl Functional Analysis and Its Application} {\bf #1} (#2)
#3}
\def\BAMS#1#2#3{{\sl Bull. Am. Math. Soc.} {\bf #1} (#2) #3}
\def\TAMS#1#2#3{{\sl Trans. Am. Math. Soc.} {\bf #1} (#2) #3}
\def\InvM#1#2#3{{\sl Invent. Math.} {\bf #1} (#2) #3}
\def\LMP#1#2#3{{\sl Letters in Math. Phys.} {\bf #1} (#2) #3}
\def\IJMPA#1#2#3{{\sl Int. J. Mod. Phys.} {\bf A#1} (#2) #3}
\def\AdM#1#2#3{{\sl Advances in Math.} {\bf #1} (#2) #3}
\def\RMaP#1#2#3{{\sl Reports on Math. Phys.} {\bf #1} (#2) #3}
\def\IJM#1#2#3{{\sl Ill. J. Math.} {\bf #1} (#2) #3}
\def\APP#1#2#3{{\sl Acta Phys. Polon.} {\bf #1} (#2) #3}
\def\TMP#1#2#3{{\sl Theor. Mat. Phys.} {\bf #1} (#2) #3}
\def\JPA#1#2#3{{\sl J. Physics} {\bf A#1} (#2) #3}
\def\JSM#1#2#3{{\sl J. Soviet Math.} {\bf #1} (#2) #3}
\def\MPLA#1#2#3{{\sl Mod. Phys. Lett.} {\bf A#1} (#2) #3}
\def\JETP#1#2#3{{\sl Sov. Phys. JETP} {\bf #1} (#2) #3}
\def\CAG#1#2#3{{\sl  Commun. Anal\&Geometry} {\bf #1} (#2) #3}
\def\JETPL#1#2#3{{\sl  Sov. Phys. JETP Lett.} {\bf #1} (#2) #3}
\def\PHSA#1#2#3{{\sl Physica} {\bf A#1} (#2) #3}
\def\PHSD#1#2#3{{\sl Physica} {\bf D#1} (#2) #3}
\def\PJA#1#2#3{{\sl Proc. Japan. Acad.} {\bf #1A} (#2) #3}
\def\JPSJ#1#2#3{{\sl J. Phys. Soc. Japan} {\bf #1} (#2) #3}
\def\SJPN#1#2#3{{\sl Sov. J. Part. Nucl.} {\bf #1} (#2) #3}

\begin{titlepage}
\vspace*{-1cm}
\noindent
December, 1995 \hfill{US-FT/27-95}\\
\phantom{bla}
\hfill{IFT-P.048/95} \\
\phantom{bla}
\hfill{LPTENS/95-33} \\
\phantom{bla}
\hfill{hep-th/9512105}
\\
\vskip 0.2cm
\begin{center}
{\Large\bf AFFINE TODA SYSTEMS\\
\medskip
 COUPLED TO MATTER FIELDS\footnote{Supported in part by the European
Community HCM network \# CHRXCT920069.}}
\vglue 1  true cm
{\bf Luiz A. FERREIRA\footnote{On leave from Instituto de F\'\i sica Te\'orica,
IFT/UNESP -  S\~ao Paulo - SP - Brasil.}}\\
{\footnotesize Facultad de F\'\i sica, Universidad de Santiago
de Compostela,\\
15706 Santiago de Compostela, Spain.}\\
{\bf Jean--Loup GERVAIS}\\
{\footnotesize Laboratoire de Physique Th\'eorique de
l'\'Ecole Normale Sup\'erieure\footnote{Unit\'e Propre du
Centre National de la Recherche Scientifique,
associ\'ee \`a l'\'Ecole Normale Sup\'erieure et \`a l'Universit\'e
de Paris-Sud.},\\
24 rue Lhomond, 75231 Paris C\'EDEX 05, ~France.}\\
{\bf Joaqu\'\i n S\'ANCHEZ GUILL\'EN}\\
{\footnotesize Facultad de F\'\i sica, Universidad de Santiago
de Compostela,\\
15706 Santiago de Compostela, Spain.}\\
{\bf Mikhail V.  SAVELIEV}\\
{\footnotesize Institute for High Energy Physics,\\
142284, Protvino, Moscow region, Russia.}

\medskip
\end{center}

\normalsize
\vskip 0.2cm

\vfill
\begin{center}
{\large {\bf ABSTRACT}}\\
\end{center}


\noindent
{\footnotesize
We investigate higher grading integrable generalizations of the
affine Toda systems, where the flat connections defining the models take
values in eigensubspaces of an integral gradation
of an affine Kac-Moody algebra,
with grades varying from $l$ to $-l$ ($l>1$).
The corresponding target space possesses nontrivial vacua and
soliton configurations, which can be interpreted as particles of the theory,
on the same footing as those associated to fundamental fields.
The models can also be formulated by a Hamiltonian reduction procedure from
the so called two--loop WZNW models. We construct the general solution and
show the classes corresponding to the solitons.
Some of the particles and solitons
become massive when the conformal symmetry is spontaneously broken by a
mechanism with an intriguing topological character and leading to a very
simple mass formula.
The massive fields associated to non zero grade
generators  obey field equations of the Dirac type
and may be regarded as matter fields. A special class of
models is  remarkable. These theories  possess a
$U(1)$ Noether current which, after  a special gauge fixing of the conformal
symmetry, is proportional  to a topological current. This
leads to the confinement of the matter field inside the solitons,
which can be regarded as a one dimensional bag model for QCD.
These models are also relevent  to  the study of
 electron self--localization in (quasi)-one-dimensional
electron--phonon systems. }
\vglue 1 true cm

\end{titlepage}

\sect{Introduction}

Integrable theories in low dimensions, besides their intrinsic beauty, have
become a very important tool in the understanding of basic non-perturbative
aspects of physical theories. They constitute always a laboratory to
test ideas on confinement, quantum physics of solitons and many others.
In some cases  they provide realistic models for very
interesting phenomena in condensed matter physics,
statistical mechanics, and  in high energy physics under special kinematical
conditions. More recently, they unexpectedly reappeared as describing the
dependence upon coupling constants of the low-energy effective actions
for supersymmetric Yang-Mills theories in
four dimensions (see e.g. ref.\cite{MW}).

In the present paper we introduce a new class of integrable theories in
$1+1$ dimensions, presenting very interesting physical properties, and
which we hope, will help understanding the role of solitons in
quantum field theories. The models generalize   the
abelian and non abelian affine Toda theories, in the sense that they contain
matter fields coupled  to the (gauge) Toda fields. The arising equations are
affine specialisations  of some general system in two dimensions
\ct{LS81} which becomes integrable when associated with a Lie algebra of
finite growth; see \ct{LS92} and references therein.
They represent  affine (non-abelian) extensions  of the
corresponding finite system \ct{GS95}, which, as  compared  with \ct{fms},
contain matter fields.

We introduce the models through a zero curvature condition, where the
flat connections take values on a affine Kac-Moody algebra $\cgh$ endowed
with an integral gradation. The connection has components not only on the
$0$, $\pm 1$  grades, as for  the usual Toda fields, but also on
eigensubspaces of grades varying from $l$ to $-l$, with $l$ being a positive
integer greater than unity. The components of the connection with grades
$\pm l$, denoted by $E_{\pm l}$, are constant (field independent) and play a
crucial role in specifying
the physical properties of the theory. Following
\ct{AFGZ,bb}, the models are made conformally invariant by the introduction
of fields in the direction of the central term and grading operator of
$\cgh$. They can also be obtained by Hamiltonian reduction from the
so--called two--loop WZNW models \ct{AFGZ,fms}.

An initial physical motivation for studying such dynamical
systems is the same as the one for finite systems \ct{GS95}.  Namely, one
describes a nontrivial, not necessarily
Riemannian target space created by the Toda type fields in the presence of
some additional matter fields. The latter are related with
higher flows of the
corresponding flat connection in the trivial holomorphic principal  fibre
bundle ${\cal M}\times\hat{G}\longmapsto {\cal M}$, where
${\cal M}$ is  a two--dimensional manifold and $\hat{G}$ is an exponential
mapping of  an affine Lie algebra $\hat{\cal G}$.
In the same way as for the finite systems, using a relevant specialisation
of the In\"on\"u--Wigner contraction \ct{IW56},
it is possible, for certain models,
 to  eliminate the back  reaction of some matter fields
to the Toda type fields. As a result, when such a
procedure is applied to all matter fields, the latter will  simply
propagate in the field
of a given  Toda solution.

Here, in contrast with the finite Lie algebra case,
where nonabelian Toda systems lead to the exactly solvable
conformal systems in
the presence of a black hole, the corresponding target space possesses
nontrivial vacua and  soliton configurations, which can be interpreted as
particles of the theory  on the same footing as those associated  to the
fundamental fields.

The conformal symmetry is spontaneously broken like in the usual abelian and
non abelian conformal affine Toda systems \ct{AFGZ,fms}, generating masses
for some particles and solitons through a Higgs like mechanism. The masses
of the fundamental particles are determined by the eigenvalues of the
constant operators $E_{\pm l}$, appearing in the flat connection. The masses
of the solitons have a topological character in the sense
that they are determined
by the asymptotic value of the field in the direction of the central term of
$\cgh$. It turns out that this is also related to the eigenvalues of
$E_{\pm l}$, and
for these reasons we are able to obtain a very simple and suggestive mass
formula for solitons and particles.

The massive fields associated to non zero grade generators
will be found to satisfy
two dimensional Dirac type equations. They are interpreted as matter
fields, and at first sight they are c--number fields, so that they would be
dubbed as bosons. However, the issue of their statistics can only been solved
by considering the corresponding quantum field theory. It is well known that
in two dimensions the statistics of fields depends upon the coupling
constant, and perhaps that could be quantized such these matter fields become
anticommuting operators. An argument in this direction will indeed be
given for a special class of models.

The general solution of the system is constructed following the methods of
reference \ct{LS92}, based on representation theory of affine Lie algebras.
Some new features appear here, due to the higher grade fields, which require
more delicate techniques to obtain the expression of their general solution.
We also use the dressing tranformations \ct{dress1,dress2,dress3,dress4}, as
an alternative, to construct the solutions in the orbit of the vacuum.
The soliton solutions are obtained through the so--called solitonic
specialization \ct{OTU93,fms}, see also \ct{OSU93} for the
nonabelian case. According to that, the
one--soliton solutions are determined by choosing the constant group element,
parametrizing the solutions in the orbit of the vacuum, as an
exponentiation of an eigenvector of the operators $E_{\pm l}$, and
the multi--solitons, by taking it as a product of such exponentials.

There is a special class of models which present some remarkable physical
properties. Any integral gradation of an affine Kac-Moody algebra \ct{kac1}
possesses a period such that the eigensubspaces, with grades differing by a
multiple of that period, have the same structure.
{}For the principal gradation, for instance, that period is equal to the
Coxeter number. In addition, subspaces of grade $n$ and $-n$ are always
isomorphic. By choosing the operators $E_{l}$ and $E_{-l}$, such that one is
the image of the other under such isomorphism, and in addition,
taking $l$ to be equal to the period associated to the gradation, one
obtains models possessing a special $U(1)$ Noether current depending only on
the matter fields. It is then possible, under some circunstances, to choose
one solution in each orbit of the conformal group, such that for these
solutions, that $U(1)$ current is equal  to a topological current
depending only on the (gauge) zero grade fields. The submodel obtained by
such special gauge fixing of the conformal symmetry, presents some very
interesting properties due to this equivalence. We show for instance, in the
case of a model associated to $sl(2)$, that the matter fields get confined
inside the solitons. In addition, the masses of solitons and particles are
shown to be proportional to their $U(1)$ charges, in a manner very similar to
what occurs in four dimensional gauge theories with Higgs in the adjoint
representation and in the BPS limit.
We  believe that this equality between topolgical and Noether currents
 will play an important role in the understanding of the quantum
theory of the solitons. We also point out that such type of models are
related to several interesting phenomena in (quasi)-one-dimensional
electron-phonon physical systems.

The paper is organized as follows. In Section 2 we define the models,
presenting their equations of motion and discussing their symmetries.
The general solution, in terms of highest weight representations is worked
out in detail in Section 3 and illustrated with many examples. The
holomorphic factorizable representation of the general solution given in
Section 3,  admits a remarkable
specialization \ct{OTU93}, which is used in Section 4 for the calculation
 of the soliton solutions. In Section 5, we perform the related
dressing procedure, which gives all solutions in the orbit of the vacuum,
including the solitonic ones. The dressing method gives
further insight into the soliton properties and it is very useful for the
applications which follow. In Section 6, an
alternative formulation of our system as two--loop WZNW model,
 provides an  improved energy--momentum
 tensor obtained via the Sugawara construction \ct{suga,go}.
The  masses of solitons and particles are calculated in Section
7 with the help of that tensor, through the spontaneous breakdown of the
conformal symmetry by a Higgs like mechanism.
In Section 8 we discuss the physical properties of the higher grading
fields establishing that the massive ones satisfy Dirac like
equations  and  discussing some of
their peculiarities, related to parity and complex conjugation.
Section 9 is devoted to a specially interesting class of models possessing
the $U(1)$ Noether current mentioned above.
In Section 10, we treat in great detail two models associated to the
principal gradation of $sl(2)^{(1)}$, discussing their
physical applications. Section 11 is devoted to the conclusions and
perspectives for future investigations.

\sect{Formulation of the System}
\label{sec:formulation}

Consider an untwisted affine Kac-Moody algebra $\cgh$ endowed with an integral
gradation $\cgh = \bigoplus_{n\in \IZ} \cgh_n$, and denote
\be
\cgh_{+} = \bigoplus_{n >0} \cgh_n\, , \qquad
\cgh_{-} = \bigoplus_{n >0} \cgh_{-n}.
\lab{grad2}
\ee

Notice that by an affine Lie algebra we mean a loop algebra corresponding to a
finite dimensional simple Lie algebra $\cg$ of rank $r$, extended by the center
$C$ and the derivation $D$.
According to \ct{kac1}, integral gradations of $\cgh$ are labelled by a
set of co-prime integers  ${\bf s}=\( s_0,  s_1, \ldots s_r \)$, and
the grading operators are given by
\be
Q_{{\bf s}} \equiv H_{{\bf s}} + N_{{\bf s}}\, D - {1\o {2 N_{{\bf s}}}}
\Tr \( H_{{\bf s}} \)^2 \, C \, .
\lab{gradop}
\ee
Here
\be
H_{{\bf s}} \equiv \sum_{a=1}^{r} s_a \l^v_a \cdot H^0 \, , \qquad
N_{{\bf s}} \equiv \sum_{i=0}^{r} s_i m_i^{\psi} \, , \qquad
\psi = \sum_{a=1}^{r}  m_a^{\psi} \a_a \, , \quad m_0^{\psi} = 1\, ;
\ee
$H^0$ is an element of the Cartan subalgebra of $\cg$; $\a_a$,
$a=1,2,\ldots r$, are its simple roots; $\psi$ is its maximal root;
$m_a^{\psi}$ the integers in  expansion $\psi = \sum_{a=1}^r m_a^{\psi} \a_a$;
and $\l^v_a
$ are the fundamental co--weights satisfying the relation $\a_a \cdot \l^v_b =
\d_{ab}$.

Let ${\cal M}$ be a two dimensional manifold with local coordinates
$x_+$ and $x_-$; $\hat{\cal G}$ be an affine Lie algebra corresponding
to a finite dimensional complex simple Lie algebra ${\cal G}$ with
the Lie group $G$; ${\cal A}$ be a flat connection in the trivial holomorphic
principal fibre bundle ${\cal M}\times \hat{G}\longmapsto {\cal M}$.
Specify the connection in such a way that its $(1,0)$-component takes values in
the subspaces $\bigoplus_{n=0}^{l} \cgh_{+n}$, and $(0,1)$-component takes
values in  $\bigoplus_{n=0}^{l} \cgh_{-n}$, with $l$ being a fixed positive
integer. In other words, up to a relevant gauge tranformation, these
components, satisfying the zero curvature condition
\be
\pa_{+} A_{-} - \pa_{-} A_{+} + \lb A_{+}\, , \, A_{-} \rb = 0,
\lab{zc}
\ee
are of the form
\br
A_{+} = - B\, F^{+} \, B^{-1} \, , \qquad
A_{-} = - \pa_{-} B \,  B^{-1} + F^{-}.
\lab{gp}
\er
Here $B$ is a mapping from ${\cal M}$ to the Lie group $\hat{G}_0$ with
the Lie algebra  $\cgh_0$; $F^{\pm}$ are mappings to $\bigoplus_{n=1}^{l}
\cgh_{\pm n}$ of the form
\be
 F^{+} = E_{l} + \sum_{m=1}^{l-1} F^{+}_m \, ,
\lab{fp}\,\,  \qquad
 F^{-} = E_{-l} + \sum_{m=1}^{l-1} F^{-}_m\, ,
\ee
with $E_{\pm l}$ being some fixed elements of $\cgh_{\pm l}$; and $F^{\pm}_m$,
$1\leq m\leq l-1$, take values in $\cgh_{\pm m}$.

Substituting the gauge potentials \rf{gp} into \rf{zc}, one gets the equations
of motion
\br
\pa_{+}\( \pa_{-} B\, B^{-1}\) &=& \lb E_{-l} \, , \, B \, E_{l}\, B^{-1}\rb
+ \sum_{n=1}^{l-1} \lb F^{-}_{n} \, , \, B \, F^{+}_{n}\, B^{-1}\rb \, ,
\lab{em1}\\
\pa_{-} F^{+}_{m} &=& \lb E_{l} \, , \, B^{-1} \, F^{-}_{l-m}\,B \rb
+ \sum_{n=1}^{l-m-1} \lb F^{+}_{n+m} \, , \, B^{-1} \, F^{-}_{n}\, B\rb \, ,
\lab{em2}\\
\pa_{+} F^{-}_{m} &=& -\lb E_{-l} \, , \, B \,  F^{+}_{l-m}\,B^{-1} \rb
- \sum_{n=1}^{l-m-1} \lb F^{-}_{n+m} \, , \, B \, F^{+}_{n}\, B^{-1}\rb \, .
\lab{em3}
\er

Note that a consideration of the systems generated by the flat connection
with the components $A_{\pm}$ taking values in the subspaces
$\oplus_{n=0}^{l_{\pm}}\hat{\cal G}_{\pm n}$ where $l_{\pm}$ can be
different positive integers, follows completely the same line. The systems
of that type, even without appealing to their explicit formulation, are
integrable, whenever the corresponding Lie algebra is finite dimensional or
affine; see \ct{LS81}, and also \ct{L83}, \ct{LS92}. We restrict
ourselves here only to the case with $l_+=l_-=l$; however, other (asymmetric)
possibilities can be attractive as well, and their investigation follows the
same arguments as presented here \footnote{Finishing the present
paper, we were acquainted with  paper \ct{CH95}
 where the authors have studied some special non--left--right
symmetric, as they called heterotic conformal Toda system, corresponding to
the case of the series $A_r$ endowed with the principal gradation, and a
choice when $l_+=2,\, l_-=1$. (In fact, their consideration is pretty valid
for an arbitrary finite dimensional Lie algebra, and, with a minor
modification, for an affine Lie algebra.) They also used some additional
restrictions, in particular that, in our notations, $E_2=\sum_{\alpha} [E^0_{
\alpha}, E^0_{(\alpha +1)}]$, $F^+_1=[E_2, \tilde{F}^-_1]$, $E_{-2}=0$,
$F^-_1=\sum_{\alpha} E^0_{-\alpha}$, where $\tilde{F}^-_1$ is a mapping
to ${\cal G}_{-1}$.}.

Since $Q_{{\bf s}}$ and $C$ are in $\cgh_0$, we parametrise $B$ as
\be
B = b\, e^{\eta \, Q_{{\bf s}}} \, e^{\nu \, C}\, ,
\lab{bdef}
\ee
where $b$ is a mapping to $G_0$, the subgroup of $\hat{G}_0$ generated by all
elements of $\cgh_0$ other than $Q_{{\bf s}}$ and $C$. The fields $\eta$ and
$\nu$ correspond to the extension of the loop algebra, and, as we will show
below, are responsible for making the system conformally invariant
\ct{AFGZ,bb}. Clearly, the order of the three factors in \rf{bdef} is
irrelevant, since they commute. In addition, we will use a special basis for
the generators of $\cgh_0$ such that they are all orthogonal to
$Q_{{\bf s}}$ and
$C$. From \rf{gradop} one observes that the generators of $\cgh_0$ are, besides
$C$ and $Q_{{\bf s}}$, the elements $H^0_a$, $a=1,2,\ldots r$, of the
 Cartan subalgebra,
and step operators $E_{\pm\a}^0$ and $E_{\pm\b}^{\mp 1}$, such that
$\sum_{a=1}^{r} s_a \l^v_a \cdot \a =0$, and
$\sum_{a=1}^{r} s_a \l^v_a \cdot \b = N_{{\bf s}}$.
There can be no step operators $E_{\gamma}^{n}$, with $\mid n
\mid >1$, as explained in appendix C of ref. \ct{fms}. Therefore,
shifting the Cartan elements as
\br
\widetilde{H}_{a}^0\, =\, H_{a}^0\, -\, {1\over N_{\bf s}}\, \Tr\left(H_{\bf
s}\, H_{a}^0\right)\, C =\, H_{a}^0\, -\, {2\over
\alpha_{a}^2}{s_a\over N_{\bf s}}\, C,
\lab{cartanplus}
\er
one gets
\br &&\Tr\left(C^2\right)= \Tr\left(C\,\widetilde{H}_{a}^0\right) = \Tr\left(
Q_{\bf s}^2 \right)=
\Tr\left( Q_{\bf s}\, \widetilde{H}_{a}^0\right)=0,
 \quad
\Tr\left( Q_{\bf s}\, C\right) = N_{\bf s},\nonu \\
&&\Tr\left(
\widetilde{H}_{a}^0\, \widetilde{H}_{b}^0\right)= \Tr\left(H_{a}^0\,
H_{b}^0\right) = 4\a_a \cdot \a_b/\a_a^2 \a_b^2\equiv \eta_{ab},
\lab{bilbasis1}
\er
for all $a,b=1\ldots,r$. Here we have used $H_a^0= 2 \a_a\cdot H^0/\a_a^2$,
$\Tr\( x\cdot H^0\, y\cdot H^0\) = x \cdot y$, and $\Tr \( C\,D\) =1$. For more
detail of such a special basis, see appendix C of ref. \ct{fms}.

Substituting \rf{bdef} into the equations of motion \rf{em1}--\rf{em3},
one has
\br
\pa_{+} \( \pa_{-} b b^{-1}\) &+&
\pa_{+}\pa_{-} \, \nu \, C = e^{l\eta} \lb E_{-l} \, , \, b\, E_{l}\,
b^{-1}\rb + \sum_{n=1}^{l-1} e^{n\eta}\, \lb F^{-}_n\, ,\, b\,
F^{+}_{n}b^{-1}\rb \, ,
\lab{eqm1}\\
\pa_{-} F^{+}_m &=& e^{(l-m)\eta}\, \lb E_{l} \, , \, b^{-1}\, F^{-}_{l-m}\, b
\rb + \sum_{n=1}^{l-m-1} e^{n\eta} \lb F^{+}_{m+n} \, , \, b^{-1}\, F^{-}_n\,
b\rb \, ,
\lab{eqm2}\\
\pa_{+} F^{-}_m &=& -e^{(l-m)\eta}\, \lb E_{-l} \, , \, b\, F^{+}_{l-m}\,
b^{-1} \rb
- \sum_{n=1}^{l-m-1} e^{n\eta} \lb F^{-}_{m+n} \, , \, b\, F^{+}_n\, b^{-1}\rb
\, ,
\lab{eqm3}\\
\pa_{+}\pa_{-}\, \eta \,  Q_{{\bf s}}  &=& 0 \, ,
\lab{eqm4}
\er
where the last equation is a consenquence of the fact that $D$, and hence
 $Q_{{\bf
s}}$, can not be obtained as the Lie bracket of any two elements of $\cgh$.

Let us discuss briefly the meaning of system (2.13)--(2.16).
First, it is clear that for $l=1$, when all the mappings $F^{\pm}_n,\,
n>1$, are absent in the game, these equations  coincide
with the standard conformally affine Toda system. Second, let us
conventionally call the fields, parametrising the mapping $b$,
the Toda type fields; and the fields, entering a parametrisation
of the mappings $F^{\pm}_n$, the matter fields coupled, in accordance
with equations (2.13)--(2.16), to the Toda type fields; the
reason for that will be clear from the following observation.
Namely, using a relevant specialisation of the In\"on\"u--Wigner contraction
\ct{IW56},
one can bring to zero the back reaction to the Toda type fields
for some or all of the matter fields. Recall that under  the In\"on\"u--Wigner
contraction of a simple Lie algebra ${\cal G}$, its elements are
multiplied by constant parameters, some of which tend to zero in
a consistent way. Roughly speaking, with this procedure one multiplies
by a contraction parameter, say $\kappa$, the elements of a subspace
${\cal P}\subset{\cal G}$, complementary to some subalgebra
${\cal H}$ of ${\cal G}$;
${\cal P}\rightarrow\kappa{\cal P}={\cal P}^{(\kappa )}$,
${\cal H}\rightarrow{\cal H}$. Then it is clear that
there exists a limit $\kappa\rightarrow 0$, when the
corresponding algebra ${\cal G}^{(0)}$ contains ${\cal P}^{(0)}$ as an ideal;
in other words, ${\cal G}$ becomes the semi--direct sum ${\cal G}^{(0)}$
of ${\cal H}$ and ${\cal P}^{(0)}$, and, hence, we end up with a
non--semisimple
Lie algebra. In particular, if a simple Lie algebra ${\cal G}$
is endowed with a ${\bf Z}$-gradation, supply the grading
subspaces ${\cal G}_n$, corresponding to the subspace ${\cal
P}$, with the parameter $\kappa$, and then tend it to zero in a consistent
way, having in mind the grading property $[{\cal G}_m, {\cal G}_n]\subset
{\cal G}_{m+n}$. The same scheme takes place for the affine
algebras. Then, we can eliminate contributions coming from
some appropriate mappings $F^{\pm}_n$ in (2.13)--(2.15), and,
of course, have a possibility to arrive at the case when
equation (2.13) does not contain the sum of the last $l-1$ terms
in the r.h.s. As a result, we come to an equation which looks similar
to those for the standard affine Toda system, however, in
general, with a different meaning of the elements $E_{\pm l}$ which
belong here to the subspaces $\hat{\cal G}_{\pm l}$. Evidently,
there are many other meaningful possibilities.
Note that an analogous In\"on\"u--Wigner procedure, albeit for
the corresponding representations of the algebras,
can be applied to obtain the solution of the contracted systems,
starting from the general solution to system (2.13)--(2.16) obtained
in the next section.

The structure of the vacuum of the system \rf{eqm1}--\rf{eqm4} is rather
complicated. We will discuss some aspects of it below. However, there is a
simple condition that guarantees the existence of static (vacuum) solutions.
If the elements $E_{\pm l}$ satisfy the relation
\be
\lb E_{l}\, , \, E_{-l}\rb = \b \, C \, , \qquad
\mbox{\rm where \quad  $\b = {l\o  N_{\bf s}}\, \Tr \( E_{l}\, E_{-l}\)$},
\lab{vaccond}
\ee
then
\be
b=1 \, , \qquad F^{\pm}_{m} = 0 \, , \qquad \eta = 0 \, , \qquad
\nu = -\b x_{+} x_{-},
\lab{vacuum1}
\ee
is a (vacuum) solution of \rf{eqm1}--\rf{eqm4}.

Another possibility for vacuum solutions arises when $E_{\pm l}$, $l>1$, belong
to a Heisenberg subalgebra of $\cgh$, see \ct{kac1,kacpet},
\be
\lb E_{M}\, , \, E_{N}\rb = \Tr \( E_{M}E_{-M}\) \, M \, \d_{M+N,0}\, C,
\lab{heis}
\ee
where $M, N$ belong to some (infinite) subset $\IZ_E$ of the integer numbers
$\IZ$. In such cases one has that
\br
b=1 \, , \quad
\eta = 0 \, , \quad  F^{\pm}_M =c^{\pm}_M\, E_{\pm M} \, , \quad
F^{\pm}_m = 0 \, , \quad \mbox {\rm if $m \notin \IZ_E$}\, , \quad
\nu = - \Omega \, x_{+}\, x_{-},
\lab{vacuum2}
\er
is a solution of \rf{eqm1}--\rf{eqm4} with $c^{\pm}_M$ being constants, and
\be
\Omega \equiv  \b + \sum_{M=1}^{l-1} \Tr\( E_{M}E_{-M}\)\,  M\, c^{+}_M\,
c^{-}_M.
\lab{omega}
\ee

Obviously, the system \rf{eqm1}--\rf{eqm4} may have many more vacuum solutions
besides \rf{vacuum1} and \rf{vacuum2}. However,  the condition \rf{vaccond}
guarantees the existence of at least one vacuum solution. Such a fact, as we
will see below, favors the existence of soliton solutions.

The models introduced above are completely characterised by the data
$\{\cgh , Q_{{\bf s}}, l, E_{\pm l}\}$; and we have a quite large class of
systems with physical properties crucially depending on a choice of
those data.

Equations \rf{eqm1} -- \rf{eqm4} are invariant under the conformal
transformation
\be
x_{+} \ra f(x_{+}) \, , \qquad x_{-} \ra g(x_{-}),
\lab{ct}
\ee
with $f$ and $g$ being analytic functions; and with the fields transforming as
\br
b(x_{+}\, , \, x_{-}) &\ra&
{\tilde b}({\tilde x}_{+}\, , \,  {\tilde x}_{-}) = b(x_{+}\, , \, x_{-}) \, ,
\lab{ctf1}\\
e^{-\nu (x_{+}\, , \, x_{-})} &\ra& e^{-{\tilde \nu}({\tilde x}_{+}\, , \,
{\tilde x}_{-})} = \( f^{\pr}\)^{\d} \, \( g^{\pr}\)^{\d}
e^{-\nu (x_{+}\, , \, x_{-})} \, ,
\lab{ctf2}\\
e^{-\eta (x_{+}\, , \, x_{-})} &\ra& e^{-{\tilde \eta}({\tilde x}_{+}\, , \,
{\tilde x}_{-})} = \( f^{\pr}\)^{1/l} \, \( g^{\pr}\)^{1/l}  e^{-\eta (x_{+}\,
, \, x_{-})} \, ,
\lab{ctf3}\\
F^{+}_m (x_{+}\, , \, x_{-}) &\ra & {\tilde F}^{+}_m ({\tilde x}_{+}\, , \,
{\tilde x}_{-}) =   \( f^{\pr}\)^{-1+m/l}\, F^{+}_m (x_{+}\, , \, x_{-}) \, ,
\lab{ctf4}\\
F^{-}_m (x_{+}\, , \, x_{-}) &\ra & {\tilde F}^{-}_m ({\tilde x}_{+}\, , \,
{\tilde x}_{-}) =   \( g^{\pr}\)^{-1+m/l}\, F^{-}_m (x_{+}\, , \, x_{-}) \, ,
\lab{ctf5}
\er
where the conformal weight $\d$, associated to $e^{-\nu}$, is arbitrary.

Notice that the Lorentz transformation $ x_{\pm} \ra \l^{\mp 1} x_{\pm}$ is
obtained from \rf{ct} by taking $f(x_{+})=x_{+}/\l$ and $g(x_{-}) = \l x_{-}$.
Therefore, from \rf{ctf1}--\rf{ctf5} we get the fields transforming as
\br
B \ra B\, , \quad
 F^{+}_{m} \ra \l^{1-m/l}\, F^{+}_{m}\, , \quad
 F^{-}_{m} \ra \l^{-1+m/l}\, F^{-}_{m}.
\lab{ltf}
\er

Equations \rf{eqm1}--\rf{eqm4} are also invariant under the
transformations
\br
b\( x_{+}\, ,\, x_{-}\) &\ra& h_L\(x_{-}\)\, b\( x_{+}\, ,\, x_{-}\)\,
h_R\(x_{+}\),
\lab{gs1}\\
F^{+}_m\( x_{+}\, ,\, x_{-}\) &\ra& h_R^{-1}\(x_{+}\)\, F^{+}_m\( x_{+}\, ,\,
x_{-}\)\,  h_R\(x_{+}\),
\lab{gs2}\\
F^{-}_m\( x_{+}\, ,\, x_{-}\) &\ra& h_L\(x_{-}\)\, F^{-}_m\( x_{+}\, ,\,
x_{-}\)\,  h_L^{-1}\(x_{-}\),
\lab{gs3}
\er
where $h_L\(x_{-}\)$ and $h_R\(x_{+}\)$ are elements of subgroups
$\ch_0^L$ and $\ch_0^R$ of $G_0$, respectively, satisfying the conditions
\br
h_R\(x_{+}\)\, E_{l} \, h_R^{-1}\(x_{+}\) = E_{l}\, , \qquad
h_L^{-1}\(x_{-}\)\, E_{-l} \, h_L\(x_{-}\) = E_{-l}.
\lab{hlr}
\er
The left and right gauge transformations commute, and so the gauge group is
$\ch_0^L \otimes \ch_0^R$. Whenever $\ch_0^L$ and $\ch_0^R$ have a set of
common generators, we get an important subgroup of the gauge group, namely
$\ch_D \equiv \ch_0^L \cap \ch_0^R$. These are global gauge transformations,
where the fields are transformed under conjugation ($h_L=h_R^{-1}\equiv h_D
= {\rm const.}$),
\be
b \ra h_D\, b h_D^{-1} \, , \qquad F^{\pm}_m \ra h_D\, F^{\pm}_m h_D^{-1},
\lab{diagonalgauge}
\ee
and $E_{\pm l} = h_D\, E_{\pm l} h_D^{-1}$. We discuss the relevance
of these transformations below.

\sect{General Solution}
\label{sec:gensol}

The flat connection \rf{gp} and the  equations of motion themselves are written
in a completely analogous way as those for the corresponding
finite dimensional systems \ct{GS95}; and, as we have already mentioned in
the Introduction, they
represent a specialisation  of a more general scheme discussed in \ct{LS92}
on the basis of some original papers refered therein. To obtain the general
solution, we have to look at the problem of a similarity between the structure
of the highest weight representation spaces of ${\cal G}$ and  $\hat{\cal G}$,
annihilated by the action of the subspaces ${\cal G}_{+m}$ and $\hat{\cal
G}_{+m}$, respectively. To clarify this point, let us recall, following
\ct{GS95}, some results
concerning the corresponding finite dimensional case,
however, on some more general level, applicable for the affine systems.

First, give some definitions and notations. The flat connection $A$ in
question, is represented in the gradient form,
\be A_{\pm}=g^{-1}\pa_{\pm}g,\lab{cm}
\ee with $g: {\cal M}\longmapsto G$; and we write $A_{\pm}$ with the help
of the modified Gauss decomposition for $g$,
\be g=\mu_-\nu_+\gamma_{0-} \mbox{ and } g=\mu_+\nu_-\gamma_{0+},\lab{gauss}
\ee respectively, in accordance with the Lie algebra decomposition ${\cal G}=
{\cal G}_-\oplus{\cal G}_0\oplus{\cal G}_+$; $\gamma_{0\pm}: {\cal M}
\longmapsto
G_0$, $\mu_{\pm}, \nu_{\pm}: {\cal M} \longmapsto G_{\pm}$.
The grading conditions provide the
holomorphic property of $\mu_{\pm}$, namely that  they
satisfy the initial value problem
\be
\pa_{\pm}\mu_{\pm}(z_{\pm})=\mu_{\pm}(z_{\pm}){\tilde{\cal
E}}_{\pm}(z_{\pm}),\lab{ivp}
\ee where
\be {\tilde{\cal E}}_{\pm}(z_{\pm})=\sum_{m=1}^M{\tilde{\cal
E}}^{\pm}_m(\Phi^{\pm});
\quad {\tilde{\cal E}}^{\pm}_m(\Phi^{\pm})=\sum_{\alpha\in\Delta^+_m}\Phi^{\pm
m}_{\alpha} (z_{\pm})X_{\pm\alpha},\lab{arb}
\ee with arbitrary functions $\Phi^{\pm m}_{\alpha}(z_{\pm})$ determining the
general solution to the system under consideration; $\Delta^+_m$ is the  set of
the positive roots of ${\cal G}=\sum_{m\in{\bf Z}}{\cal G}_m$ corresponding  to
the root vectors $X_{\alpha}$ in the subspace ${\cal G}_m$. Denote by
$\vert h >^{(m)}$ and the dual, ${}^{(m)}< h'\vert$, to  $\vert h >^{(m)}$,
the basis vectors in a representation space, annihilated by all the subspaces
${\cal G}_{+n}$ and ${\cal G}_{-n}$, $n\geq m$, respectively. Then the general
solution for the Toda type fields contained in $B$, is expressed in terms of
\be
{}^{(1)}<h' \vert (\gamma^+_0)^{-1}\mu_+^{-1}\mu_-\gamma^-_0\vert  h
>^{(1)}={}^{(1)}<h' \vert B^{-1}\vert h >^{(1)};\lab{hve}
\ee cf. \ct{LS92}, while the others, say matter fields from $V^{\pm}_m$, are
determined via the matrix elements
\be {}^{(1)}<h' \vert \hat{g}_0^{-1}\mu_+^{-1}\mu_-\vert h >^{(m)},
\; \mbox{and }\;  {}^{(m)}<h' \vert \mu_+^{-1}\mu_-\hat{g}_0^{-1}\vert  h
>^{(1)}.\lab{hvem}
\ee
Here $\hat{g}_0\equiv \gamma_{0+}\gamma_{0-}^{-1}$; $\gamma^{\pm}_0
\equiv \gamma^{\pm}_0(x_{\pm}): {\cal M}\longmapsto G_0$ are arbitrary
mappings; and $B^{-1}=(\gamma_0^+(x_+))^{-1}\hat{g}_0\gamma_0^- (x_-)$.
The whole set of arbitrary functions which determine the general
solution to the problem, consists of those parametrising the mappings
$\gamma_0^{\pm}$, and the functions $\Phi^{\pm m}_{\alpha},\; \alpha
\in\Delta^+_m,\, 1\leq m\leq M-1$; while $\Phi^{\pm M}$ are expressed
in terms of corresponding parameters of $\gamma_0^{\pm}$.
The mappings $F^{\pm}_m$ are determined via the matrix elements
\rf{hvem} in a similar way to the construction discussed very briefly in
\ct{GS95}. Here we give this construction with more details.

Namely, comparing decompositions \rf{gp} with those obtained by substitution of
\rf{gauss} in \rf{cm} with account of the holomorphic property of
$\mu_{\pm}$, one sees that
\be
F^{\pm}=\mp (\gamma^{\pm}_0)^{-1}B^{\mp 1}(\nu^{-1}_{\pm}\partial_{\pm}
\nu_{\pm})B^{\pm 1}\gamma^{\pm}_0. \lab{F}
\ee
Consider, for example, the first matrix element in \rf{hvem}, which  can be
identically rewritten as
\be {}^{(1)}<h' \vert \hat{g}_0^{-1}\mu_+^{-1}\mu_-\vert h >^{(m)}= {}^{(1)}<h'
\vert (\hat{g}_0^{-1}\nu_-\hat{g}_0)\nu_+^{-1}\vert h >^{(m)}= {}^{(1)}<h'
\vert \nu_+^{-1}\vert h >^{(m)}.\lab{ge}
\ee
Then, differentiating \rf{ge} over $x_+$ with the help of \rf{F}, we have
\be
\partial_+ [{}^{(1)}<h' \vert \hat{g}_0^{-1}\mu_+^{-1}\mu_-\vert h >^{(m)}]=
{}^{(1)}<h' \vert (\tilde{B}F^+\tilde{B}^{-1})\nu_+^{-1}\vert h >^{(m)},
\lab{dge}
\ee
where $\tilde{B}\equiv B\gamma^+_0$. Now, with $m=2$, relation \rf{dge}
takes the form
\be
\partial_+ [{}^{(1)}<h' \vert \hat{g}_0^{-1}\mu_+^{-1}\mu_-\vert h >^{(2)}]=
{}^{(1)}<h' \vert \tilde{B}F^+\tilde{B}^{-1}\vert h >^{(2)},
\lab{dgeF1}
\ee which determines the mapping $F^+_1$. To make the next step and calculate
$F^+_2$, recall that the mapping $\nu_+^{-1}$ entering equation \rf{F}, is
expressed in terms of ${\cal L}\equiv \tilde{B}F^+\tilde{B}^{-1}$ as
the Volterra type decomposition
\be
\nu_+^{-1}=\sum_{n=0}^{\infty}\int_{\Omega^n(x_+)}dy{\cal L}(y^1) {\cal
L}(y^2)\cdots {\cal L}(y^n), \lab{bsh}
\ee where $\Omega^n(x_+)$ is the simplex in ${\bf R}^n$ given by
$\Omega^n(x_+)=\{ y\in {\bf R}^n:a\leq y^n\leq y^{n-1}\leq
\cdots\leq y^1\leq x_+\}$; $a$ is some constant determining the initial value
problem; $\nu_+(a)$ is taken to be unity,
with an appropriate choice of the initial value problem for $\mu_{\pm}$.
(Note that here, of
course, the mappings ${\cal L}(y)$ depend also on the variable $x_-$.)
Then, with account of formula
\rf{bsh}, it follows from \rf{dge} that for $m=3$ there is a relation
\br &&\partial_+ [{}^{(1)}<h' \vert
\hat{g}_0^{-1}\mu_+^{-1}\mu_-\vert h >^{(3)}]= \lab{dgeF2}\\
&&{}^{(1)}<h' \vert
\tilde{B}F^+_1\tilde{B}^{-1}\int^{x_+}dy (\tilde{B}F^+_1\tilde{B}^{-1})\vert h
>^{(3)}+ {}^{(1)}<h' \vert \tilde{B}F^+_2\tilde{B}^{-1}\vert h >^{(3)},
\nonumber
\er
and hence, knowing already $F^+_1$, we obtain an expression for $F^+_2$.
Continuing the procedure, we determine all the mappings $F^+_m$
by the recurrent formula
\br
&&{}^{(1)}<h' \vert \tilde{B}F^+_m\tilde{B}^{-1}\vert h >^{(m+1)}=
\partial_+[{}^{(1)}<h' \vert\hat{g}_0^{-1}\mu_+^{-1}\mu_--\lab{rec}\\
&&\sum_{s=2}^m\sum_{k_1,\cdots ,k_s}
\int_{\Omega^s(x_+)}dy{\cal L}_{k_1}(y^1) {\cal
L}_{k_2}(y^2)\cdots {\cal L}_{k_s}(y^s)\vert h>^{(m+1)}],\nonumber
\er
where ${\cal L}_{k_i}\equiv \tilde{B}F^+_{k_i}\tilde{B}^{-1}$ with
integers $k_i\geq 1$ entering the sum in \rf{rec} only for
$\sum_{i=1}^sk_i=m$ and only once if $k_i=k_{i-1}$.
An analogous scheme allows to calculate the mappings $F^-_m$.
Note in this context that, in comparison with the method for constructing the
Toda type fields containing in $B$, which is, of course, a direct repeatition
of those in \ct{LS92}, the proposed construction of the matter fields
$F^{\pm}_m$ represents a novel feature of the integration scheme.

The matrix elements \rf{hve} and \rf{hvem} realise, in general, a matrix
version
of the tau--function for the standard Toda system. As it was already noted in
\ct{S95}, from the point of view of the Lie algebra  representation theory,
they are closely related to the Shapovalov form defined  on a Lie algebra
${\cal G}$.  Such a relationship seems to be quite general and natural;
it takes place
for a wide class of nonlinear integrable systems, including, in particular,
abelian and nonabelian (conformal and affine) Toda systems associated with  the
simple Lie algebras ${\cal G}$. First recall some known definitions,  see e.g.
\ct{Zh94}, concerning the Shapovalov form.

Let  $\wp$ be a Cartan subalgebra of ${\cal G}$, $\wp^*$ be an algebra  dual to
$\wp$;   $U(\cal G)$ be the universal enveloping algebra for ${\cal G}$. The
Shapovalov form defines the  linear mapping $U({\cal G})\otimes_{\bf C} U({\cal
G})\longmapsto U({\wp })$  and is realised as a bilinear form $(x^{\vee} y)_0$
for any two elements $x,  y\in U({\cal G})$. Here $x\rightarrow x^{\vee}$ is
the
Chevalley involution  for $x^{\vee}=x'$, and the hermitean Chevalley involution
for $x^{\vee}=x^*$;  the subscript $``0"$ means the projection of
$U({\cal G})$
on $U({\wp })$ which is parallel to ${\cal G}_-U({\cal G})+U({\cal G}){\cal
G}_+$. Note that the given  definition is naturally extended for the case of
the algebra
$U'({\cal G})= U({\cal G})\otimes_{U({\wp })}R({\wp }^*)$, where $R({\wp }^*)$
is the algebra  of rational functions over ${\wp }^*$. It is very important
that  the form $(x^{\vee} y)_0$ is degenerated on the left ideal $U'({\cal
G}){\cal  G}_+$, and is not degenerated on the subalgebra $U({\cal G}_-)$; and
hence is  not degenerated on the space $U'({\cal G})/U'({\cal G}){\cal G}_+$
which is a rational span of the corresponding Verma module.

One can get convinced that the matrix elements \rf{hve} and \rf{hvem}  are
nothing but the Shapovalov type forms $(x^{\vee}y)_0$ for some special  two
elements $x, y\in U_h({\cal G}_-)$ of the Lie algebra ${\cal G}$, with the
coefficients being holomorphic functions. And, moreover,
since this consideration is valid for the affine case as well, the  general
solution to equations \rf{em1} -- \rf{em3} which are the conformally  affine
analogues of the corresponding conformal system \ct{GS95}, also can be
written in terms of these forms. The reason is that, in accordance with the
general
construction, see \ct{LS92} and references therein, the  grading conditions,
realised in the form of decomposition  \rf{gp} for the  connection $A$,
provide the holomorphic factorisability  of the solution; a differential
geometry formulation of this fact is given in \ct{RS94}.

An explicit formulation of solutions like \rf{hve} and \rf{hvem} as the series
(infinite for an affine case, while absolutely convergent in accordance with
the
arguments given in \ct{LS92} and references therein) in arbitrary functions
$\Phi^{\pm m}_{\alpha}$ can be done, when it is needed, by the  following
purely
algebraic scheme, standard for the Lie algebra representations  theory.
Construct a basis in $U({\cal G})$ with the help of the monomials
$\hat{X}_{\bf m}=X_{m_n}\cdots X_{m_1}$ in the basis elements
$X_{m}=X_m(\Phi_m)$ of ${\cal G}$. In particular, let
$\hat{X}_{\pm{\bf m}}$, $\hat{X}_{-{\bf m}}=\hat{X}_{+{\bf m}}^{\vee}$,  be
such
a basis in $U({\cal G}_{\pm})$ with the weight $\mu_{\bf m}$. Then the
elements
$\hat{X}_{+{\bf m}}\hat{X}_{-{\bf m}}$ generate a basis of $U'({\cal G})$ over
$R({\wp }^*)$, and this procedure gives for any weight $\mu\in {\wp }^*$ a
vector space $F_{\mu}({\cal G})$ of all formal series $\sum_{{\bf  m}, {\bf
n}}c_{{\bf m}, {\bf n}}\hat{X}_{+{\bf m}}\hat{X}_{-{\bf n}}$ with
$c_{{\bf m}, {\bf n}}\in R({\wp }^*)$, where the sum runs over all
monomials of
the weight $\mu =\mu_{\bf m} -\mu_{\bf n}$. The subspaces $F_{\mu} ({\cal G})$
are in turn the subspaces of the algebra $F({\cal G})$ graded by the  weights
$\mu$. Finally, one needs to transform the elements $\hat{X}_{+{\bf m}}
\hat{X}_{-{\bf n}}$ entering \rf{hve} and \rf{hvem} to the series of the
monomials $\hat{X}_-\hat{X}_0 \hat{X}_+$, with $\hat{X}_0\in U({\wp})$, by
using the commutation relations of the algebra ${\cal G}$. Note that for the
abelian systems corresponding to the principal gradation of ${\cal G}$,
the most
suitable choice is the Verma type basis when the monomials $\hat{X}_{\bf m}$
are constructed in terms of the Chevalley generators satisfying the defining
relations. For other gradations, an adequate basis has not been discovered
yet,  and
this is why the known explicit expressions for the general solutions of the
nonabelian Toda type systems \ct{LS92} are written in a rather complicated
form, though their holomorphic factorisability in terms of the form like
\rf{hve}, related in this case with the generalised Verma modules \ct{L77},
is quite clear.

Let us now show how the structure of the matrix
elements \rf{hve} and \rf{hvem} is made concrete for the case of
the principal gradation. Here it is suitable to take as the basis vectors
$\vert h >^{(1)}$ the highest weight vectors $\vert i >,\, 1\leq i\leq r$,
of the fundamental representations of $\hat{\cal G}$. It seems more
illustrative
to discuss the formulation in question for the finite Toda systems; for the
corresponding affine deformations, due to the similarity between the structure
of the highest weight representation spaces of ${\cal G}$ and $\hat{\cal G}$,
it is practically the same.

Since for the principal gradation of ${\cal G}$, the subalgebra ${\cal G}_0$ is
abelian, parametrise the mapping $b(x_+,x_-)$ and $\gamma^{\pm}_0(x_{\pm})$ as
\[
b=e^{(\phi (x_+,x_-), H^0)},\qquad \gamma^{\pm}_0(x_{\pm})=
e^{(\phi^{\pm} (x_{\pm}), H^0)};\]
where $(\phi , H^0)\equiv \sum_{i=1}^r\phi_iH^0_i$.

To write down expressions \rf{hve} and \rf{hvem} in an explicit way, we use
the Verma type basis for the spaces of the fundamental representations
of ${\cal G}$. There, the highest vector $\vert i >$ of the $i$-th
fundamental representation satisfies the relations
\begin{equation}
E^0_j \vert  i >  =  0; \quad  H^0_j \vert  i >  = \delta _{ij} \vert  i > ;
\quad  E^0_{-j} \vert  i >  =  0,\quad i\neq j;
\lab{alpha}
\end{equation}
and the basis we are looking for, are constructed only with the help
of the Chevalley generators $E^0_{-j}$; namely we choose as the basis vectors
those of the set
\begin{eqnarray}
\vert  j_m \ldots j_1;i >\; &\equiv& E^0_{-j_m}\cdots E^0_{-j_1} \vert  i >,
\quad 1\leq j\leq r, \quad 0\leq m\leq N_i - 1; \lab{beta} \\
\vert  j_0 \ldots j_1; i > &\equiv&  \vert  i >,\nonumber
\end{eqnarray}
with nonzero norm; $N_i$ is the dimension of the representation.
There take place the following useful formulas:
\begin{eqnarray}
E^0_{-j} \vert j_m\ldots j_1; i >  &=& \vert j j_m \ldots j_1; i >,
\nonumber\\
H^0_j \vert  j_m \ldots j_1; i >  &=&
(\delta_{ij} - \sum \limits_{q=1}^m k_{j_qj}) \vert  j_m \ldots j_1; i >,
\lab{ad2} \\
E^0_{+j} \vert j_m \ldots j_1; i >  &=& \sum_{q=1}^m \delta_{jj_q}
(\delta_{ij} - \sum\limits_{n=1}^{q-1} k_{j_nj}) \vert j_m\ldots
\hat{j}_q \ldots j_1; i >, \hspace{2.em} \nonumber
\end{eqnarray}
which are evident in virtue of the defining relations for the Cartan and
Chevalley elements, namely,
\begin{equation}
{}[H^0_i, H^0_j]=0,\quad [H^0_i, E^0_{\pm j}]=\pm k_{ji}E^0_{\pm j},\quad
[E^0_{+i}, E^0_{-j}]=\delta_{ij}H^0_i,
\lab{gamma}
\end{equation}
with $k$ being the Cartan matrix of ${\cal G}$.
 Here $\hat{j}_q$ means the absence of the root vector
$E^0_{-j_q}$ in the corresponding formula for the basis vector $\vert
j_m \ldots j_1; i >$. If the norm of the vector $\vert  jj_m \ldots j_1; i >$
or $\vert j_m\ldots\hat{j}_q \ldots j_1; i >$ is equal to zero, the
corresponding term in the r.h.s. of formulas \rf{ad2} is naturally absent.

For example, let us illustrate the structure of the fundamental representation
space by an example of the simple Lie algebras of rank 2, i.e., the
algebras $A_2\equiv {\it sl} (3, {\bf C})$, $B_2\equiv {\it o} (5, {\bf C})$,
$C_2\equiv {\it sp} (4, {\bf C})$, and $G_2$. Namely, the basis vectors of
the $1$-st and $2$-nd fundamental representations of these algebras
respectively are

${\bf A_2}$:
\begin{eqnarray}
& & | 1\rangle,\; | 1;1\rangle, \; | 21;1\rangle;\nonumber\\
& & | 2\rangle,\; | 2;2\rangle, \; | 12;2\rangle;\nonumber
\end{eqnarray}
${\bf B_2}$:
\begin{eqnarray}
& & | 1\rangle,\; | 1;1\rangle, \; | 21;1\rangle,
\; | 221;1\rangle, \; | 1221;1\rangle;\nonumber\\
& & | 2\rangle,\; | 2;2\rangle, \; | 12;2\rangle, \; | 212;2\rangle;\nonumber
\end{eqnarray}
${\bf C_2}$:
\begin{eqnarray}
& & | 1\rangle,\; | 1;1\rangle, \; | 21;1\rangle, \; | 121;1\rangle;\nonumber\\
& & | 2\rangle,\; | 2;2\rangle, \; | 12;2\rangle,
\; | 112;2\rangle, \; | 2112;2\rangle;\nonumber
\end{eqnarray}
${\bf G_2}$:
\begin{eqnarray}
& & | 1\rangle,\; | 1;1\rangle, \; | 21;1\rangle,\; | 121;1\rangle, \;
| 1121;1\rangle,\; | 21121;1\rangle,\; | 121121;1\rangle;\nonumber\\
& & | 2\rangle,\; | 2;2\rangle, \; | 12;2\rangle,\; | 112;2\rangle, \;
| 1112;2\rangle,\; | 2112;2\rangle,\; | 21112;2\rangle;\nonumber\\
& & | 221112;2\rangle,\; | 121112;2\rangle,\; | 2121112;2\rangle,\;
| 1221112;2\rangle;\nonumber\\
& & | 12121112;2\rangle,\; | 112121112;2\rangle,\; | 2112121112;2\rangle.
\nonumber
\end{eqnarray}

In the case under consideration here, one can take
$\vert j_{m-1}\ldots j_1; i >$ as the basis vectors $\vert h >^{(m)}$.
Then, using the relation
$H^0_j\vert i >=\delta_{ij}\vert i >$, expression \rf{hve} is written in the
form
\begin{equation}
e^{-\phi_i}=e^{\phi^+_i-\phi^-_i}< i \vert\mu_+^{-1}\mu_-\vert i >;
\lab{ad1}
\end{equation}
while the first formula in \rf{hvem} reads as
\[
{}^{(1)}< h'\vert\gamma^-_0b(\gamma^+_0)^{-1}\mu_+^{-1}\mu_-\vert h >^{(m)}=
e^{\phi_i+\phi^+_i-\phi^-_i}< i\vert\mu_+^{-1}\mu_-\vert j_{m-1}\ldots j_1;
i>.\]
Finally, the recurrent formula \rf{rec} is reduced to the expression
\begin{eqnarray}
&&< i\vert F^+_m\vert j_m\ldots j_1; i >=\lab{ad3}\\
&&e^{\sum_{s=1}^m
\sum_{l=1}^rk_{j_sl}(\phi^--\phi )_l}\partial_+\{ e^{\phi_i+\phi^+_i-
\phi^-_i}< i\vert\mu_+^{-1}\mu_-\vert j_m\ldots j_1; i> -\nonumber\\
&&\sum_{s=2}^m\sum_{k_1,\cdots ,k_s}
\int_{\Omega^s(x_+)}dy< i\vert {\cal L}_{k_1}(y^1) {\cal
L}_{k_2}(y^2)\cdots {\cal L}_{k_s}(y^s)\vert j_m\ldots j_1; i>\} .\nonumber
\end{eqnarray}
Here, with the decomposition $F^+_m=\sum_{\alpha\in\Delta^+_m}f^+_{\alpha}
E^0_{+\alpha}$, the mappings ${\cal L}_{k_i}$ read as
\[
{\cal L}_{k_i}=\sum_{\alpha\in\Delta^+_{k_i}}
e^{(\alpha ,\phi -\phi^-)}f^+_{\alpha}E^0_{+\alpha},\]
where the product in the exponential is defined by the commutation relation
$[(\phi , H^0), E^0_{+\alpha}]=(\alpha ,\phi ) E^0_{+\alpha}$.
All the matrix elements of type
\[
< i\vert\mu_+^{-1}\mu_-\vert j_m\ldots j_1; i> \mbox{ and } < i\vert {\cal
L}_{k_1}(y^1) {\cal L}_{k_2}(y^2)\cdots {\cal
L}_{k_s}(y^s)\vert j_m\ldots j_1; i>\]
can be calculated in the explicit form of finite polynomials, using the
defining relations \rf{gamma} and the properties \rf{alpha}--\rf{beta} of
the basis vectors. Technically, it is the same task as for the
standard Toda systems, see \ct{LS92}.

In particular, for $m=1$, since $F^+_1=\sum_jf^+_jE^0_j$, one has
\begin{equation}
< i\vert F^+_1\vert i; i >=f^+_i=e^{\sum_{l=1}^rk_{il}(\phi^--\phi )_l}
\partial_+\{\frac{< i\vert\mu_+^{-1}\mu_-\vert i; i>}{< i\vert\mu_+^{-1}
\mu_-\vert i>}\}.\lab{ad4}
\end{equation}
For $m=2$ one can represent the mapping $F^+_2$ as
\[
F^+_2=\sum_{j_1j_2}A_{j_1j_2}E^0_{j_1}E^0_{j_2},\]
where the matrix $A_{j_1j_2}$ is obtained from $2\delta_{j_1j_2}-k_{j_1j_2}$
by replacing all nonzero entries by $f^+_{j_1j_2}(x_+, x_-)\equiv
-f^+_{j_2j_1}(x_+, x_-)$, so that
\[
< i\vert F^+_2\vert j_2j_1; i >=\delta_{ij_1}(2\delta_{ij_2}-k_{ij_2})
A_{ij_2}.\]
Remembering that ${\cal L}_1=\sum_je^{(k, \phi -\phi^-)_j}f^+_jE^0_j$,
formula \rf{ad3} for $m=2$ is read as
\begin{eqnarray}
& & -k_{ij}f^+_{ij}=e^{(k, \phi^--\phi )_i+(k, \phi^--\phi )_j}\{\partial_+
\frac{< i\vert\mu_+^{-1}\mu_-\vert ji; i>}{< i\vert\mu_+^{-1}\mu_-
\vert i>}\nonumber\\
&+& k_{ij}e^{(k, \phi -\phi^-)_i}f^+_i\int^{x_+}dy e^{(k, \phi -\phi^-)_j(y)}
f^+_j(y)\}\lab{for2}\\
&=&e^{(k, \phi^--\phi )_i+(k, \phi^--\phi )_j}\{\partial_+
\frac{< i\vert\mu_+^{-1}\mu_-\vert ji; i>}{< i\vert\mu_+^{-1}\mu_-
\vert i>}\nonumber\\
&+&k_{ij}\frac{< j\vert\mu_+^{-1}\mu_-\vert j; j>}{< j\vert\mu_+^{-1}\mu_-
\vert j>}\partial_+\frac{< i\vert\mu_+^{-1}\mu_-\vert i; i>}{< i\vert
\mu_+^{-1}\mu_-\vert i>}\}, \; i\neq j.\nonumber
\end{eqnarray}
Note that the matrix element $< i\vert\mu_+^{-1}\mu_-\vert ji; i>$ also
vanishes when $k_{ij}=0$.

To obtain explicit solution for the coefficient functions $f^+_{\alpha}$
entering the mappings $F^+_m$ for higher values of $m$, rewrite the above
given decomposition as
\begin{equation}
F^+_m=\sum_{i_1\cdots i_m}f^+_{i_1\cdots i_m}E^0_{\sum_{s=1}^m\alpha_{i_s}},
\qquad \sum_{s=1}^m\alpha_{i_s}\in\Delta^+_m,\lab{Ff}
\end{equation}
with $\alpha_i$ being the simple roots of ${\cal G}$; thereof
\begin{equation}
{\cal L}_{k_j}(y^j)=\sum_{i_1^{(j)}\cdots i_{k_j}^{(j)}}
e^{\sum_{l=1}^{k_j}(k, \phi -\phi^-)_{i_l^{(j)}}(y^j)}
f^+_{i_1^{(j)}\cdots i_{k_j}^{(j)}}(y^j) E^0_{\sum_{s=1}^{k_j}
\alpha_{i_s^{(j)}}},\lab{Lf}
\end{equation}
where $\sum_{s=1}^{k_j}\alpha_{i_s^{(j)}}\in\Delta^+_{k_j}$. Introduce now
the notation
\begin{equation}
{\cal D}^{\{ i \}}_{I_{k_1}\cdots I_{k_s}; J_m}\equiv
<i\vert E^0_{\sum_{l=1}^{k_1}\alpha_{i_l^{(1)}}}\cdots
E^0_{\sum_{l=1}^{k_s}\alpha_{i_l^{(s)}}}\vert j_m\cdots j_1;i>;\lab{not}
\end{equation}
with which
\begin{eqnarray}
& & <i\vert F^+_m\vert j_m\cdots j_1;i>=\sum_{\alpha\in\Delta^+_m}f^+_{\alpha}
<i\vert E^0_{+\alpha}\vert j_m\cdots j_1;i>\equiv\nonumber\\
& & \sum_{\alpha\in\Delta^+_m}{\cal D}^{\{ i \}}_{\alpha ; J_m}f_{\alpha}
\delta_{\alpha ,\sum_{s=1}^m\alpha_{j_s}}\equiv
{\cal D}^{\{ i \}}_{I_m; J_m}f^+_{j_1\cdots j_m};\nonumber
\end{eqnarray}
and
\begin{eqnarray}
& & < i\vert {\cal L}_{k_1}(y^1) {\cal L}_{k_2}(y^2)\cdots
{\cal L}_{k_s}(y^s)\vert j_m\ldots j_1; i>\nonumber\\
& = & \sum_{I_{k_1}\cdots I_{k_s}}\prod_{j=1}^se^{\sum_{l=1}^{k_j}
(k, \phi -\phi^-)_{i_l^{(j)}}(y^j)}f^+_{i_1^{(j)}\cdots
i_{k_j}^{(j)}}(y^j) {\cal D}^{\{ i \}}_{I_{k_1}\cdots I_{k_s}; J_m}.\nonumber
\end{eqnarray}
Here the sum over $I_{k_j}$ means the sum over all indices $i_1^{(j)},\cdots
,i_{k_j}^{(j)}$. Hence, the recurrent formula \rf{ad3} takes the form
\begin{eqnarray}
& & {\cal D}^{\{ i \}}_{I_m; J_m}f^+_{j_1\cdots j_m}=\lab{fff}\\
& & e^{\sum_{s=1}^m(k, \phi^--\phi )_{j_s}}\partial_+
\left\{\frac{< i\vert\mu_+^{-1}
\mu_-\vert j_m\cdots j_1; i>}{< i\vert\mu_+^{-1}\mu_-\vert i>}-
\nonumber\right.\\
& & \left.\sum_{s=2}^m\sum_{k_1,\cdots ,k_s}\int_{\Omega^s(x_+)}dy
\sum_{I_{k_1}\cdots I_{k_s}}\prod_{j=1}^se^{\sum_{l=1}^{k_j}
(k, \phi -\phi^-)_{i_l^{(j)}}(y^j)}f^+_{i_1^{(j)}\cdots
i_{k_j}^{(j)}}(y^j) {\cal D}^{\{ i \}}_{I_{k_1}}\cdots I_{k_s}; J_{m}\right\}
\nonumber
\end{eqnarray}
Here the matrix elements \rf{not} can be calculated explicitly. For this
goal one needs to rewrite the root vectors $E^0_{\alpha}$ as monomials
over the Chevalley generators, and then use the corresponding formula
from \ct{LS92} for the matrix elements of type $<i|E^0_{+i_1}\cdots
E^0_{+i_l}E^0_{-j_l}\cdots E^0_{-j_1}|i>$
written in terms of the Cartan matrix.

\sect{Solitons from the general solution}
\label{sec:solitons}

To obtain soliton solutions to the system \rf{em1}--\rf{em3} from
the general solutions
described in the previous section, we use the method  suggested in \ct{OTU93}
for the abelian affine (periodic) Toda system, and then discussed in
\ct{OSU93},
\ct{U93} and \ct{fms} for their nonabelian generalisations. In the cases
when $E_{\pm l}$, $l>1$, live in a Heisenberg subalgebra of $\cgh$, see
\rf{heis},  consider the initial value problem \rf{ivp} with ${\tilde{\cal
E}}_{\pm}$ in \rf{arb}  taken to be ${\cal E}_{\pm}$, which are
\be
\ce_{\pm} \equiv E_{\pm l} + \sum_{N=1}^{l-1} c^{\pm}_N E_{\pm N}\, , \quad
\mbox{\rm and so} \quad
\lb \ce_{+} \, , \, \ce_{-} \rb = \Omega \, C;
\lab{cepm}
\ee
where $c^{\pm}_N$ and $\Omega$ were introduced in \rf{vacuum2} and \rf{omega},
respectively. In other words, choose
all  arbitrary functions $\Phi^{\pm m}_{\alpha}$ there to be zero except those
standing in the direction of the Heisenberg subalgebra generators $E_{\pm M}$
and $E_{\pm l}$. If $E_{\pm l}$ do not lie in a Heisenberg
subalgebra, take ${\tilde{\cal E}}_{\pm}$ in \rf{arb} to be just $E_{\pm l}$
($c^{\pm}_n=0$).   Then the mappings $\mu_{\pm}$ are
\begin{equation}
\mu_{\pm}=\mu_{\pm}^s = \mu_{\pm}^0e^{x_{\pm}{\cal E}_{\pm}},\lab{musol}
\end{equation}
where $\mu_{\pm}^0$ are some fixed mappings independent on the
local coordinates $x_{\pm}$. It is quite clear that then one ends up with the
following expressions for the matrix elements \rf{hve},
\rf{hvem} determining, with such a choice of the mappings, a particular
solution to system \rf{em1}--\rf{em3}:
\begin{equation}
\tau_{mn}(\mu^0 )= {}^{(m)}<h'\vert
e^{-x_+{\cal E}_{+}}\mu^0 e^{x_-{\cal E}_{-}}\vert h>^{(n)},\lab{mu0}
\end{equation}
with $\mu^0\equiv (\mu_+^0)^{-1}\mu_-^0$ being a fixed mapping
${\cal M}\longmapsto \hat{G}$. Next step in the calculation is to remove
the dependence
on the elements ${\cal E}_{\pm}$ in \rf{mu0}, which can be easily done by
replacing the exponential with ${\cal E}_{+}$ (${\cal E}_{-}$) to the extreme
right (left) position  where it gives unity under the action on the
corresponding basis state which is annihilated by  ${\cal E}_{+}$ (${\cal
E}_{-}$). Note that this procedure is, of course, as it was mentioned
in the previous section, common for all methods for
obtaining an explicit series for the matrix elements under consideration, and
is traditional in the representation theory, see e.g.
\ct{Zh94}. Now we make use of the second  ingredient of the scheme which
consists in an assumption that the mapping
$\mu^0$ is an eigenvector with respect to ${\cal E}_{+}$ and ${\cal E}_{-}$.
Finally, following \ct{OTU93}, we write $\mu^0$ as a product of the
exponentials,
$\mu^0 =\prod_{i=1}^N\exp{\cal V}_i$, where ${\cal V}_i$ are eigenvectors under
the adjoint action of the elements ${\cal E}_{\pm}$, namely $[{\cal E}_{\pm},
{\cal V}_i]=\omega_{\pm }^{(i)}{\cal V}_i$. Note that different orders of the
exponentials in the product, in general, can give different soliton solutions.
These two assumptions concerning a
choice of the fixed mapping $\mu^0$, clearly are equivalent to imposing some
initial conditions on \rf{ivp}, which are relevant  for obtaining $N$-soliton
solutions of our system. The solutions are  characterised by soliton
parameters $\omega^{(i)}\equiv \omega_{+}^{(i)}-
\omega_{-}^{(i)}$ and $v^{(i)}\equiv-\frac{\omega_{+}^{(i)}+
\omega_{-}^{(i)}}{\omega_{+}^{(i)}-\omega_{-}^{(i)}}$, and, if one
interpretes $v^{(i)}$ as velocity of the $i$-th soliton, $\omega_{+}^{(i)}
\omega_{-}^{(i)}$ must be negative, otherwise $\vert v^{(i)}\vert >1$;
moreover, velocities $v^{(i)}$ can be taken to be equal one to each other.
To be more precise,
\begin{equation}
\tau_{mn}(\mu^0 )=  {}^{(m)}<h'\vert \prod_{i=1}^Ne^{{\cal
V}_i\exp [\omega^{(i)}(x-v^{(i)}t)]}
\vert h>^{(n)},\lab{Nsol}
\end{equation} where we are back to the spatial and time variables,
$x_{\pm}=t\pm x$. The final step in the construction of the soliton solutions
is to use an appropriate basis vectors
$\vert h >^{(n)}$ and ${}^{(m)}< h'\vert$
in the representation space, annihilated by all the subspaces
${\cal G}_{+s}$ and ${\cal G}_{-t}$, $s\geq n$, $t\geq m$, respectively.

\sect{Dressing transformations}

Now we present an alternative way of constructing  solutions given by the
solitonic specialisation of the general solution discussed above.
They are obtained by the  dressing transformations which are non local gauge
transformations acting on the gauge potentials $A_{\mu}$ and leaving their
grading structure invariant \ct{dress1,dress2,dress3,dress4}. The dressing
procedure requires existence of two gauge tranformations mapping a given
$A_{\mu}$ to the potential $A_{\mu}^{\rho}$ with the same grading spectrum.
Namely, there must exist $\Theta_{+}$ and $\Theta_{-}$ such
that \footnote{By convention, here and in what follows
derivatives act only on the
first term of products, unless parentheses indicate
otherwise.}
\be
 A_{\mu} \ra  A_{\mu}^{\rho} \equiv \Theta_{\pm}  A_{\mu} \Theta_{\pm}^{-1} -
\pa_{\mu} \Theta_{\pm} \Theta_{\pm}^{-1},
\lab{dressing}
\ee with $A_{\mu}$ and $A_{\mu}^{\rho}$ having the same grading structure as
\rf{gp}. Here $\Theta_{\pm}$ are mappings from ${\cal M}$ to ${\hat G}$. Since
$A_{\mu}$ satisfy the zero curvature condition, they are of the form
\be
A_{\mu} = - \pa_{\mu} T T^{-1}.
\lab{puregauge}
\ee Equivalence of these two transformations implies that
\be
T^{\rho}\equiv \tpp T = \tmm T \rho ,\quad  {\rm or,}\quad {\rm
equivalently,}\quad \tmm^{-1} \tpp = T \rho T^{-1},
\lab{reltptm}
\ee
with $\rho$ being a constant mapping from ${\cal M}$ to ${\hat G}$, and
$A_{\mu}^{\rho} = - \pa_{\mu} T^{\rho} {(T^{\rho})}^{-1}$.

The dressing transformations are defined, in fact, in terms of a
modified Gauss decomposition of the element $T \rho T^{-1}$,
\be
T \rho T^{-1} = \( T \rho T^{-1}\)_{<0}\, \( T \rho T^{-1}\)_{0} \,
\( T \rho T^{-1}\)_{>0}
\ee
with $\( T \rho T^{-1}\)_{0}$, $\( T \rho T^{-1}\)_{>0}$ and  $\( T \rho
T^{-1}\)_{<0}$ belonging to the subgroups of ${\hat G}$, whose Lie algebras,
defined in \rf{grad2}, are  $\cgh_0$, $\cgh_{+}$ and $\cgh_{-}$, respectively.
Then, we define the element $T^{\rho}$ as
\be
T^{\rho} \equiv \tp0 \,\( T \rho T^{-1}\)_{>0}\,  T =
\tm0\, \( T \rho T^{-1}\)_{<0}^{-1}\, T \, \rho ,
\ee
where $\tp0$ and $\tm0$ are introduced in such a way that
\be
{\tm0}^{-1}\, \tp0  = \( T \rho T^{-1}\)_{0}.
\ee

Comparing with \rf{reltptm}, we write the elements $\Theta_{\pm}$ in
\rf{dressing} as
\be
\tpp = \tp0 \,\tpg \, , \qquad \qquad \tmm = \tm0 \,\tms ;
\lab{tpm}
\ee
with
\be
\tpg = \( T \rho T^{-1}\)_{>0} \, , \qquad
\tms = \( T \rho T^{-1}\)_{<0}^{-1}.
\lab{tpms2}
\ee

Therefore, the dressing transformations provide a mapping between solutions.
Hence, the space of solutions to the system can be splitted into orbits
of such transformations. The knowledge of a given simple solution is then
enough to generate a whole orbit of solutions. In the case of the models under
consideration, interesting solutions, namely solitons, fortunately are
on the orbit of a vacuum solution. That is the fact we will explore;
it is related to the solitonic specialisation discussed above.

Perform now the dressing transformation by taking as an initial
configuration a vacuum solution of \rf{eqm1}--\rf{eqm4}. As we have said, the
model under consideration may have several type of vacuum solutions. However,
here we will deal with the solutions of type \rf{vacuum1} or \rf{vacuum2}.

For the vacuum solutions \rf{vacuum2}, the gauge potentials \rf{gp} become
\br
A_{+}^{(0)} = - \ce_{+}\, , \qquad A_{-}^{(0)} =  \ce_{-} + \Omega x_{+} C,
\lab{vacgp}
\er
with ${\cal E}_{\pm}$ given by \rf{cepm}. They can be written as
\be A_{\pm}^{(0)} = - \pa_{\pm} T_0 \, T_0^{-1},
\ee  with
\be T_0 = e^{x_{+}\, \ce_{+}}\, e^{-x_{-}\, \ce_{-}}.
\lab{t0}
\ee

The gauge potentials for the vacuum solution \rf{vacuum1} are
obtained from \rf{vacgp} by taking $c^{\pm}_n =0$. In fact, they are connected
by the gauge transformation
\[
A_{\pm}^{(0)} = {\tilde T}_0
A_{\pm}^{(0)}\mid_{c^{\pm}_n=0}\, {\tilde T}_0^{-1} - \pa_{\pm} {\tilde T}_0
{\tilde T}_0^{-1}, \mbox{ with } {\tilde T}_0 = \exp [ x_{+} \(\ce_{+} -
E_{l}\) ] \exp [ -x_{-} \(\ce_{-} - E_{-l}\) ].\]
However, in general, the vacuum solutions \rf{vacuum1}
and \rf{vacuum2} may not be connected by any dressing transformation,
and, in such a case, the
existence of two elements of  form \rf{tpm}, is not always possible.
Consequently, one can have soliton solutions lying on different
orbits under the dressing transformations.

In order to perform the dressing procedure, we  take \rf{vacgp}
as  initial gauge potentials.
Then, we obtain, under the dressing procedure, the solutions on
the orbit of vacuum \rf{vacuum2}, and for $c^{\pm}_n=0$ those on the orbit
of the vacuum  \rf{vacuum1}.  From the structure of the dressing
transformations and from the fact that the grading operator
\rf{gradop} is never the result of any commutation, since it contains $D$, it
follows that the dressing transformations do not excite the field $\eta$.
Therefore, from  \rf{gp}, \rf{bdef}, \rf{vacgp} and  \rf{dressing} we get
\be
b\, E_{l} \, b^{-1} + \sum_{m=1}^{l-1} b\, F^{+}_{m} \, b^{-1} =
\Theta_{\pm}\, E_{l} \, \Theta_{\pm}^{-1} + \sum_{n=1}^{l-1} c^{+}_{n}
\,\Theta_{\pm}\, E_{n} \, \Theta_{\pm}^{-1}  + \pa_{+} \Theta_{\pm}\,
\Theta_{\pm}^{-1},
\lab{work1}
\ee
\br
& & -\pa_{-}\, b b^{-1} - \pa_{-} \,\( \nu +  \Omega x_{+}
x_{-}\)
\, C + E_{-l} +
\sum_{m=1}^{l-1}  F^{-}_{m} \nonu\\
& &=\Theta_{\pm}\, E_{-l} \, \Theta_{\pm}^{-1} + \sum_{m=1}^{l-1} c^{-}_{m}
\,\Theta_{\pm}\, E_{-m} \, \Theta_{\pm}^{-1}- \pa_{-} \Theta_{\pm}\,
\Theta_{\pm}^{-1}.
\lab{work2}
\er
Note that in the above relations, the fields $b$, $\nu$ and
$F^{\pm}_{m}$ stand  for the solutions on the orbit of the vacuum solution
\rf{vacuum2}. The procedure to construct the solution requires  to split the
above equations into the eigensubspaces of the grading operator \rf{gradop}.
It is convenient to write
\br
\tpg = \exp \( \sum_{s>0} t^{(s)}\) \, , \qquad
\tms = \exp \( \sum_{s>0} t^{(-s)}\) \, , \mbox{ where }
\; t^{(\pm s)} \in \cgh_{\pm s}.
\lab{tpms}
\er
The mappings $t^{(\pm s)}$, for each choice of $\rho$, are determined from
\rf{tpms2} with $T$ being $T_0$ given in \rf{t0}. Then, the components of
\rf{work1} and \rf{work2} in each eigensubspace, give an equation connecting
the fields with $t^{(\pm s)}$. Thus the solutions for the  fields $b$, $\nu$
and $F^{\pm}_{m}$ are determined from $t^{(\pm s)}$. Such a procedure is rather
cumbersome, but fortunately, one needs to know very few $t^{(\pm s)}$'s to get
the solution. For instance, taking relations \rf{work1} and \rf{work2} for
$\Theta_{+}$ ($\Theta_{-}$) with grade components $0$ and $-l$ ($l$ and $0$),
one gets
\be
\tp0 = h_L^{-1} (x_{-})
\, , \qquad
\tm0 = b\, e^{\( \nu + \Omega x_{+} x_{-}\)\, C}\, h_R (x_{+}),
\lab{tpm0}
\ee
with $h_L (x_{-})$ and $h_R (x_{+})$ defined in \rf{hlr}.

{}From \rf{reltptm}, \rf{t0} and \rf{tpm0}  it follows that
\be
{\tms}^{-1}\, \( h_L(x_{-})\, b\, e^{\( \nu + \Omega x_{+} x_{-}\)\,
C}\, h_R(x_{+})\)^{-1}\, \tpg =  e^{x_{+}\, \ce_{+}} \, e^{-x_{-}\, \ce_{-}} \,
\rho \, e^{x_{-}\, \ce_{-}}\, e^{-x_{+}\, \ce_{+}}.
\lab{nice}
\ee

The space--time dependence of the r.h.s. of the above relation is given
explicitly. One can extract the solutions out of \rf{nice} by taking the
expectation  value of its both sides between suitable states of a given
representation of $\cgh$, in a  similar way to that one explained in section
\ref{sec:gensol}.

The solitons solutions are obtained from \rf{nice} by choosing the
fixed group element $\rho$, characterising the dressing
transformation, as the exponential of an eigenvector of $\ce_{\pm}$, i.e.
\be
\rho = e^V.
\ee
That is the {\em solitonic specialization} discussed in section
\ref{sec:solitons}. Indeed, if $V$ satisfies the relations
\be
\lb \ce_{\pm} \, , \, V \rb = \omega_{\pm} \, V,
\ee
then \rf{nice} reads as
\be
\exp \( e^{x_{+}\, \omega_{+} - x_{-}\,
\omega_{-}} \, V\) \equiv
\exp \( e^{\gamma \( x - vt\)}\, V\) ,
\lab{nicesoliton}
\ee
with $\gamma = \omega_{+} + \omega_{-}$, and $v = \( \omega_{-} - \omega_{+}\)/
\( \omega_{+} + \omega_{-}\)$, since $x_{\pm} = t \pm x$.

Therefore, for each eigenvector $V$, expression \rf{nicesoliton} corresponds
to a solution
that travels with a constant velocity $v$ without dispersion.  Depending upon
the properties of $V$, as we will see below in the examples, such solutions
correspond to one--soliton solutions.

The multi--soliton solutions are obtained by taking $\rho$ to be the product of
several one--soliton $\rho$'s, i.e.,
\be
\rho = e^{V_1}\, e^{V_2}\, e^{V_3}\, \ldots e^{V_N},
\ee
with each $V_i$ satisfying $\lb \ce_{\pm} \, , \, V_i \rb =
\omega_{\pm}^i \, V_i$.

Notice that, under the global gauge transformations \rf{diagonalgauge}, the
gauge potentials \rf{gp} are transformed as $A_{\pm} \ra h_D \, A_{\pm} \,
h_D^{-1}$. Therefore, from \rf{puregauge} one has $T \ra h_D T$, and
consequently \rf{tpms2} implies $\Theta_{+}^{>} \ra h_D\, \Theta_{+}^{>} \,
h_D^{-1}$ and $\Theta_{-}^{<} \ra h_D\, \Theta_{-}^{<} \, h_D^{-1}$.
Hence, with solution \rf{nice} corresponding to a fixed element
$\rho$, a solution, obtained from that by a global gauge
transformation \rf{diagonalgauge}, is given by \rf{nice} with the replacement
\be
\rho \ra  h_D\, \rho \, h_D^{-1},
\lab{symrho}
\ee
if the condition $h_D\, \ce_{\pm} \, h_D^{-1} = \ce_{\pm}$ is satisfied.
For the solutions on the orbit of the vacuum \rf{vacuum1}, that is indeed
true, since $\ce_{\pm} = E_{\pm l}$; see \rf{cepm}. For the solitonic case,
one then obtains for each eigenvector $V$ of $\ce_{\pm}$, an orbit
of equivalent one--soliton (or multi--soliton) solutions generated by $h_D\, V
\, h_D^{-1}$.

\sect{Stress tensor and Hamiltonian  reduction}

The two--loop WZNW model was introduced in \ct{AFGZ}; it
leads, under the Hamiltonian reduction procedure, to the conformally affine
abelian
Toda systems. The structure of the two--loop WZNW model was further studied in
\ct{schw}; in \ct{fms} the Hamiltonian reduction was extended to
non abelian affine Toda models. The procedure we discuss here is, in fact,
simpler than those in \ct{fms}, since, due to the extra higher grade
fields of the reduced model, all the constraints are of the first class.

We use the Hamiltonian reduction to obtain, from the Sugawara
construction for the two--loop WZNW model, the energy--momentum tensor for
the systems defined in section \ref{sec:formulation}.
For such systems it is not easy
to get the  canonical
energy--momentum tensor in a direct way, because, in general,
the expression for the  Lagrangian   is  not known yet; and, in fact, in
terms of the fields $F^{\pm}_m$ introduced in  \rf{fp}, it is non local.

The action for the two--loop WZNW model is the same as that for the
ordinary WZNW theory, with the fields $\gh$ being mappings from ${\cal M}$ to
${\hat G}$. Therefore, the equations of motion for the
two--loop WZNW model \ct{AFGZ} are
\br
\pa_{+} \, \( \pa_{-}\, \gh \, \gh^{-1}\) = 0 \, \, , \qquad
\pa_{-} \, \(  \gh^{-1}\,\pa_{+}\, \gh \) = 0.
\lab{wzeq}
\er
We are interested in those mappings which can be represented in the form
\be
\gh = N\, B\, M,
\lab{gaussred}
\ee
where $N$, $B$ and $M$ are generated by the subalgebras $\cgh_{+}$,
$\cgh_0$ and $\cgh_{-}$, respectively, defined in \rf{grad2}. Introduce the
mappings $K_{L/R}$ as
\br
\pa_{-}\, \gh \, \gh^{-1} &=& N\, K_L \, N^{-1} =
N\,\(  N^{-1} \pa_{-} N + \pa_{-} B\, B^{-1} + B\pa_{-} M M^{-1} B^{-1}\)
N^{-1},\nonu\\
\gh^{-1}\,\pa_{+}\, \gh &=& M^{-1} K_R M =
M^{-1}\( B^{-1} N^{-1} \pa_{+} N B + B^{-1} \pa_{+} B + \pa_{+}M \, M^{-1}\) M,
\lab{klr}
\er
and so, from \rf{wzeq}, one obtains
\br
\pa_{-} K_R &=& - \lb K_R \, ,\, \pa_{-} M M^{-1}\,  \rb ,\nonu\\
\pa_{+} K_L &=&  \lb K_L \, ,\, N^{-1}\,\pa_{+} N  \rb\, ;
\lab{klreq}
\er
cf. those in \ct{LS92,AFGZ}.

We show now that  the system \rf{em1} -- \rf{em3} can be obtained from the
Hamiltonian reduction of the two--loop WZNW model by imposing the constraints
\br
\( \pa_{-}\, \gh \, \gh^{-1}\)_{-l} &=& E_{-l} \, \, , \qquad
\( \pa_{-}\, \gh \, \gh^{-1}\)_{<-l} = 0\, ,\nonu\\
\(  \gh^{-1}\,\pa_{+}\, \gh \)_{l} &=& E_{l} \, \, , \qquad
 \(  \gh^{-1}\,\pa_{+}\, \gh \)_{>l} = 0\, .
\lab{const}
\er
{}From \rf{klr} one sees that constraints \rf{const} are equivalent to
the following ones:
\br
\( \pa_{-} M M^{-1}\)_{-l} &=& B^{-1} E_{-l} B \, \, , \qquad
\( \pa_{-} M M^{-1}\)_{<-l} = 0\, ,\nonu\\
\( N^{-1} \pa_{+} N \)_{l} &=& B E_{l} B^{-1} \, \, , \qquad
\( N^{-1} \pa_{+} N \)_{>l} = 0 \, .
\lab{constmn}
\er

Therefore, the mappings $\pa_{-} M M^{-1}$ and $N^{-1} \pa_{+} N$ have
components in the subspaces $\cgh_{-n}$ and $\cgh_n$, respectively, with $1\leq
n\leq l-1$. To provide a relation to system \rf{em1} -- \rf{em3}, denote
\br
B\,\pa_{-} M M^{-1}\, B^{-1} = E_{-l} + \sum_{m=1}^{l-1} F^{-}_m \equiv
F^{-}\, , \quad
B^{-1}\, N^{-1} \pa_{+} N\, B = E_{l} + \sum_{m=1}^{l-1} F^{+}_m \equiv F^{+}.
\lab{fpm}
\er

Substituing the mappings $K_{L/R}$, defined in \rf{klr}, into \rf{klreq}, and
using  constraints \rf{constmn}, one can easily check that $B$, $F^{+}$ and
$F^{-}$ satisfy \rf{em1}--\rf{em3}. Hence, the  system can indeed
be obtained by the Hamiltonian
reduction from the two--loop WZNW model.

Similarly to the finite dimensional case of the standard Toda system
\ct{GRRS93} and its higher grading generalizations \ct{GS95}, one can
prove that the mappings $N$, $M$ and $B$ entering the
decomposition $\gh = NBM$, are determined by the
holomorphic mappings $\mu_{\pm}$, see \rf{ivp}, via the formulas
\begin{equation}
N = \sigma_+ (x_- ) (\gamma^{-}_0)^{-1} \nu_{+}\gamma^{-}_0\, , \qquad
M = \sigma_- (x_+ ) (\gamma^{+}_0)^{-1} \nu_{-}\gamma^{+}_0,
\end{equation}
and the equality
\[
\mu_+^{-1}\mu_-=\nu_-\hat{g}_0\nu_+^{-1}.\]
So, the WZNW field $\gh$ is represented in a holomorphic
factorisable form
\begin{equation}
\gh^{-1} = \gh_L(x_+) \gh_R(x_-), \lab{hf}
\end{equation}
with
\begin{equation}
\gh_R= \mu_-(x_-)\gamma^-_0(x_-)\sigma_+^{-1}(x_-),\qquad
\gh_L^{-1}= \mu_+(x_+)\gamma^+_0(x_+)\sigma_-^{-1}(x_+).
\lab{RL}
\end{equation}
Here $\gamma^{\pm}_0$ are arbitrary mappings determining the
general solution to our system; $\sigma_{\pm}$ are
additional arbitrary mappings generated by the subspaces
$\sum_{m>0}{\cal G}_{\pm m}$, specific for the WZNW theory, and they
 become absent (or, in a sense, hidden) after the Hamiltonian reduction
to the Toda type theories.

It follows from  equations \rf{wzeq} that the corresponding chiral
currents,
\be J_R (X;x_{+}) \equiv - k\,\Tr \( X  \gh^{-1}\,\pa_{+}\, \gh
\) ,\qquad
J_L (X;x_{-}) \equiv  k\,\Tr \( X \pa_{-}\, \gh \, \gh^{-1}  \),
\lab{cur}
\ee
with $X\in \cgh$, satisfy the equations $\partial_-J_R=\partial_+J_L=0$.

The two--loop WZNW model is conformally invariant, and its energy--momentum
tensor is given by the Sugawara construction \ct{suga,go}. On the classical
level, we have its components given by \ct{AFGZ}
\be
T_{++} = {k\o {2}} \Tr \( \gh^{-1}\,\pa_{+}\, \gh \)^2, \qquad
T_{--} = {k\o {2}} \Tr \( \pa_{-}\, \gh \, \gh^{-1} \)^2,
\lab{suga}
\ee
and $T_{+-}=T_{+-}=0$. Here $k$ is the central term of the two--loop current
algebra,
\be
\lb J (X,x) \, , \, J (X^{\pr},y) \rb_{\rm pb} = J ( \lb X\, , \,
X^{\pr} \rb , x) \d
(x-y) + k \Tr \( X\, X^{\pr}\) \d^{\pr}(x-y)\, ;
\ee
$X$ and $X^{\pr}$ are elements of an affine Kac-Moody algebra $\cgh$.
The above relation is satisfied by the left and right currents \rf{cur};
and currents of different chiralities commute.

The components $T_{++}$ and $T_{--}$ have vanishing Poisson bracket
denoted $[*, *]_{\rm pb}$, and each
of them generates a copy of the centerless Virasoro algebra,
\be
\lb T(x)\, , \, T(y) \rb_{\rm pb} = 2 T(y) \d^{\pr}(x-y) -
T^{\pr}(y) \d (x-y)\, .
\ee

The left (right) currents have vanishing Poisson bracket with $T_{++}$
($T_{--}$), and are transformed under $T_{--}$ ($T_{++}$) as  primary fields of
conformal weight $1$,
\be
\lb T(x)\, , \, J(X;y) \rb_{\rm pb} =  J(X;y) \d^{\pr}(x-y) -
J^{\pr}(X;y) \d (x-y)\, .
\ee

Under  constraints \rf{const}, the currents $J_R(X^{(-l)};x)$ and
$J_L(X^{(l)};x)$, such that \\ $\Tr \( X^{(-l)} E_{l}\) \neq 0$, and $\Tr \(
X^{(l)} E_{-l}\) \neq 0$, take fixed non vanishing values. Therefore, the
Virasoro generators do not weakly commute with such currents. In other words,
since the currents are not scalars under the conformal transformations
generated by $T_{++}$ and $T_{--}$,  it means that  constraints \rf{const}
break the conformal invariance. However, one can use the currents in the
direction of the grading operator $Q_{\bf s}$ given in \rf{gradop}, to improve
the Virasoro generators. Define
\be
L_{++}(x_{+}) \equiv T_{++}(x_{+}) + {1\o l} J^{\pr}_R( Q_{\bf s},x_{+}) \,\, ,
\qquad
L_{--}(x_{-}) \equiv T_{--}(x_{-}) - {1\o l} J^{\pr}_L( Q_{\bf s},x_{-}).
\lab{improved}
\ee
One can easily verify that these components generate two commuting copies of
the Virasoro algebra, which are centerless because $\Tr ( Q_{\bf
s})^2=0$, for the particular grading operator in \rf{gradop}. In addition,
these generators weakly commute with  constraints \rf{const}, and, therefore,
the reduced model is conformally invariant. Substituting  constraints
\rf{const} in \rf{improved}, we obtain the energy--momentum tensor for the
reduced model, which generate two commuting copies of the Virasoro algebra
under the Dirac bracket. Indeed, one can easily check, using
\rf{klr}, \rf{suga}, \rf{cur}  and \rf{const}, that
\br
{1\o k}\, L_{++}^{\rm red.}(x_{+})  &=& \h \Tr \( B^{-1}\pa_{+}B \)^2
- {1\o l} \Tr \( Q_{\bf s}\, \pa_{+} \( B^{-1}\pa_{+}B\)\) \nonu\\
&-& {1\o l} \sum_{n=1}^{l-1}
\Tr\(\( \pa_{+} F^{+} \)_n \(M\, Q_{\bf s} M^{-1}\)_{-n}\) \nonu\\
&+&
\sum_{n=1}^{l-1} \(1-{n\o l}\) \Tr \( \(  M^{-1} \pa_{+} M \)_{-n} \(  M^{-1}
F^{+} M\)_n\),
\lab{lpp}
\er
and
\br  {1\o k}\, L_{--}^{\rm red.}(x_{-})  &=& \h \Tr \( \pa_{-} B \, B^{-1} \)^2
- {1\o l}\Tr \( Q_{\bf s} \, \pa_{-}\(\pa_{-} B \, B^{-1}\)\) \nonu\\ &-& {1\o
l} \sum_{n=1}^{l-1}
\Tr\(\( \pa_{-} F^{-} \)_{-n} \(N^{-1}\, Q_{\bf s} N\)_{n}\) \nonu\\    &+&
\sum_{n=1}^{l-1} \(1-{n\o l}\) \Tr \( \(   \pa_{-} N\, N^{-1} \)_{n} \(
N F^{-} N^{-1}\)_{-n}\).
\lab{lmm}
\er

Representing
\be
N= \exp \( \sum_{s>0} \zeta_s \) \,\, , \qquad M=\exp \(\sum_{s>0} \xi_{-s}\),
\lab{zetaxi}
\ee
one sees that the mappings $F^{+}$ and $F^{-}$ defined in \rf{fpm}, depend
on $\zeta_s$ and $\xi_{-s}$, respectively, for $1\leq s\leq l-1$. In
addition, due to the grading structure of the terms entering  \rf{lpp}
and \rf{lmm}, the fields $\zeta_s$ and $\xi_{-s}$, for $s\geq l$, do
not contribute to $L_{++}^{\rm red.}$ and $L_{--}^{\rm red.}$. Therefore, the
components \rf{lpp} and \rf{lmm} of the energy--momentum tensor, are
local functions of the fields $B$, $\zeta_s$ and $\xi_{-s}$, $1\leq s\leq
l-1$, of the reduced model. If one wishes to express those
components in terms of $F^{\pm}$, one gets a non local
expression.  That is, in fact, a difficulty for obtaining the canonical
energy--momentum tensor for system \rf{em1}--\rf{em3}. The corresponding
Lagrangian, written in terms of $F^{\pm}$, is also non local.

The canonical energy--momentum tensor \rf{lpp} -- \rf{lmm} is conserved and
traceless as a consequence of the fact that it is the reduced two--loop WZNW
stress tensor, i.e.,
\be
\pa_{-}\, L_{++}^{\rm red.}=0 \, , \qquad  \pa_{+}\, L_{--}^{\rm red.}=0 \, ,
\qquad L_{+-}^{\rm red.} = L_{-+}^{\rm red.} =0 \, .
\lab{conserve}
\ee
Notice that such a tensor has a part which is a total derivative, namely \\
$\Tr \( Q_{\bf s}\, \pa_{+} \( B^{-1}\pa_{+}B\)\)$ in \rf{lpp}, and
$\Tr \( Q_{\bf s}\, \pa_{-}\(\pa_{-} B \, B^{-1}\)\)$ in \rf{lmm}.
They do not correspond
to the full reduced improvement terms $J^{\pr}_{R/L}( Q_{\bf s})$, but only
to part of it. We can use them to construct a trivially conserved
tensor, namely\footnotemark{\footnotetext{The metric is $g_{++}=g_{--}=0$,
$g_{+-}=g_{-+}=1/2$, and since $B$ commutes with $Q_{\bf s}$, one has
$\pa_{\mu} \Tr \( Q_{\bf s} B^{-1} \pa_{\nu} B\) = \pa_{\nu}
\Tr \( Q_{\bf s}  \pa_{\mu} B \, B^{-1}\)$.}}
\be
S_{\mu\nu} \equiv -{k\o l} \Tr \( Q_{\bf s} \( \pa_{\mu}\, \( B^{-1}
\pa_{\nu} \, B \) - g_{\mu\nu} \pa_{\rho} \( B^{-1} \pa^{\rho} \, B \) \)\)
= - {k\, N_{\bf s}\o l} \( \pa_{\mu} \pa_{\nu} -  g_{\mu\nu} \pa^2\) \nu ,
\lab{smunu}
\ee
where, in the last equality, we have made use of the fact that, thanks to the
basis chosen to parametrise $B$, see \rf{bdef} -- \rf{bilbasis1},
$Q_{\bf s}$ is orthogonal to  all generators of $B$, except the
central term $C$. So, such a tensor is symmetric and
conserved,\footnotemark{\footnotetext {Due to the fact that $B$ commutes with
$Q_{\bf s}$, it follows that $S_{\mu\nu}$ is symmetric and conserved
independently of  the basis we use.}}
\be
\pa^{\mu}\,S_{\mu\nu}=0\, .
\lab{scons}
\ee
Then, introduce an energy--momentum tensor $\Theta_{\mu\nu}$ as
\be
\Theta_{\mu\nu} = L_{\mu\nu}^{\rm red.} - S_{\mu\nu}\, .
\lab{teta}
\ee
Due to \rf{conserve} and \rf{scons}, it is also symmetric and conserved,
\be
\pa^{\mu}\,\Theta_{\mu\nu}=0 \, ,
\lab{tetacons}
\ee
but it is not traceless.

\sect{Masses of fundamental particles and solitons}
\label{sec:masses}

As we have seen above, the system under consideration is conformally invariant.
Therefore, since we do not have a continuum mass spectrum, its fundamental
particles have to be massless.  However, such a symmetry can be spontaneoulsy
broken by choosing a particular constant solution for the field
$\eta$, say $\eta=\eta_0$. The resulting theory is then massive.
Representing the mapping $B$ as $B \equiv \exp T$, and
considering only the linear  field approximation, i.e., the free part
of the equations of motion  \rf{em1}--\rf{em3}, one gets
\br
\pa_{+}\pa_{-} \, T &=& - v_{\eta}\lb E_{-l} \, , \, \lb E_{l}\, , \, T \rb \rb
\, ,
\lab{mass1}\\
\pa_{+}\pa_{-} F^{+}_{m} &=& - v_{\eta}\lb E_{-l} \, , \, \lb E_{l}\, , \,
F^{+}_{m}\rb \rb \, ,
\lab{mass2}\\
\pa_{+}\pa_{-} F^{-}_{m} &=& - v_{\eta} \lb E_{-l} \, , \, \lb E_{l}\, , \,
F^{-}_{m}\rb \rb \, ,
\lab{mass3}
\er
where $v_{\eta} = e^{l\, \eta_0}$.

Therefore, the masses of  fundamental particles in such a theory are given by
the eigenvalues of the operator $\lb E_{-l}\, , \, \lb E_{l}\, , \, * \rb
\rb$  in the subspaces $\cgh_n$, $n=0,\pm 1, \pm 2,\ldots \pm (l-1)$, i.e.,
\be
\lb E_{-l} \, , \, \lb E_{l}\, , \, X \rb \rb = \l \, X.
\lab{fundeigen}
\ee
Since $\pa_{+}\pa_{-} = {1\o 4}(\pa_t^2 - \pa_x^2)$, we obtain
the masses from the Klein--Gordon type equations \rf{mass1} -- \rf{mass3} as
\be
m_{\l}^2 = 4\,\l \, v_{\eta}.
\lab{massf}
\ee
That result constitute a generalization of the arguments used in the abelian
and non abelian affine Toda models \ct{fring,fms}.
Of course, we are interested in those cases where the eigenvalues of the
operator $\lb E_{-l}\, , \, \lb E_{l}\, , \, * \rb \rb$ are real and
positive on the subspaces under consideration.  That will be, in fact, one of
the conditions we use to select the data $\{ \cgh , Q_{{\bf s}}, l, E_{\pm
l}\}$ for defining  physical models through \rf{gp}.

Notice that the field $e^{l\,\eta}$ plays the role of a Higgs field, since it
not only spontaneously breaks  the conformal symmetry, but also because its
vacuum expectation value sets the mass scale of the theory. We have here the
same mechanism as in non abelian affine Toda theories \ct{acfgz,fms}.

The masses of the fields $\zeta_m$ and $\xi_{-m}$, introduced
in \rf{zetaxi} for $1\leq m \leq l-1$, are the same as those of the fields
$F^{\pm}_m$. Indeed, substituting \rf{fpm} in \rf{eqm2}--\rf{eqm3}, and
taking the linear field approximation  with $\eta = \eta_0$, we get
\br
\pa_{+}\pa_{-} \zeta_m &=& v_{\eta} \lb E_{l}\, , \, \lb E_{-l} \, , \,
\zeta_m\rb\rb ,\nonu\\
\pa_{+}\pa_{-} \xi_{-m} &=& v_{\eta} \lb E_{l}\, , \, \lb E_{-l} \, , \,
\xi_{-m}\rb\rb ,
\er
where a trivial integration is performed to eliminate one derivative.

Let us explain now, following the reasonings of \ct{acfgz}
and \ct{fms}, that
the masses of solitons are also generated by the spontaneous breakdown of
the conformal symmetry.  The energy of  classical solutions are given by the
space integral of the $(0,0)$ component of energy--momentum tensor
$L_{\mu\nu}^{\rm red.}$ defined in \rf{lpp}--\rf{lmm}. In the Lorentz frame
where the classical soliton  solution is static, the energy should be
interpreted as the mass of the soliton. However, since the  theory is
conformally
invariant, it has no mass scale, and the soliton mass  should vanish. When the
conformal symmetry is spontaneously broken by choosing  a particular constant
solution for the field $\eta$,  we obtain a massive theory. Construct the
energy--momentum tensor of such a theory as follows. Clearly, the tensor
$\Theta_{\mu\nu}$, introduced in \rf{teta} and evaluated at any classical
solution, satisfies \rf{tetacons}. Therefore, the tensor defined by
\be
\Theta^{\rm broken}_{\mu\nu} \equiv \Theta_{\mu\nu}\mid_{\eta ={\rm constant}}
\, ,
\ee
is symmetric and conserved,
\be
\pa^{\mu}  \Theta^{\rm broken}_{\mu\nu} =0 \, ,
\ee
since $\eta ={\rm constant}$ is a solution of the equations of motion. Then,
let the energy in the massive theory be proportional to the space
integral of $\Theta^{\rm broken}_{00}$. Using \rf{smunu} and \rf{teta}, we
obtain the soliton mass in the form
\br
{M \o{\sqrt{1-v^2}}} &\equiv& -\(\int_{-\infty}^{\infty} dx \, \Theta^{\rm
broken}_{00} - E_{\rm vac.} \) = -{k\, N_{\bf s}\o l} \pa_x \(\nu + \Omega
x_{+}x_{-}\) \mid_{-\infty}^{\infty}\, ,
\lab{solmass}
\er
because the integral of $L_{00}^{\rm red.}$ vanishes by the above arguments.
Here
$v$ is the soliton velocity in the units of the speed of the light. Notice
that we have subtracted the energy $E_{\rm vac.}$ of the vacuum solution which
is, in fact,  divergent. Of course, the vacuum solution is not unique, and it
is not  clear which one provides the absolute minimum of the energy. We will
use the following prescription for the soliton mass formula. For the
soliton solutions lying, under the dressing transformations, on the orbit of
the vacuum solution \rf{vacuum2}, we take $\Omega$ in \rf{solmass} to be that
one given in \rf{omega}. However, for those soliton solutions lying on the
orbit of the vacuum \rf{vacuum1}, we take $\Omega$ in \rf{solmass} to be
equal to the parameter $\b$
introduced in \rf{vaccond}. Such a prescription guarantees the finiteness of
the soliton masses.

The soliton masses are determined solely by the behaviour at $x=\pm
\infty$ of the space derivative of the field $\nu$. That is quite a remarkable
fact. In addition, as we now explain, it is very easy to obtain
such a behaviour in the general case from the solitonic solutions \rf{mu0} or
\rf{nicesoliton}.

Consider the integrable highest weight representation of $\cgh$ with highest
weight state $\mid\l_{\bf s}\rangle$, satisfying the relations \ct{kac1,fms}
\be
H_a^0\, \mid\l_{\bf s}\rangle = s_a \, \mid\l_{\bf s}\rangle \, , \quad
C \,\mid\l_{\bf s}\rangle = {\psi^2 \o 2}\( \sum_{i=0}^{r} l_i^{\psi} s_i\) \,
\mid\l_{\bf s}\rangle \, , \quad
f_i \, \mid\l_{\bf s}\rangle = 0 \, , \; \mbox{\rm if $s_i=0$};
\lab{integrable}
\ee
cf. \rf{alpha}, where $f_a \equiv E_{-\a_a}^0$, $a=1,2,\ldots r$, $f_0 \equiv
E_{\psi}^{-1}$,
and  $s_i$ are the co-prime integers labeling a given integral representation
of $\cgh$ \rf{gradop}, $l_i^{\psi}$ are the integers in the expansion ${\psi \o
{\psi^2}} = \sum_{a=1}^r l_a^{\psi} {\a_a\o {\a_a^2}}$ and $l_0^{\psi}=1$. Such
a highest weight state can be realised by
\be
\mid\l_{\bf s}\rangle =
\bigotimes_{i=0}^{r} \,  \mid {\hat{\l}}_i\rangle^{\oplus s_i},
\ee
where $\mid {\hat{\l}}_i\rangle$ are the highest weight states of the
fundamental representations of $\cgh$, and ${\hat{\l}}_i$ are the
corresponding fundamental weights of $\cgh$.

Consider now an integral gradation of $\cgh$, with $s^{\pr}_i = {\psi^2
\o{\a_i^2}}\, s_i$, $\alpha_0\equiv - \psi$, and $s_i$ labeling the gradation
that
defines the model \rf{eqm1}--\rf{eqm4}.  Therefore, it follows that the Cartan
 generators of the special basis introduced in
\rf{cartanplus}--\rf{bilbasis1} provide
\be
\widetilde{H}_{a}^0\, \mid\l_{\bf s^{\pr}}\rangle = 0.
\ee
 From \rf{integrable} one has that $\mid\l_{\bf s^{\pr}}\rangle$ is annihilated
by all negative root step operators with ${\bf s^{\pr}}$-zero grade, and,
consequently, with ${\bf s}$-zero grade too. Since it is a highest weight
state,
$\mid\l_{\bf s^{\pr}}\rangle$ is annihilated by all generators of the subgroup
$G_0$. Therefore, taking the expectation value of both sides of \rf{nice} in
such state, one gets (with the gauge choice $h_L(x_{-})=h_R(x_{+})=1$)
\be
e^{-\( \nu + \Omega x_{+} x_{-}\) N_{\bf s} {\psi^2\o 2}} = \langle \l_{\bf
s^{\pr}} \mid e^{x_{+}\, \ce_{+}} \, e^{-x_{-}\, \ce_{-}} \, \rho \, e^{x_{-}\,
\ce_{-}}\, e^{-x_{+}\, \ce_{+}} \mid \l_{\bf s^{\pr}} \rangle .
\ee
Now, choosing $\rho$ to be the exponential of an eigenvector of $\ce_{\pm}$,
\be
\lb \ce_{\pm} \, , \, V \rb = \omega_{\pm} V,
\lab{eigenvalue}
\ee
we obtain a soliton solution
\be
e^{-\( \nu + \Omega x_{+} x_{-}\) N_{\bf s} {\psi^2\o 2}} = \langle
\l_{{\bf s}^{\pr}}  \mid e^{e^{\Gamma} \, V} \mid \l_{{\bf s}^{\pr}} \rangle
\ee
with $\Gamma = \omega_{+} x_{+} - \omega_{-} x_{-} \equiv \gamma \( x - vt\)$.

Suppose $V$ is an operator in such a representation for wich  there is
a positive integer $N_V^{\pr}$, such that
\be
\langle \l_{{\bf s}^{\pr}} \mid V^n \mid \l_{{\bf s}^{\pr}} \rangle =0
\qquad \mbox{\rm for $n>N_V^{\pr}$}.
\lab{truncation}
\ee
Then the soliton mass is easily obtained from \rf{solmass},
where for $\gamma > 0$
($\gamma < 0$) only the upper (lower) limit $x=\infty$ ($x=-\infty$)
contributes in the integral\footnotemark{\footnotetext{We point out that the
soliton mass formula \rf{solitonmass} could be equally obtained by defining the
mass  through the momentum formula, instead through the energy like
in \rf{solmass}, as ${M\, v\o{\sqrt{1-v^2}}} \equiv \int \, dx \Theta_{01}^{\rm
broken}$. In this case, we do not have to subtract the vacuum momentum, since
it vanishes.}} ,
\be
M = {2\o {\psi^2}}{k\, N_V^{\pr}\o l} \mid \gamma \mid {\sqrt{1-v^2}}=
 {2\o {\psi^2}}\, {2\,k\,\,N_V^{\pr}\o l} \, \sqrt{\omega_{+}\, \omega_{-}}.
\lab{solitonmass}
\ee

Notice that we must have $\omega_{+}\omega_{-}>0$ in order to have the soliton
velocity $v = (\omega_{-}-\omega_{+})/(\omega_{-}+\omega_{+})$,
not  exceeding the light velocity ($c=1$).

The soliton mass formula \rf{solitonmass} has some remarkable properties. One
of them concerns the relation  particle--soliton in the theory,
indicating some sort of duality similar to the electromagnetic duality of some
four dimensional gauge theories possessing the Bogomolny (monopole) limit
\ct{duality}.
As we have seen, the soliton solutions are created by the eigenvectors $V$ of
$\ce_{\pm}$. From \rf{eigenvalue} one has
$\lb \ce_{+} \, , \,\lb \ce_{-} \, , \, V \rb\rb  = \omega_{+}\omega_{-} V$.
Expanding $V$ over the eigenvectors of the grading operator $Q_{\bf s}$ as
$V = \sum_n V^{(n)}$, one observes that
$\lb \ce_{+} \, , \,\lb \ce_{-} \, , \, V^{(n)} \rb\rb  = \omega_{+}\omega_{-}
V^{(n)}$. Therefore, if some $V^{(n)}\in \cgh_n$, $n=0,\pm 1,\pm 2,\ldots \pm
(l-1)$, does not vanish, it implies that $V^{(n)}$ must be one of
the eigenvectors $X$ in \rf{fundeigen}. Then we associate a soliton with a
fundamental particle. In addition, we have $\lambda
\equiv \omega_{+}\omega_{-}$, and, consequently, from \rf{massf} and
\rf{solitonmass}, the masses of the corresponding soliton and fundamental
particle are determined by  the same eigenvalue. In fact, we have from
\rf{massf} and \rf{solitonmass}, with $v_{\eta}=1$, that
\be
M_{\rm sol.} =  {2\o {\psi^2}}\, {k\,\,N_V^{\pr}\o l}\, m_{\l}^{\rm part.}.
\ee
Of course, in the expansion of $V$, we may have more than one non
vanishing $V^{(n)}$, with $n=0,\pm 1,\pm 2,\ldots \pm (l-1)$. Then we would
associate a one--soliton solution to more than one fundamental particle. The
counting of one--soliton solutions has to be better analysed in each particular
case. We discuss this issue in the section devoted to the examples.

\sect{Physical properties of the higher grading fields}
\lab{sec:spinors}

It is clear from \rf{ctf1}-\rf{ctf5}, that the
 massive fields associated with non
vanishing grade (namely $F^{\pm}_m$),  are chiral fields with
non vanishing spins, in contrast with the Toda type fields.
In fact, we show that the free equations for such fields take the form of
the massive Dirac equation, as could be expected from
general
covariance arguments.

Consider the subspace $\cgh_m$ for $0<m<l$. Let $\cgh_m^{(F)}$ be the subspace
of $\cgh_m$, generated by the eigenvectors of $\lb E_{-l}\, , \, \lb E_{l}\, ,
\, \cdot \rb \rb$ with non zero eigenvalues, i.e.,
\be
\cgh_m^{(F)} \equiv \{ T^{(m)} \in \cgh_m  \mid  \l^{(m)} \neq 0 \} ,
\ee
where $\l^{(m)}$ is defined as
\be
\lb E_{-l}\, , \, \lb E_{l}\, , \, T^{(m)} \rb \rb =
\lb E_{l}\, , \, \lb E_{-l}\, , \, T^{(m)} \rb \rb = \l^{(m)} \, T^{(m)}.
\ee
Decompose the subspace $\cgh_m$, as a vector space, into the sum
\be
\cgh_m = \cgh_m^{(F)} + \cgh_m^{(K)},
\ee
where $\cgh_m^{(K)}$ is the complement of $\cgh_m^{(F)}$ in $\cgh_m$.
Define now
\be
T^{(-l+m)} \equiv \lb E_{-l}\, , \, T^{(m)} \rb \, \in \, \cgh_{-l+m}.
\ee
Since $\l^{(m)} \neq 0$, it follows that
$T^{(-l+m)}$ is necessarily non vanishing.
Therefore, we have
\br
 \lb E_{l}\, , \, \lb E_{-l}\, , \, T^{(-l+m)} \rb \rb &=&
\lb E_{-l}\, , \, \lb E_{l}\, , \, T^{(-l+m)} \rb \rb\nonu\\
&=& \lb E_{-l}\, , \, \lb E_{l}\, , \, \lb E_{-l} \, , \, T^{(m)} \rb \rb\rb
= \l^{(m)} \, \lb E_{-l} \, , \, T^{(m)} \rb  = \l^{(m)} \,  T^{(-l+m)}.
\er
So, whenever we have an eigenvector in $\cgh_m$ with eigenvalue
$\l^{(m)}\neq 0$, we also have a corresponding eigenvector in $\cgh_{-l+m}$
with the same eigenvalue. Notice that if $\l^{(m)}$ is degenerate, then the
corresponding eigenvectors are mapped bijectively. Suppose $T^{(m)}_1$ and
$T^{(m)}_2$ have the same eigenvalue, and that they are mapped onto the same
element in $\cgh_{-l+m}$, i.e.,
\be
\lb E_{-l}\, , \, T^{(m)}_1 \rb = \lb E_{-l}\, , \, T^{(m)}_2 \rb \, ,
\quad \mbox{\rm and so} \quad
\lb E_{-l}\, , \, T^{(m)}_1 - T^{(m)}_2  \rb = 0.
\ee
However, this relation is in contradiction with the following ones:
\be
\lb E_{l}\, , \, \lb E_{-l}\, , \, T^{(m)}_1  - T^{(m)}_2  \rb \rb =
\l^{(m)} \, \( T^{(m)}_1  - T^{(m)}_2  \), \quad \mbox{\rm and} \quad
T^{(m)}_1  - T^{(m)}_2 \neq 0 \, ; \quad \l^{(m)} \neq 0.
\ee
Analogously, we write
\be
\cgh_{-l+m} = \cgh_{-l+m}^{(F)} + \cgh_{-l+m}^{(K)},
\ee
with
\be
\cgh_{-l+m}^{(F)} \equiv \{ T^{(-l+m)} \in \cgh_{-l+m}  \mid  \l^{(m)}
\neq 0 \} ,
\ee
and $\cgh_{-l+m}^{(K)}$ being the complement of $\cgh_{-l+m}^{(F)}$ in
$\cgh_{-l+m}$.
Therefore, from the considerations given above, we conclude that the subspaces
$\cgh_{-l+m}^{(F)}$ and $\cgh_{m}^{(F)}$ are isomorphic. The mapping is given
by the action of $E_{-l}$ on $\cgh_{m}^{(F)}$.
Our arguments are applied equally well in the reversed direction,
the mapping can also be given by the action of $E_{l}$ on
$\cgh_{-l+m}^{(F)}$; and hence
\br
\lb E_{-l} \, , \, \cgh_m^{(F)} \rb &=& \cgh_{-l+m}^{(F)} \, \, , \qquad
\lb E_{-l} \, , \, \cgh_m^{(F)} \rb \cap \cgh_{-l+m}^{(K)} = \{\emptyset\} ,\\
\lb E_{l} \, , \, \cgh_{-l+m}^{(F)} \rb &=& \cgh_m^{(F)} \, \, , \qquad
\lb E_{l} \, , \, \cgh_{-l+m}^{(F)} \rb \cap \cgh_m^{(K)} = \{\emptyset\} .
\er
One has
\be
\lb E_{-l} \, , \, \cgh_m^{(K)} \rb \cap \cgh_{-l+m}^{(F)} = \{\emptyset\}
 \, , \qquad
\lb E_{l} \, , \, \cgh_{-l+m}^{(K)} \rb \cap \cgh_{m}^{(F)} = \{\emptyset\} ,
\ee
because, otherwise, it would be in a contradiction with the relations
\be
\lb E_{l} \, , \, \lb E_{-l} \, , \, \cgh_m^{(K)} \rb\rb =0 \, , \qquad
\lb E_{l} \, , \, \lb E_{-l} \, , \, \cgh_{-l+m}^{(K)} \rb\rb =0.
\lab{consist}
\ee
Notice that $\cgh_{m}$ ($\cgh_{-l+m}$) does not lie necessarily in the
kernel of $E_{-l}$ ($E_{l}$). However, if a given element $J^{(m)}$ of
$\cgh_m^{(K)}$ is in the image of $E_{l}$;
then it follows from \rf{consist} that it must be
in the kernel of $E_{-l}$,
i.e.,
\be
J^{(m)} = \lb E_{l} \, , \, J^{(-l+m)} \rb \Longrightarrow
\lb E_{-l} \, , \, J^{(m)} \rb = 0 \, \, , \qquad
J^{(n)} \in \cgh_n^{(K)}.
\ee
Analogously,
\be J^{(-l+m)} = \lb E_{-l} \, , \, J^{(m)} \rb \Longrightarrow
\lb E_{l} \, , \, J^{(-l+m)} \rb = 0 \, \, , \qquad  J^{(n)} \in \cgh_n^{(K)}.
\ee
In the same way as in section \ref{sec:masses}, we consider the linear
approximation where  $\eta=\eta_0$, $b=1$, $\nu=0$ (recall equation \rf{bdef}).
This gives  the free part of
equations \rf{em2} and \rf{em3},
\be
\pa_{-} F^{+}_{m} = e^{(l-m)\eta_0}
\lb E_{l} \, ,   \, F^{-}_{l-m} \rb  ,\qquad
\pa_{+} F^{-}_{l-m} = -e^{m\eta_0}
\lb E_{-l} \, ,  \,  F^{+}_{m} \rb .
\lab{linear1}
\ee
Denote the generators of $\cgh_{m}^{(F)}$ and $\cgh_{-l+m}^{(F)}$,
corresponding to the eigenvalue $\l^{(m)}\neq 0$, as
$T^{(m)}_{\l^{(m)},i}$ and $T^{(-l+m)}_{\l^{(m)},i}$, respectively, where the
index $i$ stands for a possible degeneracy of $\l^{(m)}$. The basis is chosen
in such a way that
\be
\lb E_{-l}\, , \, T^{(m)}_{\l^{(m)},i} \rb = \sqrt{\l^{(m)}} \,
T^{(-l+m)}_{\l^{(m)},i}, \qquad
\lb E_{l}\, , \, T^{(-l+m)}_{\l^{(m)},i} \rb = \sqrt{\l^{(m)}} \,
T^{(m)}_{\l^{(m)},i}.
\lab{basisdef}
\ee
We also use the notations
\be
F^{+}_{m} =  F^{{+},(F)}_{m} +  F^{{+},(K)}_{m}\, , \qquad
F^{-}_{l-m} =  F^{{-},(F)}_{l-m} +  F^{{-},(K)}_{l-m},
\ee
with
\be
F^{{+},(F/K)}_{m} \in \cgh_m^{(F/K)}\, , \qquad
F^{{-},(F/K)}_{l-m} \in \cgh_{-l+m}^{(F/K)};
\ee
and
\be
F^{{+},(F)}_{m} = \sum_{\l^{(m)},i} \psi_R^{\l^{(m)},i}\,\,
T^{(m)}_{\l^{(m)},i},\quad
F^{{-},(F)}_{l-m} = \sum_{\l^{(m)},i} \psi_L^{\l^{(m)},i}\,\,
T^{(-l+m)}_{\l^{(m)},i}.
\lab{psidef}
\ee
Since the action of $E_{\pm l}$ does not mix the subspaces of indices $F$
and $K$, we can split  equations \rf{linear1} to get
\be
\pa_{-} \psi_R^{\l^{(m)},i} =
e^{(l-m)\eta_0} \sqrt{\l^{(m)}} \, \psi_L^{\l^{(m)},i},\quad
\pa_{+} \psi_L^{\l^{(m)},i} =
-e^{m\eta_0}\sqrt{\l^{(m)}} \, \psi_R^{\l^{(m)},i}.
\lab{predirac}
\ee
Introduce
\br
\psi^{\l^{(m)},i} = \(
\begin{array}{r}
\psi_R^{\l^{(m)},i}\\
\psi_L^{\l^{(m)},i}
\end{array}\),
\lab{psidef2}
\er
and
\br
\gamma_0 = -i \(
\begin{array}{rr} 0&-1\\ 1&0
\end{array}\) , \qquad
\gamma_1 = -i \(
\begin{array}{rr}  0&1\\  1&0
\end{array}\),
\lab{gammas}
\er
satisfying anticommutation relations
\br
\{ \gamma_{\mu}\, , \, \gamma_{\nu}\} = 2 g_{\mu\nu} \one ,
\er
with $g_{00}=-g_{11}=1$, $g_{01}=0$.
Therefore,  equations \rf{predirac} can be written as  the Dirac type
equations
\be
i\gamma^{\mu} \pa_{\mu} \psi^{\l^{(m)},i} =
m_{\l^{(m)}}\, \left [ {1+\gamma_5 \over 2} e^{m\eta_0}
+ {1-\gamma_5 \over 2} e^{(l-m) \eta_0} \right ]  \psi^{\l^{(m)},i},
\lab{dirac}
\ee
where, following \rf{massf}, we put
\be
m_{\l^{(m)}} \equiv 2 \sqrt{\l^{(m)}}.
\ee
Hence, the massive degrees of freedom of the model, corresponding to
generators of grade $\pm m$ with $0<m<l$, are, in fact, Dirac  fields.
However, in general, the mass term involves a $\gamma_5$ term, so that
parity in violated. A noticeable exception is when $l=2$, $m=1$.

Notice that $\psi_{R/L}^{\l^{(m)},i}$ are the projections of
$\psi^{\l^{(m)},i}$ under ${\( 1 \pm \gamma_5 \)\o 2}$, with $\gamma_5 \equiv
\gamma_0\, \gamma_1$.
Under the gauge transformations \rf{gs2}--\rf{gs3},
$\psi_{R}^{\l^{(m)},i}$ and $\psi_{L}^{\l^{(m)},i}$ are transformed
independently
under the action of $h_R \( x_{+}\)$ and  $h_L \( x_{-}\)$, respectively.

At this point we have seen that the mapping between $\cgh_m$ and
$\cgh_{-l+m}$ displayed at the beginning of the section precisely ensures the
existence of both
chirality components of the Dirac fields. The next question is whether
we may write a free action.
Suppressing indices for a while,
a   free field actions  would  take the form
\be
{\cal L}=i \overline \psi \gamma^\mu\partial_\mu \psi
-\overline \psi (a+b\gamma_5)\psi .
\lab{freeL}
\ee
 Writing
\be
\psi=\left ( \begin{array}{c} \psi_R \\ \psi_L\end{array} \right), \quad
\overline \psi = (\widetilde \psi_R,  \> \widetilde \psi_L)\gamma_0,
\ee
we obtain
\be
{\cal L}/i=2 \left( \widetilde  \psi_R \partial_- \psi_R
 +\widetilde  \psi_L \partial_+ \psi_L\right )
 -m_R \widetilde  \psi_L  \psi_R
 +m_L \widetilde \psi_R  \psi_L
\lab{freeL2}
\ee
where $m_R=a+b$, $m_L=a-b$. Next,  equation \rf{dirac} shows that the masses
vanish  in the limit $\eta_0\to \infty$. Then, we may discuss  the conformal
properties of the $\psi$ fields in the usual language, where a field
$A(z_+, z_-)$ is primary with weights $\Delta_+, \Delta_-$ if
$A(dz_+)^{\Delta_+}  (dz_-)^{\Delta_-}$
is invariant by conformal transformations.
The kinetic term of equation \rf{freeL2} is conformal invariant if we have
the following weight assignements.
\be
\begin{array}{c} \> \\ \Delta_+ \\ \Delta_- \end{array} \qquad
\begin{array}{llll} \psi_R & \widetilde \psi_R& \psi_L & \widetilde \psi_L \\
J_R \>  & 1-J_R & 0 \>  & 0\> \\
0\>  & 0 \>  & J_L \>  & 1-J_L \end{array}
\ee
Consider next the mass term.
With the Lorentz transformation  \rf{ltf}, $x_{\pm} \ra \d^{\mp 1}
x_{\pm}$, $ \widetilde  \psi_L  \psi_R$ and $\widetilde \psi_L  \psi_R$ will
be invariant if $J_L+J_R=1$.  Returning to our Dirac fields, one finds the
identification
\be
\left \{ \begin{array}{ll}
\psi_R\Rightarrow  \psi_R^{\l^{(m)},i}, \quad &
\psi_L\Rightarrow  \psi_L^{\l^{(m)},j}, \\
\widetilde \psi_R\Rightarrow  \psi_R^{\l^{(l-m)},k}, \quad &
\widetilde  \psi_L\Rightarrow  \psi_L^{\l^{(l-m)},\ell	},
\end{array} \right.
\lab{assign2}
\ee
for some choice of $i, j, k, \ell$. Thus the fields of type $R$ ($L$) are to
be split between $\psi$ and $\widetilde \psi$ type fields.
The spectrum of weights is just right to form the needed
quadruplets provided the
number of $R$ ($L$) fields is even. This is not always true as we shall see.
Another  difficulty is that
 in general $\widetilde \psi$ is not the complex conjugate
of $\psi$. Thus the above free action is not real.
This was already the case for the usual affine
Toda theories beyond sine--Gordon, and we may expect that this will not be
a problem. Another concern is the statistics. At first sight   our $\psi$
fields
are c-number fields so that  they seem to describe bosons. However, it is well
known that in two dimensions the statistics of fields depends upon the
coupling
constant. Perhaps the latter
  should be fixed so that the $\psi$ fields
become anticommuting operators. We will have more to say on this below.

\sect{A special class of models}
\label{sec:special}
We now describe a class of models possessing a $U(1)$ Noether current,
which, under some circunstances, is proportional
 to a topological current. That
 occurs for those models where the grade $l$ of the operator $E_l$, introduced
 in \rf{fp}, is equal to the integer $\ns$ defined in \rf{gradop}.  So,
throughout this section we have $l=\ns$.

The calculations  become simpler if we realize the generators of the
 affine Kac-Moody algebra $\cgh$, in terms of those of the finite simple Lie
algebra $\cg$ as
\be
H^n_a \equiv z^n \, H_a  \, , \qquad E_{\a}^n \equiv z^n \, E_{\a}
\, , \qquad D \equiv z {d \,\, \o dz}
\ee
with $z$ being a complex variable. Then, the Lie bracket and trace form on
$\cgh$ are given by
\be
\lb A(z) \, , \, B(z) \rb = \lb A(z) \, , \, B(z) \rb_{\cg} +
C\,  \oint {dz \o 2\pi i} \Tr \( B(z) {d \,\, \o dz} A(z)\)
\ee
and
\be
\Tr \( A(z) \, B(z) \) = \oint {dz \o 2\pi iz} {\rm tr} \( A(z) \, B(z) \)
\ee
where $\lb \cdot \, , \, \cdot \rb_{\cg}$ and ${\rm tr} \( \cdot \, \cdot \)$
are the Lie bracket and trace form, respectively, on $\cg$.

As a first step, we construct two chiral currents associated to the elements
$z^{-1} E_{\ns}$ and $z E_{-\ns}$ of $\cgh_0$.  Multiply eq. \rf{em2} by
 $z^{-1} F^{+}_{\ns -m}$, sum over $m$ and
take the trace. Then, take the same equation with $m$ replaced by $\ns -m$,
 multiply by $z^{-1} F^{+}_m$, sum over $m$ and take the trace. Add both to
get
\br
\sum_{m=1}^{\ns -1} \pa_{-} \Tr \( z^{-1} F^{+}_m F^{+}_{\ns -m} \) =
-2  \sum_{m=1}^{\ns -1} \Tr \( z^{-1} E_{\ns} \lb F^{+}_m \, , \,
B^{-1} F^{-}_m B \rb \) + X^{+}
\lab{first}
\er
where
\br
X^{+} &=&  \sum_{m=1}^{\ns -2}\sum_{n=1}^{\ns -m-1} \Tr \( z^{-1}
F^{+}_{\ns -m} \lb F^{+}_{n+m} \, , \, B^{-1} F^{-}_n B \rb \) \nonu\\
&+& \sum_{m=2}^{\ns -1}\sum_{n=1}^{m-1} \Tr \( z^{-1}
F^{+}_{m} \lb F^{+}_{\ns +n-m} \, , \, B^{-1} F^{-}_n B \rb \)
\er
Notice, the first sum in $m$ ends at $\ns -2$ and the second starts at $2$,
 because the quadratic term in eq. \rf{em2} does not exist for $m=\ns -1$.

Similarly, multiply \rf{em3} by
$z F^{-}_{\ns -m}$, sum over $m$, and take the trace. Then, take the same
equation with $m$ replaced by $\ns -m$, multiply by $z F^{-}_m$, sum over $m$
 and take the trace. Add them to get
\br
\sum_{m=1}^{\ns -1} \pa_{+} \Tr \( z F^{-}_m F^{-}_{\ns -m} \) =
2  \sum_{m=1}^{\ns -1} \Tr \( z E_{-\ns} \lb F^{-}_m \, , \,
B F^{+}_m B^{-1} \rb \) - X^{-}
\lab{second}
\er
where
\br
X^{-} &=&  \sum_{m=1}^{\ns -2}\sum_{n=1}^{\ns -m-1} \Tr \( z
F^{-}_{\ns -m} \lb F^{-}_{n+m} \, , \, B F^{+}_n B^{-1} \rb \) \nonu\\
&+& \sum_{m=2}^{\ns -1}\sum_{n=1}^{m-1} \Tr \( z
F^{-}_{m} \lb F^{-}_{\ns +n-m} \, , \, B F^{+}_n B^{-1} \rb \)
\er
Using the fact that
\be
\sum_{m=1}^{\ns -2}\sum_{n=1}^{\ns -m-1} =
\sum_{m^{\pr}=2}^{\ns -1}\sum_{n=1}^{m^{\pr}-1}
\, , \qquad \mbox{\rm where $m^{\pr} = m+n$}
\lab{nicesum}
\ee
one gets\footnote{In fact, each double sum in $X^{\pm}$ vanishes
separately. Use that $\sum_{m=2}^{\ns -1}\sum_{n=1}^{m-1} =
\sum_{n=1}^{\ns -2}\sum_{m=n+1}^{\ns -1}$, and notice that the terms for (fixed
$n$) $m=n+k$ and $m=\ns -k$ cancel.}
\be
X^{+} = X^{-} = 0
\ee
Projecting \rf{em1} onto $z^{-1} E_{\ns}$ and then onto $z E_{-\ns}$, and
 comparing with \rf{first} and \rf{second}, one gets\footnote{In the
 projection onto $z^{-1} E_{\ns}$, it is easier to use the equivalent form
 of \rf{em1},
$\pa_{-}\( B^{-1} \pa_{+} B\) = -\lb E_{\ns} \, , \, B^{-1} \, E_{-\ns}\,
B\rb  - \sum_{n=1}^{\ns -1} \lb F^{+}_{n} \, , \, B^{-1} \, F^{-}_{n}\, B\rb$}
\be
\pa_{-} {\cal J} = 0 \, , \qquad
\pa_{+} {\bar {\cal J}} = 0
\lab{rcons}
\ee
where
\br
{\cal J}(x_{+}) &=& \Tr \( z^{-1} E_{\ns} B^{-1} \pa_{+} B \) -
\h \sum_{m=1}^{\ns -1}  \Tr \( z^{-1} F^{+}_m F^{+}_{\ns -m} \)
\nonu\\
{\bar {\cal J}}(x_{-}) &=& \Tr \( z E_{-\ns} \pa_{-} B B^{-1} \) -
\h \sum_{m=1}^{\ns -1}  \Tr \( z F^{-}_m F^{-}_{\ns -m} \)
\lab{curr}
\er

We now consider those models where $z^{-1}E_{\ns}$ and $z E_{-\ns}$ are
 parallel, and lie in the center of $\cgh_0$, i.e.\footnote{Obviously, $E_0$
can not have any component in the direction of the central term $C$ or
 the derivation $D$, since
it is the projection of elements belonging to $\cgh_{\pm \ns}$.}
\be
z E_{-\ns} = \mu z^{-1} E_{\ns} \equiv E_0 \in \mbox{\rm center of $\cgh_0$}
\lab{parallel}
\ee
where $\mu$ is some constant independent of $z$.
Using such condition one observes, from \rf{first} and
\rf{second}, that the
current\footnote{We use symbols with a   tilde
 since the currents we are defining right now are not
yet properly normalised  (see below).}
\br
{\tilde J}_{+} = -{\mu \o 2}\sum_{m=1}^{\ns -1}
\Tr \( z^{-1} F^{+}_m F^{+}_{\ns -m}\)
\, , \qquad
{\tilde J}_{-} = \h \sum_{m=1}^{\ns -1}  \Tr \( z F^{-}_m F^{-}_{\ns -m} \)
\lab{ncurr}
\er
is conserved
\be
\pa_{\mu} {\tilde J}^{\mu} =0
\ee
In addition, the condition \rf{parallel} implies that the current
\br
{\tilde \jmath}_{+} = -  \Tr \( E_{0}\, B^{-1} \pa_{+} B \) \, , \qquad
{\tilde  \jmath}_{-} =  \Tr \( E_{0}\, \pa_{-} B B^{-1} \)
\lab{tcurr}
\er
is a topological current, i.e. it is conserved
independently of the equations
of motion
\be
\pa_{\mu} {\tilde  \jmath}^{\mu} =0
\ee
We now come to a very interesting property of these models. Under the
conformal
transformations \rf{ct}-\rf{ctf5}, the chiral currents \rf{curr} transform as
\br
{\cal J}(x_{+}) &\ra & \( f^{\pr}(x_{+})\)^{-1} \( {\cal J}(x_{+}) -
{1\o {\mu \ns}} \Tr \( E_0 \, Q_{\bf s}\) \( \ln f^{\pr}(x_{+})\)^{\pr}
\) \nonu\\
{\bar {\cal J}}(x_{-}) &\ra & \( g^{\pr}(x_{-})\)^{-1}
\( {\bar {\cal J}}(x_{-})
 - {1\o {\ns}} \Tr \( E_0 \, Q_{\bf s}\)
\( \ln g^{\pr}(x_{-})\)^{\pr} \)
\lab{confcurr}
\er
Therefore, if
\be
\Tr \( E_0 \, Q_{\bf s}\) \neq 0
\lab{nonzerotr}
\ee
one concludes that, given a solution of the model, one can always map it,
under
a conformal transformation, into a solution where\footnote{The exceptions
occur
when ${\cal J}(x_{+})$ and/or ${\bar {\cal J}}(x_{-})$ present singularities.}
\be
{\cal J}(x_{+}) =  {\bar {\cal J}}(x_{-}) = 0
\lab{zerocur}
\ee
Such a procedure amounts to a gauge fixing of the conformal symmetry. We are
choosing one solution in each orbit of the conformal group.
Another way of saying it, is that \rf{zerocur} are constraints and we are
performing a Hamiltonian reduction. The degree of freedom eliminated
 corresponds to the field $\eta$.
So, in the submodel defined by \rf{zerocur}
one observes from \rf{curr}, that
the Noether and topological currents \rf{ncurr} and \rf{tcurr} are
equal
\be
{\tilde J}_{\mu} ={\tilde  \jmath}_{\mu}
\lab{Noetopequiv}
\ee
as a result of the field equations.
As we will see on explicit examples, the true Noether
and topological currents
differ from the one just defined by overall constants:
\be
J_{\mu}=c_{\rm Noether} {\tilde J}_\mu, \quad
j_{\mu}=c_{\rm topol.} {\tilde \jmath}_\mu
\ee
Thus the topologican and Noether currents are proportional
\be
J_{\mu} ={c_{\rm Noether}\over c_{\rm topol.}} j_{\mu}.
\lab{Noetopprop}
\ee
This is a very important property of such models, which can
lead, in some cases
as we will discuss below, to the confinement of the matter fields.
In general, that is if ${\cal J}\not=0$, or
${\bar {\cal J}}\not=0$,   one gets
from eqs.\rf{curr}, \rf{ncurr} and \rf{tcurr},
\be
\mu {\cal J} = {\tilde J}_{+} - {\tilde \jmath}_{+} \, \qquad
{\bar {\cal J}} = - {\tilde J}_{-} + {\tilde \jmath}_{-}
\ee
Therefore the Noether and topological charges
\be
q_{\rm Noether} \equiv \int_{-\infty}^{\infty} dx \, J_t \, \qquad
q_{\rm topol.} \equiv \int_{-\infty}^{\infty} dx \, j_t
\lab{ntcharges}
\ee
satisfy
\be
{q_{\rm Noether}\over c_{\rm Noether}}
 - {q_{\rm topol.}\over c_{\rm topol.}} = \h \( q_R - q_L\)
\ee
where we have denoted
\be
q_R \equiv \mu \int_{-\infty}^{\infty} dx \, {\cal J}(x_{+}) \, ;\qquad
q_L \equiv \int_{-\infty}^{\infty} dx \, {\bar {\cal J}}(x_{-})
\lab{lrcharge}
\ee

We now comment on the relationship  between the masses of particles and
solitons of the theory and symmetries associated to $E_0$.
{}From \rf{parallel} one gets that $E_0$ commutes with $E_{\pm \ns}$ and
therefore it generates a diagonal $U(1)$ global gauge symmetry of the type
described in  \rf{diagonalgauge}, namely
\be
B \ra e^{i\theta\, E_0} \, B \, e^{-i\theta\, E_0} = B \, , \qquad
F^{\pm}_m \ra e^{i\theta\, E_0} \, F^{\pm}_m  \, e^{-i\theta\, E_0}
\lab{u1symm}
\ee
with $\theta = {\rm constant}$.
The charges of the fields associated to such $U(1)$ symmetry are, of course,
given by the eigenvalues of $E_0$ in the subspace  $\cgh_{\rm fields} \equiv
\bigoplus_{n=-\ns +1}^{\ns -1} \, \cgh_n$, i.e.
\be
\lb E_0 \, , \, T \rb = q \, T \, , \qquad T \in \cgh_{\rm fields}
\lab{u1charge}
\ee
The masses of the fields are determined by the eigenvalue of $\lb E_{\ns}\, ,
\lb E_{-\ns} \, , \cdot \rb\rb$ (see \rf{massf}). So, using \rf{parallel}
\be
\lb E_{\ns}\, , \lb E_{-\ns} \, , \, T \rb\rb =
{1\o\mu} \lb E_{0}\, , \lb E_{0} \, , \, T \rb\rb = {q^2\o\mu}\, T
\ee
and so
\be
m_T^2 =  {4\o\mu}\, q^2
\ee
The solitons of such theory are created by the operators
\be
V_T(\zeta ) = \sum_{n=-\infty}^{\infty} \zeta^{-n} T^n
\ee
since they are eigenvectors of
$E_{\pm \ns}$, (see \rf{nice}-\rf{nicesoliton},
$T^n \equiv z^n T$)
\br
\lb E_{\ns} \, , \, V_T(\zeta ) \rb\ &=& {\zeta \o \mu} \, q \, V_T(\zeta )\\
\lb E_{-\ns} \, , \, V_T(\zeta ) \rb\ &=& \zeta^{-1} \, q \, V_T(\zeta )
\er
Therefore, if the expansion of the exponential of $V_T(\zeta )$ truncates, we
 get, from \rf{solitonmass}, that the one-soliton masses are
\be
M_{\rm sol.} \sim {q\o \sqrt{\mu}}
\ee

Therefore, the masses of the fundamental particles and solitons of such theory
 are proportional to the $U(1)$ charge. We have here a situation similar to
four dimensional gauge theories with Higgs in the adjoint and in the BPS limit,
where the masses of particles and monopoles (dyons) are given by
$mass \sim \sqrt{ q_{\rm elect.}^2 + q_{\rm mag.}^2}$. That point has to be
further investigated, because we believe it indicates these systems possess
some sort of duality involving particles and solitons \ct{duality}.

\subsection{The example of the principal gradation}

Some special examples of models which possesses the structures described above
can  be constructed as follows. Let $\cgh$ be any untwisted affine Kac-Moody
algebra  furnished with the principal gradation, where ${\bf s}=(1,1,\ldots
,1)$, and corresponding grading operator, $Q_{{\rm ppal}}$, given by
\rf{gradop} with  $\ns = h \equiv  \mbox{\rm Coxeter number}$.
Therefore
\br
\cgh_0 &=& \{ H_a^0 \, , a=1,2, \ldots r\, ; C ; Q_{{\rm ppal}} \} \nonu\\
\cgh_m &=& \{ E_{\a^{(m)}}^0,  E_{-\a^{(h-m)}}^{1} \} \nonu\\
\cgh_{-m} &=& \{ E_{-\a^{(m)}}^0,  E_{\a^{(h-m)}}^{-1} \}
\er
where $0<m<h$, and $\a^{(m)}$  are positive roots of height $m$, i.e. they
 contain $m$  simple roots in their expansion.
 Following \rf{bdef} we parametrize $B$ as
\be
B = e^{\vp \cdot {\tilde H}^0}\, e^{\nu \, C} \, e^{\eta Q_{{\rm ppal}}} =
e^{\vp \cdot H^0}\, e^{{\tilde \nu} \, C} \, e^{\eta Q_{{\rm ppal}}}
\ee
where ${\tilde H}^0$ is defined in \rf{cartanplus}, and ${\tilde \nu} = \nu -
{2\o h} {\hat \d} \cdot \vp$, with
${\hat \d}= \sum_{a=1}^r {\l_a \o \a_a^2}$, and $\l_a$ being the
fundamental weights of $\cg$.
We then choose $E_{\pm h}$ to be parallel
\be
E_{\pm h} = {\bf m} \cdot H^{\pm 1}
\lab{epmh}
\ee
where we shall choose ${\bf m}$ to be a vector inside the Fundamental Weyl
chamber (FWC), and so ${\bf m}\cdot \a >0$ for $\a >0$. Consequently
\be
E_0 \equiv {\bf m} \cdot H^{0}
\ee
satisfies \rf{parallel} (with $\mu =1$), since
$\lb E_0 \, , \, \cgh_0 \rb =0$.
Since the masses are determined by the eigenvalues of
$\lb E_{h}\, , \lb E_{-h} \, , \cdot \rb\rb$ (see \rf{massf}) we conclude
that the $\vp$'s fields are massless, and the fields in the direction
of the step operators have masses $m_{\a}^2 =  4 ( {\bf m} \cdot \a )^2$.
In addition, since ${\bf m}$ lies inside the FWC, one has $m_{\a} \neq 0$
for any $\a$. Therefore, according to the discussion in section
\rf{sec:spinors}, all the fields
in the direction of the step operators are Dirac fields. Then, following
\rf{basisdef} and \rf{psidef}, we write
\br
F^{+}_m &=& \sum_{\a^{(m)}} \sqrt{im_{\psi^{\a^{(m)}}}}
\psi^{\a^{(m)}}_R \, E_{\a^{(m)}}^0 +
\sum_{\a^{(h-m)}} \sqrt{im_{\psi^{\a^{(h-m)}}}}
{\tilde \psi}^{\a^{(h-m)}}_R \, E_{-\a^{(h-m)}}^1 \nonu\\
F^{-}_{h-m} &=& \sum_{\a^{(m)}} \sqrt{im_{\psi^{\a^{(m)}}}}
\psi^{\a^{(m)}}_L \, E_{\a^{(m)}}^{-1} -
\sum_{\a^{(h-m)}} \sqrt{im_{\psi^{\a^{(h-m)}}}}
{\tilde \psi}^{\a^{(h-m)}}_L \, E_{-\a^{(h-m)}}^0
\er
with $0<m<h$, and where we have denoted
\be
m_{\psi^{\a^{(m)}}} = m_{{\tilde \psi}^{\a^{(m)}}} = 2 {\bf m} \cdot \a^{(m)}
\ee
Consequently, associated to each positive root $\a^{(m)}$, we have two
Dirac fields
\br
\psi^{\a^{(m)}} \equiv \(
\begin{array}{c}
\psi^{\a^{(m)}}_R \\
\psi^{\a^{(m)}}_L
\end{array} \)
\, ; \qquad
{\tilde \psi}^{\a^{(m)}} \equiv \(
\begin{array}{c}
{\tilde \psi}^{\a^{(m)}}_R \\
{\tilde \psi}^{\a^{(m)}}_L
\end{array} \)
\er
Notice such notation is in accordance with \rf{assign2}, since
$\psi^{\a^{(m)}}_R $ and ${\tilde \psi}^{\a^{(m)}}_R$ are in the direction of
$E_{\a^{(m)}}^0 \in \cgh_{m}$ and $E_{-\a^{(m)}}^1 \in \cgh_{h-m}$
respectively. Similarly,
$\psi^{\a^{(m)}}_L $ and ${\tilde \psi}^{\a^{(m)}}_L$ are in the direction of
$E_{\a^{(m)}}^{-1} \in \cgh_{-h+m}$ and $E_{-\a^{(m)}}^0 \in \cgh_{-m}$
respectively.

These systems possess a $\( U(1)_L\)^r \otimes \( U(1)_R\)^r$ gauge symmetry
of the type \rf{gs1}-\rf{gs3}, with
\be
h_L(x_{-}) = e^{i\theta_L(x_{-})\cdot H^0}\, ; \qquad
h_R(x_{+}) = e^{i\theta_R(x_{+})\cdot H^0}
\ee
since these $h_{L/R}$ satisfy \rf{hlr}.
 They also possess a global gauge symmetry of the type \rf{diagonalgauge},
namely
\be
\vp \ra \vp \, ; \qquad \nu \ra \nu \, ; \qquad \eta \ra \eta \, ; \qquad
F^{\pm}_m \ra e^{i\theta \cdot H^0} \, F^{\pm}_m e^{-i\theta \cdot H^0}
\ee
with $\theta = {\rm constant}$.
 Notice that
the charges of the fields with respect to $U(1)$ global symmetry
\rf{u1symm} are (see \rf{u1charge})
\be
q_{\vp}=q_{\eta}=q_{{\tilde \nu}} = 0 \, ; \qquad
q_{\psi^{\a^{(m)}}} = - q_{{\tilde \psi}^{\a^{(m)}}} = {\bf m}\cdot \a^{(m)}
\ee
The equations of motion for these systems, obtained from \rf{eqm1}-\rf{eqm4},
are
\be
\pa^2 \vp =
-4 \sum_{m=1}^{h-1} \sum_{\a^{(m)}} {2 \a^{(m)}\o {\a^{(m)}}^2}
m_{\psi^{\a^{(m)}}} {\bar \psi}^{\a^{(m)}} \,
e^{\gamma_5 \( \a^{(m)} \cdot \vp + m \eta \)}
\( {1+\gamma_5 \o 2} - e^{h\, \eta} {\( 1-\gamma_5\)\o 2} \) \psi^{\a^{(m)}}
\lab{motion1}
\ee
\be
\pa^2 {\widetilde \nu}  =
-2 \sum_{m=1}^{h-1} \sum_{\a^{(m)}} {2 \o {\a^{(m)}}^2}
m_{\psi^{\a^{(m)}}}\, e^{h\, \eta}\,  {\bar \psi}^{\a^{(m)}} \,
e^{\gamma_5 \( \a^{(m)} \cdot \vp + m \eta \)}
\( 1-\gamma_5\)  \psi^{\a^{(m)}} - 4{\bf m}^2 \, e^{h\, \eta}
\lab{motion2}
\ee
\br
i\gamma^{\mu}\pa_{\mu} \psi^{\a^{(m)}}& =& m_{\psi^{\a^{(m)}}} \,
e^{\gamma_5 \( \a^{(m)} \cdot \vp + m \eta \)} \,
\({\( 1+\gamma_5\) \o 2} + e^{h\, \eta}
{\( 1-\gamma_5\)\o 2}\)\psi^{\a^{(m)}}
+ U^{\a^{(m)}}
\lab{motion3}\\
i\gamma^{\mu}\pa_{\mu} {\tilde \psi}^{\a^{(m)}} &=& m_{\psi^{\a^{(m)}}} \,
e^{-\gamma_5 \( \a^{(m)} \cdot \vp + m \eta \)} \,
\( e^{h\, \eta} {\( 1+\gamma_5\) \o 2} + {\( 1-\gamma_5\)\o 2} \)
{\tilde \psi}^{\a^{(m)}} + {\tilde U}^{\a^{(m)}}
\\
\pa^2 \eta &=& 0
\lab{motion4}
\er
where the gamma matrices are defined in \rf{gammas},
${\bar \psi}^{\a^{(m)}} \equiv \({\tilde \psi}^{\a^{(m)}}\)^T
\gamma_0$, and\footnote{Notice that we are not assuming
${\tilde \psi}^{\a^{(m)}} = {\psi^{\a^{(m)}}}^*$}
\br
U^{\a^{(m)}} = \(
\begin{array}{c}
U^{\a^{(m)}}_R\\
U^{\a^{(m)}}_L
\end{array} \) \, ; \qquad
{\tilde U}^{\a^{(m)}} = \(
\begin{array}{c}
{\tilde U}^{\a^{(m)}}_R\\
{\tilde U}^{\a^{(m)}}_L
\end{array} \)
\er
with\footnote{We use the following normalization for the trace form,
$\Tr \( x\cdot H^n y \cdot H^{-n}\) = x \cdot y$, and
$\Tr \( E_{\a}^n E_{-\a}^{-n}\) = {2 \o \a^2}$.}
\br
U^{\a^{(m)}}_R &=& - {{\a^{(m)}}^2 \o \sqrt{i m_{\psi^{\a^{(m)}}}}}
\sum_{n=1}^{m-1} e^{n\, \eta} \Tr \( E_{-\a^{(m)}}^1 \lb F^{-}_{h-m+n} \, , \,
e^{\vp \cdot H^0} F^{+}_n e^{-\vp \cdot H^0}\rb \)
\nonu\\
U^{\a^{(m)}}_L &=&  {{\a^{(m)}}^2 \o \sqrt{i m_{\psi^{\a^{(m)}}}}}
\sum_{n=1}^{h-m-1} e^{n\, \eta} \Tr \( E_{-\a^{(m)}}^0 \lb F^{+}_{m+n} \, , \,
e^{-\vp \cdot H^0} F^{-}_n e^{\vp \cdot H^0}\rb \)
\nonu\\
{\tilde U}^{\a^{(m)}}_R &=&  {{\a^{(m)}}^2 \o \sqrt{i m_{\psi^{\a^{(m)}}}}}
\sum_{n=1}^{h-m-1} e^{n\, \eta} \Tr \( E_{\a^{(m)}}^0 \lb F^{-}_{m+n} \, , \,
e^{\vp \cdot H^0} F^{+}_n e^{-\vp \cdot H^0}\rb \)
\nonu\\
{\tilde U}^{\a^{(m)}}_L &=&  {{\a^{(m)}}^2 \o \sqrt{i m_{\psi^{\a^{(m)}}}}}
\sum_{n=1}^{m-1} e^{n\, \eta} \Tr \( E_{\a^{(m)}}^{-1} \lb F^{+}_{h-m+n}
\, , \, e^{-\vp \cdot H^0} F^{-}_n e^{\vp \cdot H^0}\rb \)
\er
Since these systems satisfy \rf{parallel}, they possess
the conserved Noether
and topological currents \rf{ncurr} and \rf{tcurr}, respectively.
In order to correctly define these currents, we have to be more
precise about normalizations. First, concerning the topological current,
the potential terms in Eqs.\ref{motion1}--\ref{motion4}  involve $\vp$
only through
$\exp(\pm \alpha^{(m)}.\vp)$. We see that the
potential is invariant under
\be
\vp \ra \vp + 2 \pi i \mu^{v}
\lab{symphi}
\ee
where $\mu^{v}$ is a co-weight of $\cg$, i.e. $\mu^{v}= \sum_{n_a\in \IZ}
n_a {2 \l_a\o \a_a^2}$, with $\l_a$ being the fundamental weights of $\cg$,
satisfying ${2 \l_a \cdot \a_b\o \a_a^2} = \d_{ab}$, and $\a_a$ being the
simple roots of $\cg$. Therefore, $\mu^{v} \cdot \a$ is a integer for any root
$\a$ of $\cg$. We shall choose ${\bf m}= \mu_0 \sum_a \alpha_a
m_a$, with $m_a \in \IZ$, and where $\mu_0$ is an overall
scale parameter with the dimension of a mass\footnote{Notice that $\mu_0$ and
all $m_a$'s have to have the same sign in order for ${\bf m}$ to lie in the
Fundamental Weyl Chamber (see \rf{epmh})}.   Under Eq.\ref{symphi}, we have
${\bf m}.\vp/\mu_0 \to {\bf m}.\vp/\mu_0 +2\pi i \IZ$.  The appropriate
definition of the topological current is
\be
j^{\mu} = {1\over 2 \pi i\mu_0}  \epsilon^{\mu\nu} \pa_{\nu}
\( {\bf m}\cdot \vp\).
\lab{tcurabel}
\ee
Indeed, the vacua are  infinitely degenerate and for a soliton solution
the asymtotic values  of $\vp$, at $x=\pm \infty$, have  to differ
by values appearing on the right hand side
of Eq.\ref{symphi}. Thus
 it follows from the argument just given that
$Q_{\rm{topol.}}=\int dx j^0 \in {\cal Z}$.
Concerning the  Noether current,
 we have to take account of the fact that, with our
field normalization, the free part of the Lagrangian has an overall
factor $k$, where $k$ is the coupling constant.
The correct definition of the Noether current is
\be
J_{\mu} =  {k\over \mu_0}
\sum_{m=1}^{h-1} \sum_{\a^{(m)}} {2 \o {\a^{(m)}}^2}
m_{\psi^{\a^{(m)}}}  {\bar \psi}^{\a^{(m)}}\gamma_{\mu} \psi^{\a^{(m)}}
\equiv  {k\over \mu_0}
\sum_{\a>0} {2 \o {\a^2}} m_{\psi^{\a}} \, {\bar \psi}^{\a}
\gamma_{\mu}\, \psi^{\a}
\lab{ncurabel}
\ee
where $\a$ stands for any positive root of $\cg$. Using the
special form of ${\bf m}$ just introduced, we obtain
\be
J_{\mu}=\sum_{\a>0} \sum_a m_a {2 \o {\a^2}} \alpha.\alpha_a
 \, k {\bar \psi}^{\a}
\gamma_{\mu}\, \psi^{\a}.
\lab{ncurabel2}
\ee
For each term  $N^{\a}\equiv \int k {\bar \psi}^{\a}
\gamma_{0}\, \psi^{\a}$ is such that
$$
i\left \{ \psi^{\a}, \, N^{\beta} \right\}_{\rm P.B.}
= \delta_{\alpha,\, \beta} \psi^{\a}.
$$
Moreover, as is well known, ${2 \o {\a^2}} \alpha.\alpha_a \in \IZ$.
Thus, the above definition of the Noether current is such that the
Noether charge $Q_{\rm Noether}=\int dx J_0$ has
integer eigenvalues, as it should.

Next let us compare the two currents so defined.
Clearly  $\Tr \( Q_{\rm ppal} E_0\) = 2 {\hat \d} \cdot {\bf m} \neq 0$
(${\hat \d}=\sum_{a=1}^r {\l_a \o \a_a^2}$), since ${\bf m}$ lies inside the
FWC. Therefore \rf{nonzerotr} is satisfied, and one can always find one
solution in each orbit of the conformal group such that \rf{zerocur} is
true. Consequently, for such solutions we have
\be
i\sum_{\a>0} {2 \o {\a^2}} m_{\psi^{\a}} \, {\bar \psi}^{\a}
\gamma^{\mu}\, \psi^{\a}
\equiv \epsilon^{\mu\nu} \pa_{\nu} \( {\bf m}\cdot \vp\)
\ee
Consequently the Noether and topological currents are proportional:
\be
j^\mu={1\over k\pi} J^\mu,\quad
Q_{\rm topol.}={1\over k \pi} Q_{\rm Noether}
\lab{t-n.gen}
\ee
This  equation  has an important consequence.
It will certainly hold at the quantum level, after a suitable
redefinition.
Since  the eigenvalues of the
Noether charge will be integers.
and, since  the topological charge is also an integer, we obtain that
$k \pi$ should be rational.
One  may expect that this will lead to  the correct statistics for the
$\psi$ fields, following the argument given at
 the end of the previous section.
A quantification of $k$ is of course expected, since our actions are related
with the one of WZNW.
On the other hand, one may regard eqs.\ref{t-n.gen} as   classical versions
of a formulae  of the Atiyah Patodi Singer type (see e.g. ref.\ct{NS}
for a review). In pratice, Eqs.\ref{t-n.gen} mean that the Noether density
is non zero only where $\partial  \vp\not=0$, that is inside the
solitons. Thus the $\psi$ fields are confined inside the solitons.
We shall come back to this on the simplest example below.

The soliton solutions are obtained as follows. The operators diagonalizing
the adjoint action of $E_{\pm h}$, given in \rf{epmh}, are
\be
V_{\a} \( \zeta \) \equiv \sum_{n \in \IZ} \zeta^{-n} E_{\a}^n
\ee
for any root $\a$ (positive or not) of $\cg$. Indeed
\be
\lb E_{\pm h} \, , \, V_{\a} \( \zeta \) \rb = \zeta^{\pm 1}\,
\( {\bf m} \cdot \a\) \, V_{\a} \( \zeta \)
\ee
Notice that $V_{\a} \( \zeta \)$ and $V_{-\a} \(- \zeta \)$ have the same
eigenvalues, and it turns out that the one-soliton solutions are obtained
from \rf{nice}, by choosing the constant group element $\rho$ to be
an exponentiation of (see discussion below \rf{nice})
\be
V_{\a} \( a_{\pm}^{\a} , \zeta\) \equiv a_{+}^{\a}\, V_{\a} \( \zeta \) +
a_{-}^{\a}\, V_{-\a} \(- \zeta \)
\ee
We then have a one-soliton solution (with two parameters $a_{\pm}^{\a}$) for
each pair of Dirac fields  $\psi^{\a }$ and ${\tilde \psi}^{\a }$. The masses
of the solitons and Dirac particle  are proportional to ${\bf m} \cdot \a$.

The construction of the soliton solutions, as well as the physical
properties  of such models, are discussed in detail in section
\ref{sec:exsl2} for the  case of $sl(2)$.
\sect{Example of the principal gradation for $sl(2)^{(1)}$}
\label{sec:exsl2}
In this section we discuss two examples, for $l=2$ and $l=3$, associated
with the principal gradation of the untwisted affine Kac-Moody algebra
$sl(2)^{(1)}$.
Let us denote by $H^n$, $E_{\pm}^n$, $D$ and $C$ the Chevalley basis
generators of the $sl(2)^{(1)}$. The commutation relations are
\br
\lb H^m \, , \, H^n \rb &=& 2 \, m \, C \, \d_{m+n,0},
\lab{sl2a}\\
\lb H^m \, , \, E^n_{\pm} \rb &=& \pm 2 \, E^{m+n}_{\pm},
\lab{sl2b}\\
\lb E^m_{+} \, , \, E^n_{-} \rb &=& H^{m+n} + m \, C \, \d_{m+n,0},
\lab{sl2c}\\
\lb D \, , \, T^m \rb &=& m \, T^m \, , \qquad T^m \equiv H^m , E_{\pm}^m;
\lab{sl2d}
\er
all other commutation relations are trivial.
The grading operator for the principal gradation (${\bf s}=(1,1)$) is
\be
Q\equiv \h H^0 + 2 D.
\lab{ppalgop}
\ee
Then the eigensubspaces are
\br
\cgh_0 &=& \{ H^0, C, Q\} ;\nonu\\
\cgh_{2n+1} &=& \{ E_{+}^n , E_{-}^{n+1}\} \, \qquad n\in \IZ ;\nonu\\
\cgh_{2n} &=& \{ H^n\} ,\, \qquad n\in \{ \IZ - 0\} .
\lab{eigensl2}
\er
The mapping $B$ is parametrised as
\be
B= e^{\vp\, H^0}\, e^{{\tilde {\nu}} \, C}\,e^{\eta\, Q} =
e^{\vp\, {\tilde H}^0}\, e^{\nu \, C}\,e^{\eta\, Q},
\lab{bsl2}
\ee
where ${\tilde H}^0 = H^0 -\h \, C$ is the Cartan generator in the
special basis introduced in \rf{cartanplus}, and so ${\tilde {\nu}} =
\nu - \h
\vp$.
\subsection{Case $l=2$}
\label{subsec:l=2}
Consider the case $l=2$, and choose
\be
E_2 \equiv m \, H^1 \, , \qquad E_{-2} \equiv m \, H^{-1},
\lab{e2}
\ee
where $m$ is a  constant.
We then have
\be
\lb E_{2} \, , \, \lb E_{-2} \, , \, E_{\pm}^n \rb\rb =
4 m^2 \, E_{\pm}^n.
\ee
Therefore, each of the subspaces $\cgh_{\pm 1}$ has two generators with the
same eigenvalue
$ 4 m^2$.
Following \rf{basisdef} and \rf{psidef} we write
\be
F^{+}_1 = 2\sqrt{i m}\( \psi_R\, E_+^0 +
\widetilde \psi_R E_-^1\)\, ,\quad
F^{-}_1 = 2\sqrt{i m}\( \psi_L\, E_+^{-1} -
\widetilde \psi_L\, E_-^0 \) ,
\ee
and introduce the  Dirac fields
\br
\psi = \(
\begin{array}{c}
\psi_R\\
\psi_L
\end{array}\) \, ; \qquad
\widetilde \psi = \(
\begin{array}{c}
\widetilde \psi_R\\
\widetilde \psi_L
\end{array}\)
\er
{}From \rf{massf} we obtain the masses of the particles,
\be
m_{\vp} = m_{{\tilde{\nu}}} = m_{\eta} = 0;\; \qquad
m_{\psi} = 4 m;
\lab{mass}
\ee
The equations of motion derived   from \rf{eqm1}--\rf{eqm4}, are
\br
\pa^2 \,\vp &=& -4 m_{\psi}\,  \overline
\psi \gamma_5 e^{\eta+2\varphi \gamma_5} \psi,
\lab{sl2eqm1}\\
\pa^2\,{\tilde{\nu}} &=& -  2 m_{\psi}\,  \overline
\psi (1-\gamma_5) e^{\eta+2\varphi \gamma_5} \psi
 - {1\over 2}  m_{\psi}^2 e^{2\eta},
\lab{sl2eqm2}\\
\pa^2 \eta &=& 0,
\lab{sl2eqm3}\\
i \gamma^{\mu} \pa_{\mu} \psi &=& m_{\psi}\, e^{\eta+2\vp\,\gamma_5} \,
\psi ,\lab{sl2eqm4}\\
i \gamma^{\mu} \pa_{\mu} \widetilde \psi &
=& m_{\psi}\, e^{\eta-2\vp\,\gamma_5} \,
\widetilde  \psi ,
 \lab{sl2eqm5}
\er
where the gamma matrices are defined in \rf{gammas}, and $\gamma_5 =
\gamma_0\gamma_1$,
and  ${\bar{\psi}} \equiv {\widetilde \psi}^{T} \, \gamma_0$.
Recall that  $\pa^2 = \pa_t^2
- \pa_x^2$, $x_{\pm}=t \pm x$.
The corresponding Lagrangian has the form
\br
{1\o k}\, \cl &=& {1\o 4} \pa_{\mu} \vp \, \pa^{\mu} \vp
+ {1\o 4}  \pa_{\mu} \vp \, \pa^{\mu} \eta
+ \h  \pa_{\mu} {\tilde{\nu}} \, \pa^{\mu} \eta
- {1\o 8}\, m_{\psi}^2 \, e^{2\,\eta} \nonu\\
&+& i  {\bar{\psi}} \gamma^{\mu} \pa_{\mu} \psi
- m_{\psi}\,  {\bar{\psi}} \,
e^{\eta+2\vp\,\gamma_5}\, \psi.
\lab{lagrangian}
\er
It is real (for $\eta = \mbox{\rm real constant}$) if $\widetilde \psi$ is
the complex conjugate of $\psi$, and
if $\varphi $ is pure imaginary. This will be true for the soliton solution
as we shall see below.

Notice that such model is invariant under the  transformations
\be
x_{+} \leftrightarrow x_{-} \, ; \quad \psi_R \leftrightarrow
\epsilon \widetilde \psi_L \, ; \quad  \widetilde \psi_R \leftrightarrow
-\epsilon \psi_L \, ; \quad \vp \leftrightarrow \vp \, ; \quad \eta
\leftrightarrow \eta \, ; \quad \nu \leftrightarrow \nu
\ee
where $\epsilon = \pm 1$. It should be interprated as the product CP
of charge conjugation times parity.  Parity alone is clearly violated.

The generator $H^0 \in \cgh_0$ commutes with $E_{\pm 2}$, and, therefore, the
gauge symmetry \rf{gs1}--\rf{gs3} of the model is $U(1)_L \otimes U(1)_R$,
\be
h_L(x_{-}) = e^{\xi_- (x_{-}) \, H^0}\, , \qquad
h_R(x_{+}) = e^{\xi_+ (x_{+}) \, H^0}.
\lab{gaugesl2}
\ee
Since the genereators of $U(1)_L$ and $U(1)_R$ are the same, we have the
global gauge transfomations \rf{diagonalgauge} generated by $h_D \equiv h_L =
h_R^{-1}\equiv e^{i \theta \, H^0/2}$ ($\theta = {\rm const.}$). The fields
 are transformed as
\be
\psi \ra e^{i\theta} \psi \, \quad
\widetilde \psi \ra e^{-i\theta} \widetilde \psi \, \quad
\vp \ra \vp \, , \quad {\tilde{\nu}}\ra {\tilde{\nu}} \, , \quad
\eta \ra \eta ;
\lab{globalsl2}
\ee
and the corresponding Noether current is
\be
J^{\mu} = {\bar{\psi}}\, \gamma^{\mu}\, \psi \, , \qquad
\pa_{\mu}\, J^{\mu} = 0.
\lab{noethersl2}
\ee
The fields $\psi$ and $\widetilde \psi$ have charges $1$ and $-1$,
respectively; and $\vp$, ${\tilde{\nu}}$ and $\eta$ have charge zero.

Let us next see how the general arguments given above concerning
Noether and topological charges apply here. The topological
current and charges are
\be
j^{\mu} =  {1\o{2\pi i}}\epsilon^{\mu\nu} \pa_{\nu} \, \vp
\, , \qquad
Q_{\rm topol.} \equiv \int \, dx \, j^0 \, , \qquad
\ee
Indeed, the Lagrangian \rf{lagrangian} has infinitely
many degenerate vacua
due to the invariance under $\vp \ra \vp +  i\pi$.
Making use of the field equations, one easily verifies that
\be
\partial_\mu \left[ i\bar \psi \gamma_5 \gamma^\mu \psi+{1\over 2}
\partial^\mu \vp \right ]=0
\lab{PCAC}
\ee
Combining this equation with the conservation of the vector current
$\bar \psi \gamma^\mu \psi$, one deduces that there exist two charges
 defined by
$$
{\cal J}=-i \widetilde \psi_R \psi_R +{1\over 2} \partial_+\vp,
\quad
{\bar {\cal J}}=i \widetilde \psi_L \psi_L +{1\over 2} \partial_-\vp
$$
which satisfy $\partial_-{\cal J}=0$, $\partial_+{\bar {\cal J}}=0$.
Applying the general argument of the previous section, we conclude that we
may choose the solution so that Eqs.\ref{zerocur} hold, so that
${\cal J}={\bar {\cal J}}=0$.
This gives, altogether,
\be
{1\o{2\pi i}}\epsilon^{\mu\nu} \pa_{\nu} \, \vp=
{1\o \pi} \bar \psi \gamma^\mu  \psi,
\lab{currents}
\ee
so that the topological and Noether  currents
are  proportional. As discussed in section \ref{sec:special}, that is
a
consequence of the fact that $E_{\pm 2}$ satisfies \rf{parallel}.

Let us turn to the Noether charge  which here is simply the
fermion number. It should be defined such that
it satisfies the Poisson bracket relation
\be
i \left \{ \psi, Q_{\rm Noether}\right\}_{\rm P.B.}= \psi
\ee
Since the coupling constant $k$ was taken as an overall factor, this is
satisfied by
\be
Q_{\rm Noether}={k} \int dx
\bar \psi \gamma^0  \psi
\ee
so that
\be
Q_{\rm topol.}={1\over k \pi} Q_{\rm Noether}
\lab{top-noe}
\ee
As argued in general, this means that
$k$ should only take discrete values
as  expected, since our actions are related
with the one of WZNW.

Let us now construct the soliton solutions. The operators $E_{\pm 2}$ given
in \rf{e2}, lie in the homogeneous Heisenberg
subalgebra generated by $H^n$, with the commutation relations
\rf{sl2a}. Such a subalgebra has no generators of grade $\pm 1$ for the
principal gradation. Therefore, the model under consideration has no vacuum
solutions of type \rf{vacuum2}. Then, from \rf{cepm}, we get
\be
\ce_{\pm} = E_{\pm 2} = m \, H^{\pm 1}.
\ee
We perform the dressing transformation starting from the vacuum solution
\rf{vacuum1}, namely
\be
\vp = \eta = \psi=\widetilde \psi  = 0 \, ,  \qquad \, ; \quad
{\tilde{\nu}} = -{1\o 8} m_{\psi}^2 x_{+} x_{-}\equiv \nu_0.
\lab{vacsl2l=2}
\ee
Now, let $\mid\, {\hat{\l}}_0\, \rangle$ and  $\mid\, {\hat{\l}}_1\, \rangle$
be
the highest weight states of two fundamental representations of the  affine
Kac--Moody algebra $sl(2)^{(1)}$, respectively the scalar and spinor ones.
Then, from \rf{nice} with $\eta =0$, we obtain the solutions on the orbit of
the
vacuum  \rf{vacsl2l=2},
\br
e^{-\vp} &=& {{\langle \,{\hat{\l}}_1\,\mid\, G\, \mid\, {\hat{\l}}_1\,
\rangle}\over  {\langle \,{\hat{\l}}_0\,\mid\, G\, \mid\, {\hat{\l}}_0\,
\rangle}},\; \; \qquad
e^{-({\tilde{\nu}} - \nu_0)} = \langle \,{\hat{\l}}_0\,\mid\,
G\, \mid\, {\hat{\l}}_0\, \rangle ,\nonu\\
\psi_R &=&  \sqrt{ m\over i}\, {{\langle \,{\hat{\l}}_0\,\mid\, E_-^1\, G\,
\mid\,  {\hat{\l}}_0\, \rangle}\over  {\langle\,{\hat{\l}}_0\,\mid\, G\, \mid\,
{\hat{\l}}_0\, \rangle}},\;\; \;
\widetilde \psi_R =-  \sqrt{m\over i}\, {{\langle
\,{\hat{\l}}_1\,\mid\, E_+^0\, G\, \mid\, {\hat{\l}}_1\, \rangle}\over
{\langle\,{\hat{\l}}_1\,\mid\, G\, \mid\, {\hat{\l}}_1\, \rangle}}
\nonumber \\
\psi_L &=&- \sqrt{m\over i}\, {{\langle \,{\hat{\l}}_1\,\mid\,  G\,E_-^0\,
\mid\, {\hat{\l}}_1\, \rangle}\over  {\langle\,{\hat{\l}}_1\,\mid\, G\, \mid\,
{\hat{\l}}_1\, \rangle}},\;
\widetilde \psi_L = - \sqrt{m\over i}\,{{\langle \,{\hat{\l}}_0\,\mid\,
G\,E_+^{-1}\, \mid\, {\hat{\l}}_0\, \rangle}\over
{\langle\,{\hat{\l}}_0\,\mid\, G\, \mid\, {\hat{\l}}_0\, \rangle}},
\lab{solsl2}
\er
where
\be
G \equiv  e^{x_{+}\, \ce_{+}} \, e^{-x_{-}\, \ce_{-}} \, \rho \, e^{x_{-}\,
\ce_{-}}\, e^{-x_{+}\, \ce_{+}}.
\ee
In order to get the soliton solutions, we choose the fixed mapping
$\rho$ to be an exponentiation of an eigenvector of $\ce_{\pm}$ (solitonic
specialization); namely, $\rho = e^{V}$, with $\lb \ce_{\pm}\, ,\, V \rb =
\omega_{\pm} \, V$. Therefore,
\be
G = \exp \( e^{\Gamma} \, V\) \quad \mbox{ with } \;
\Gamma = \omega_{+} x_{+} - \omega_{-} x_{-} \equiv \gamma \( x
- v\,t\) .
\ee
In this case the eigenvectors of $\ce_{\pm}$ are
\be
V_{\pm} (z) = \sum_{n \in \IZ} z^{-n}\, E_{\pm}^n.
\ee
Indeed,
\br
\lb \ce_{+}\, ,\, V_{\pm}(z) \rb &=& \pm 2 m z \, V_{\pm}(z) \equiv
\omega_{+}^{\pm} V_{\pm}(z),\\
\lb \ce_{-}\, ,\, V_{\pm}(z) \rb &=& \pm {2 m\o z }\, V_{\pm}(z) \equiv
\omega_{-}^{\pm} V_{\pm}(z).
\er
The solution, associated with $V_{+}(z)$, is
\br
\nu = \nu_0,\, \quad
\vp = \widetilde \psi = 0,\, \quad
\psi = \sqrt{m\over  i}  e^{\Gamma}\, \(
\begin{array}{r}
z \\
-1
\end{array}\) ;
\lab{sol1}
\er
while those, associated with $V_{-}(z)$, is given by
\br
\nu = \nu_0,\, \quad
\vp = \psi = 0,\, \quad
\widetilde \psi = -\sqrt{m\over  i} e^{-\Gamma}\, \(
\begin{array}{r}
1 \\
1/ z
\end{array}\) ,
\lab{sol2}
\er
where
\be
\Gamma = 2 m (z x_+ - {1\o z} x_-)  \equiv \gamma \( x - vt\).
\lab{gammasl2}
\ee
The masses of these solutions are obtained from \rf{solitonmass}. Here the
relevant
state $\mid \l_{\bf s^{\pr}}\rangle$ in \rf{truncation} is
\be
\mid \l_{\bf s^{\pr}}\rangle = \mid {\hat{\l}}_{0}\rangle \otimes
\mid {\hat{\l}}_{1}\rangle .
\lab{lambdaspr}
\ee
Using level one vertex operators, one can verify that
\be
\langle {\hat{\l}}_i \mid \, \( V_{\pm}(z)\)^n \, \mid {\hat{\l}}_i \rangle = 0
\, , \quad \mbox{\rm for $n \geq 1$ and $i=0,1$}.
\ee
Therefore, $N_V^{\pr}=0$ in \rf{truncation}, and from \rf{solitonmass} one
gets that the masses of the solutions \rf{sol1} and \rf{sol2} vanish.
Such solutions correspond to the objects which travel with velocities
$v=\pm \( 1 -  z^2\)/\( 1 + z^2\)$; and keeping
$z^2 >0$, one has $\mid v \mid <1$.
Therefore, these solutions cannot be interpreted as
solitons (particles), since
they would correspond to massless particles traveling with velocity smaller
that light velocity. We should interpret them as vacuum configurations, since
they have the same energy as vacuum \rf{vacsl2l=2}.

The true soliton solutions of the system are constructed as follows.
Notice that $V_{+}(z)$ and $V_{-}(-z)$ have the same eigenvalues. Therefore,
any linear combination of them, leads  to solutions traveling with a constant
velocity without dispersion. So, we  let
\be
V(a_{\pm},z) \equiv \sqrt{i}\left ( a_+ V_{+}(z) + a_- V_{-}(-z)\right);
\lab{nicesol}
\ee
\be
\lb \ce_{+}\, ,\, V(a_{\pm},z) \rb =  2 m z \, V(a_{\pm},z) \, , \qquad
\lb \ce_{-}\, ,\, V(a_{\pm},z) \rb =  {2 m\o z }\, V(a_{\pm},z),
\ee
and so $\omega_{+}=2 m z$ and $\omega_{-}={2 m\o z }$.
The particular factor $\sqrt{i}$ is chosen such that the reality condition
will be obeyed with $a_-=a_+^*$.
Again, using level one vertex operators, one can verify that\footnote{Notice
that the truncation occurs for powers greater than $4$, and not $2$, because
$\mid \l_{\bf s^{\pr}}\rangle$ lies in the tensor
product representation, see
\rf{lambdaspr}}
\be
\langle \l_{\bf s^{\pr}} \mid \, V(a_{\pm},z)^n \, \mid \l_{\bf s^{\pr}}\rangle
= 0  \qquad \mbox{\rm for $n > 4$}.
\ee
Therefore, $N_V^{\pr}=4$ in \rf{truncation}, and from \rf{solitonmass}
with $\psi^2 =2$, and $\psi$ being the highest root of $sl(2)$,
one gets that the mass of such solutions is
\be
M= 8 k \, m = 2\, k\, m_{\psi},
\lab{sl2solmass}
\ee
where $k$ is the coupling constant appearing in the Lagrangian \rf{lagrangian},
see \rf{suga}.
The solutions generated by \rf{nicesol}, have two parameters, namely $a_{\pm}$.
One parameter is always present, because one can scale an eigenvector of
$\ce_{\pm}$ without changing the
width $\gamma$ and velocity $v$ of the soliton,
obtained from the eigenvectors $\omega_{\pm}$; see \rf{nicesoliton}. However,
in this case, the second parameter comes from a symmetry. As we have pointed
out in  \rf{symrho}, associated to the fixed element $\rho =
e^{V(a_{\pm},z)}$, we have an orbit of equivalent solutions due to the global
transformations \rf{globalsl2},
\be
V(a_{\pm},z) \ra \sqrt{i}\left (
a_+\, e^{i\theta}\, V_{+}(z) + a_-\, e^{-i\theta}\, V_{-}(-z)\right) .
\ee
The explicit form of the solutions generated by \rf{nicesol}, is obtained
using \rf{solsl2},
\br
\vp &=& \log \(  {{1 + i\sigma e^{2 \Gamma}}\o { 1 - i\sigma e^{2 \Gamma}}}\) ,
\lab{solsl2a}\\
{\tilde {\nu}} &=& - \log \( 1 + i\sigma e^{2 \Gamma}\) - {1\o 8} m_{\psi}^2
 x_{+} x_{-},
\lab{solsl2b}\\
\eta &=& 0;
\lab{solsl2c}
\er
and
\br
\psi = a_{+}\sqrt{m} \, e^{\Gamma}\, \(
\begin{array}{c}
 {z\o{1 + i\sigma e^{2 \Gamma}}}\\
{-1 \o{1 - i\sigma e^{2 \Gamma}}}
\end{array}\) \, , \qquad
\widetilde \psi = a_{-}\sqrt{m}\, e^{\Gamma}\, \(
\begin{array}{c}
 {z\o{1 - i\sigma e^{2 \Gamma}}}\\
{-1 \o{1 + i\sigma e^{2 \Gamma}}}
\end{array}\) ;
\lab{solsl2d}
\er
where $\Gamma$ is given in \rf{gammasl2}, and $\sigma = a_{+} a_{-}z/4$.
Keeping $m$ and $z$ real,  we have the
mass $M$  of the soliton, from \rf{sl2solmass},
real and positive, and also the
parameters $\gamma$ and $v$ \rf{gammasl2} are real. The reality condition is
obeyed if $a_-=a_+^*$, as anticipated.  At this point, it is useful
to re-express the expressions just given in terms of the
physical parameters of the
soliton. Using equations \rf{gammas}, and \rf{mass}, one deduces that
\be
\gamma= m_\psi\left / \sqrt{1-v^2}\right., \quad
z=  \sqrt{ (1-v)/(1+v)}.
\lab{prm}
\ee
Moreover, since $a_\pm$ are complex conjugate, we may write
\be
a_\pm =e^{\pm i\theta} 2 \sqrt {\sigma\over z}.
\lab{apm}
\ee
   The dependence upon space-time
 appears   only
through  $\sqrt {\sigma} \exp(\Gamma)$. We will
write \footnote{by convention, we choose $\sigma$ to be positive}
\be
\sqrt{\sigma} e^{\Gamma}=\exp( (\gamma(x-x_0-vt))
\ee
where $x_0$ is the position of the soliton at
time zero.
 Then  we have
\be
\vp = 2i \arctan \( \exp \( 2m_\psi \( x-x_0-vt\)/\sqrt{1-v^2}\)\) ,
\lab{solfinale}
\ee
which is the sine--Gordon soliton. The Dirac fields are given by
\be
\psi = e^{i\theta} \sqrt{m_\psi} \,
e^{m_\psi \( x-x_0-vt\)/\sqrt{1-v^2}}\, \(
\begin{array}{c}
\left( { 1-v\o 1+v}\right)^{1/4}
{1 \o 1 + ie^{2 m_\psi \( x-x_0-vt\)/\sqrt{1-v^2}}}\\
-\left( { 1+v\o 1-v}\right)^{1/4}
{1 \o 1 - ie^{2 m_\psi \( x-x_0-vt\)/\sqrt{1-v^2}}},
\end{array}\)
\lab{solsimple}
\ee
and $\widetilde \psi$ is the complex conjugate of $\psi$. Thus the only
parameters are the soliton mass and velocity,
together with the angle $\theta$ which reflects the global invariance
\rf{globalsl2}.
Notice that the sign of the tolopogical charge
 can be reversed by reversing the sign of
$z$. Therefore, the solutions \rf{solfinale}--\rf{solsimple} contain the
sine--Gordon soliton and anti--soliton.

Finally, we come to the  very important feature of the present model already
mentioned above in general, namely
it is clear from the explicit expressions Eqs.\ref{solsimple} that $\psi$
vanishes
exponentially when $x-x_0\to \pm  \infty$, so that the Dirac field is
confined inside the soliton. That this must be true is of course a general
consequence of Eq.\ref{currents} which may be verified directly on
the explicit solution. This phenomenon has been much studied for
electron phonon systems. Models of a similar type  describe  the
 electron self-localization
in quasi-one-dimensional dielectrics (for recent reviews see
\ct{BK}, \ct{HKSW}). At low temperature these systems
go over to   dielectric
states  characterized by charge density waves which can
be constructed on the basis of the Peierls model. The  continuous limits
are  described by Lagrangians
similar to Eq.\ref{lagrangian}. Discussing this important issue
is beyond the scope of the present article,
so we will not dwell upon it here.
Let us simply recall that the typical example of the polyacteline molecule
was much discussed in connection with fermion number
fractionization \ct{JR,GW}.  Clearly, on the other hand one may regard
our soliton solution  a sort of one dimensional bag model for QCD.
In this connection let us note that, if we introduce the two-by-two
matrix $U=\exp(\eta+2\varphi \gamma_5)$, we may rewrite the
Lagrangian Eq.\ref{lagrangian} as
\br
\cl &=& {1\over 16} \Bigl \{ \hbox{tr} \left [ U^{-1} \partial_\mu U
 {1+\gamma_5\over 2}  U^{-1} \partial^\mu U\right]
-{1\over 2} \hbox{tr} \left [ U^{-1} \partial_\mu U
 \right]\hbox{tr} \left [  U^{-1} \partial^\mu U\right]\Bigr \}
\nonu\\
& &
+i\bar \psi \gamma_\mu \partial_\mu \psi
- \bar \psi  U \psi
-{m_\psi^2\over 8} \hbox{det} (U),
\lab{langrU}
\er
which is similar to a two-dimensional version of the low energy effective
action for QCD (see e.g. \ct{CHT}).

\subsection{Case $l=3$}

In this case we choose
\be
E_3 = q_{+} E_{+}^1 + q_{-} E_{-}^{2}\, , \qquad
E_{-3} = q_{+} E_{+}^{-2} + q_{-} E_{-}^{-1};
\lab{e3}
\ee
and so
\be
\lb E_{3}\, , \, E_{-3} \rb = 3q_{+}q_{-} C \equiv \b C.
\ee
Introduce also the notations
\br
E_1 &=& q_{+} E_{+}^0 + q_{-} E_{-}^{1}\, , \qquad
E_{-1} = q_{+} E_{+}^{-1} + q_{-} E_{-}^{0},\nonu\\
f_1 &=& q_{+} E_{+}^0 - q_{-} E_{-}^{1}\, , \qquad
f_{-1} = q_{+} E_{+}^{-1} - q_{-} E_{-}^{0},\nonu\\
f_{2}&=& -\sqrt{q_{+}q_{-}}\, H^1 \,\, , \qquad
f_{-2}= -\sqrt{q_{+}q_{-}}\, H^{-1} .
\er

The fields of the model are those introduced in \rf{bsl2}, and the matter
fields $\psi^i_{R/L}$, $i=1,2$, and $\chi_{\pm}$ defined as
\br
F_1^+ &=& \psi_R^2 f_1 + \chi_+ E_1 \, , \qquad
F_2^+ = \psi_R^1 f_2;\nonu\\
F_1^- &=& \psi_L^1 f_{-1} + \chi_- E_{-1} \, , \qquad
F_2^- = \psi_L^2 f_{-2}.
\er
According to \rf{psidef}, $\psi_{R/L}^i$, $i=1,2$, are the components of two
Dirac fields   which we shall denote by $\psi^i$.

One can easily verify that
\br
\lb E_{3}\, , \,\lb E_{-3}\, , \, E_{\pm 1} \rb\rb &=& 0 \nonu\\
\lb E_{3}\, , \,\lb E_{-3}\, , \, {\tilde H}^0 \rb\rb &=&
4 q_{+}q_{-}{\tilde H}^0 \nonu\\
\lb E_{3}\, , \,\lb E_{-3}\, , \, f_{i} \rb\rb &=& 4 q_{+}q_{-}f_{i}\, , \;
i=\pm 1,\pm 2 .
\er

Therefore, from \rf{massf}, the masses of the particles are
\be
m_{\nu} = m_{\eta} = m_{\chi_{\pm}} = 0 \, , \qquad
m_{\vp} = m_{\psi^i} = 4 \sqrt{ q_{+} q_{-}} \, \; , \, i=1,2
\ee
and so we have to choose $q_{\pm}$, such that $q_{+}q_{-}>0$.

Then, from \rf{eqm1}-\rf{eqm4}, the equations of motion are
\br
\pa_{+}\pa_{-}\, \vp &=& - q_{+}q_{-}\( \( e^{2\vp} - e^{-2\vp}\) \( e^{3\eta}
- e^{\eta} \( \psi_L^1 \psi_R^2 - \chi_+\chi_- \)\) \right. \nonu\\
&-& \left. e^{\eta} \( e^{2\vp} +
e^{-2\vp}\) \( \psi_L^1 \chi_+ -\psi_R^2 \chi_- \) \) ,
\lab{l3eqm1}\\
\pa_{+}\pa_{-}\, \nu &=& - q_{+}q_{-}\( \( e^{2\vp} + e^{-2\vp}\) \(
{3\o 2} e^{3\eta} - \h e^{\eta} \( \psi_L^1 \psi_R^2 - \chi_+\chi_- \)\)
\right.
\nonu\\
&-& \left. \h e^{\eta} \( e^{2\vp} - e^{-2\vp}\) \( \psi_L^1 \chi_+
-\psi_R^2 \chi_- \) + 2 e^{2\eta} \psi_L^2\psi_R^1\) ,
\lab{l3eqm2}\\
\pa_{-}\, \psi_R^1 &=& \sqrt{q_{+}q_{-}} e^{\eta}\( \psi_L^1  \( e^{2\vp} +
e^{-2\vp}\) - \chi_- \( e^{2\vp} - e^{-2\vp}\)\) ,
\lab{l3eqm3}\\
\pa_{+}\, \psi_L^1 &=& 2\sqrt{q_{+}q_{-}} \( -e^{2\eta} \psi_R^1 + \h e^{\eta}
\psi_L^2 \( \psi_R^2 \( e^{2\vp} - e^{-2\vp}\) + \chi_+\( e^{2\vp} +
e^{-2\vp}\)\)\) ,
\lab{l3eqm4}\\
\pa_{-}\, \psi_R^2 &=& 2\sqrt{q_{+}q_{-}} \( e^{2\eta} \psi_L^2 + \h e^{\eta}
\psi_R^1 \( \psi_L^1 \( e^{2\vp} - e^{-2\vp}\) - \chi_-\( e^{2\vp} +
e^{-2\vp}\)\)\) ,
\lab{l3eqm5}\\
\pa_{+}\, \psi_L^2 &=& \sqrt{q_{+}q_{-}} e^{\eta}\( -\psi_R^2  \( e^{2\vp} +
e^{-2\vp}\) - \chi_+ \( e^{2\vp} - e^{-2\vp}\)\) ,
\lab{l3eqm6}\\
\pa_{-}\, \chi_+ &=& \sqrt{q_{+}q_{-}} e^{\eta} \psi_R^1 \( \chi_-  \(
e^{2\vp} - e^{-2\vp}\) - \psi_L^1 \( e^{2\vp} + e^{-2\vp}\)\) ,
\lab{l3eqm7}\\
\pa_{+}\, \chi_- &=& \sqrt{q_{+}q_{-}} e^{\eta} \psi_L^2 \( \chi_+  \( e^{2\vp}
- e^{-2\vp}\) + \psi_R^2 \( e^{2\vp} + e^{-2\vp}\)\) ,
\lab{l3eqm8}\\
\pa_{+}\pa_{-}\, \eta &=& 0.
\lab{l3eqm9}
\er
Notice that these equations are invariant under the CP  transformation
\br
x_{+} \leftrightarrow x_{-} \, , \quad \psi_R^1 \leftrightarrow \psi_L^2
\, , \quad \psi_R^2 \leftrightarrow -\psi_L^1
\, , \quad \chi_+ \leftrightarrow \chi_-
\, , \quad  \vp \leftrightarrow \vp
\, , \quad  \nu \leftrightarrow \nu
\, , \quad  \eta \leftrightarrow \eta .
\er
Since, there are no generators of $\cgh_0$, given in \rf{eigensl2}, that
commute with $E_{\pm 3}$ in \rf{e3}, we do not have any gauge symmetry of the
 type \rf{gs1}-\rf{gs3}. In the linear approximation for the $\psi$ and
$\chi$ fields with $\varphi=0$ and $\eta=$ constant, one verifies that
the $\psi$ field equations  may be deduced from a Lagrangian of the type
Eq.\ref{freeL}, with $\psi^1\leftrightarrow \psi$, and $
\psi^2\leftrightarrow \widetilde \psi$. Such is not the case for the $\chi$
fields
although they  are  massless in the linear approximation. Indeed, $\chi_\pm$
has   weights $({2\over 3}, 0)$ and $(0, {2\over 3})$ respectively,
so that we cannot
introduce a conjugate  field to write down a covariant kinetic term.
As a matter of fact, we have been unable to derive the above field equations
 from a local action.

The sine--Gordon (or sinh--Gordon) model is a submodel of the system
\rf{l3eqm1}-\rf{l3eqm9}. Indeed, $\psi^i=\chi_{\pm}=\eta =0$ is a solution of
 the equations of motion, and then $i\vp$ ($\vp$) has to satisfy the
sine--Gordon (sinh--Gordon) equation.

The operators $E_{\pm 3}$ belong to the principal Heisenberg subalgebra of
$sl(2)^{(1)}$
\be
\lb E_{2m+1}\, , \, E_{2n+1} \rb = q_{+} q_{-} (2m+1) \, C \, \d_{n+m+1,0}
\ee
where
\be
E_{2m+1} \equiv q_{+}\, E_{+}^{m} + q_{-}\, E_{-}^{m+1}\, , \qquad m \in \IZ
\ee
They are eigenvectors of the grading operator \rf{ppalgop}
\be
\lb Q \, , \, E_{2m+1} \rb = (2m+1) \, E_{2m+1}
\ee
The adjoint action of $E_{2m+1}$ is diagonalized by the  operators
 \br
V(z) &\equiv& - \sqrt{q_{+}q_{-}} \sum_{n=-\infty}^{\infty}\, z^{-2n}\, H^{n}
+ \sum_{n=-\infty}^{\infty}\, z^{-2n-1}\, \( q_{+} \,E_{+}^{n} -
q_{-}\, E_{-}^{n+1}\) - \h \, \sqrt{q_{+}q_{-}}\, C \nonu\\
&\equiv& \sum_{n=-\infty}^{\infty}\, z^{-n}\, V_n .
\lab{vl3}
\er
where
\be
\lb Q \, , \, V_n \rb = n \, V_n
\ee
Indeed, one gets
\be
\lb E_{2m+1} \, , \,  V(z)\rb = 2 \sqrt{q_{+}q_{-}}\,\, z^{2m+1}\, V(z)
\ee
Therefore, $V(z)$ are eigenvectors of $E_{\pm 3}$ with eigenvalues
\be
\omega_{\pm} = 2 \sqrt{q_{+}q_{-}}\,\, z^{\pm 3}
\ee
Notice that, if one shifts $z \ra \omega z$, with $\omega^3=1$, the
eigenvalues
do not change. Therefore
\be
V(a_j,z) \equiv a_0 V(z) + a_1 V(\omega z) + a_2 V(\omega^2 z)
\lab{degeneratev}
\ee
are also  eigenvectors of $E_{\pm 3}$.
Therefore, if one performs the dressing transformations from the vacuum
\rf{vacuum1}, namely
\be
\vp = \eta = \chi_{\pm} = \psi^1 = \psi^2 = 0 \, \, , \qquad
\nu = -3q_{+}q_{-}\, x_{+} x_{-}
\lab{vacsl21}
\ee
one obtains solutions traveling with constant velocity, without dispersion, by
 taking the constant group element $\rho$ in \rf{nice} as the exponentiation
of \rf{degeneratev} (see \rf{nicesoliton}).

The operator $V(z)$ introduced in \rf{vl3} can be realized through the
principal
vertex operator construction \ct{vertex,OTU93}. Using such contruction, and
taking into account that the highest weight state \rf{lambdaspr} lies in the
 tensor product representation, one gets
\br
\langle \l_{{\bf s^{\pr}}} \mid \, \( V (a_j, z)\)^n \, \mid
\l_{{\bf s^{\pr}}}
 \rangle
&=& 0  \, ; \qquad \mbox{\rm for $n>2$, with just one non vanishing $a_j$'s}
\lab{trunc1}\\
\langle \l_{{\bf s^{\pr}}} \mid \, \( V (a_j, z)\)^n \, \mid \l_{{\bf s^{\pr}}}
 \rangle
&=& 0  \, ; \qquad \mbox{\rm for $n>4$, with two non vanishing $a_j$'s}
\lab{trunc2}\\
\langle \l_{{\bf s^{\pr}}} \mid \, \( V (a_j, z)\)^n \, \mid \l_{{\bf s^{\pr}}}
 \rangle
&=& 0  \, ; \qquad \mbox{\rm for $n>6$, with all three $a_j$'s, non vanishing}
\lab{trunc3}
\er
Using the mass formula \rf{solitonmass}, one gets that the masses of the
solitons created by the operator $V(a_j,z)$ in \rf{degeneratev}, from the
vacuum \rf{vacsl21} are
\br
M_1 &=& {8\o 3}\, k\,\sqrt{q_{+} q_{-}} = 2\, k\, m_{\vp} \, ; \qquad
\mbox{\rm with just one non vanishing $a_j$'s} \\
M_2 &=& {16\o 3}\, k\,\sqrt{q_{+} q_{-}} = 4 \, k \, m_{\vp} \, ; \qquad
\mbox{\rm with two non vanishing $a_j$'s} \\
M_3 &=& 8\, k\,\sqrt{q_{+} q_{-}} = 6\, k \, m_{\vp} \, ; \qquad
\mbox{\rm with all three $a_j$'s non vanishing}
\er
Now, if one performs the dressing transformations from the vacumm
\rf{vacuum2},
namely
\be
\vp = \eta =  \psi^1 = \psi^2 = 0 \, \, , \qquad  \chi_{\pm} = c^{\pm}_1
\, \, , \qquad \nu = -3q_{+}q_{-}\, x_{+} x_{-}
\lab{vacsl22}
\ee
with $c^{\pm}_1 = {\rm const.}$, then the soliton type solutions are created
by eigenvectors of
\be
\ce_{\pm} \equiv E_{\pm 3} + c^{\pm}_1 E_{\pm 1}
\lab{cpml3}
\ee
Again, $V(z)$ are eigenvectors of $\ce_{\pm}$ with eigenvalues
\be
\omega_{\pm}^{\pr} = 2 \sqrt{q_{+}q_{-}}\,\,\( z^{\pm 3} +
c^{\pm}_1 z^{\pm 1}\)
\ee
Therefore from  \rf{solitonmass} one gets that the mass of the corresponding
soliton is ($N^{\pr}_V = 2$ from \rf{trunc1})
\be
M^{\pr} = {8\o 3}\, k\,\sqrt{q_{+} q_{-} \, ( 1 + c^{+}_1 c^{-}_1 +
c^{-}_1 z^2 + c^{+}_1/z^2 )}
\ee

The spectrum of the soliton solutions of such model is rather complicated.
There remains to perform a more detailed  analysis of those solutions, in
order to understand, among other things, the physical consequences of the
 existence of (at least) two classical vacuum configurations.

\sect{Outlook}
The models we introduced in this paper constitute a generalization of the
affine Toda systems, in the sense that
 they contain matter fields coupled to
the usual (gauge) Toda fields. We believe such models open the study of a
variety of massive integrable theories with very interesting physical
properties. As we have already mentioned, from the point of view of non
perturbative aspects of quantum field theories, we hope these models will be
useful, as a laboratory, in the understanding of the quantum theory of
solitons, some confinement mechanisms and also to obtain exact results on the
strong coupling regime. On the other hand, these systems can be
used to describe very interesting phenomena in condensed matter physics and
statistical mechanics like electron--phonon systems, electron
self--localization in quasi-one-dimensional dieletrics and polyacetylene
molecules.

The next step in achieving such program is to consider the quantum theory
of these models, trying to obtain exact results by exploring some of their
special physical properties. In this sense the most promising class of
models is that described in section \ref{sec:special}, which simplest
example corresponds to the model associated to $sl(2)$, described in
subsection \ref{subsec:l=2}. That class of models possesses a $U(1)$ Noether
current depending only on the matter fields which, under a special gauge
fixing of the conformal symmetry, is equivalent to a topological current
depending only on the (gauge) Toda fields. We believe the full consequences
of such equivalence have not been understood yet and, the use of it, may
shed light on several non perturbative aspects of the theory. We have
already pointed out that such equivalence leads, at the classical level, to
the localization of the matter fields inside the soliton. It is very
plausible that, at the quantum level, this gives   a confinement mechanism,
which can be regarded as a one dimensional bag model for QCD. Another
special aspect of these models concerns the mass formula. The masses of
solitons and particles are both determined by the
eigenvalues of the operators
$E_{\pm l}$ appearing in the flat connection defining the models. In
addition, the soliton masses have a topological character due to the
spontaneous breakdown of the conformal symmetry. The models presenting
the equivalence between Noether and topological currents have an additional
feature; the masses of particles and solitons are proportional to the $U(1)$
Noether charge. This situation
 resembles very much the one of four dimensional gauge theories
with Higgs in the adjoint representation and in the BPS limit. These facts
may indicate the existence of a sort of duality in these models involving
solitons and particles \ct{duality}. These points certainly deserve further
study and we hope to explore them in a future work. They may have a
direct connection with supersymmetric Yang-Mills theories along the line of
ref.\cite{MW}

In addition to those, there are still many  interesting practical
problems to be solved following the lines of the present paper
and references herein. We mention first
the study the $W$--symmetries of the system under consideration,
in particular, generalizations of the $W$--algebras of the
standard Toda systems, $W$--geometries associated with them in
the spirit of \ct{GM93}, \ct{GS93}, \ct{RS94}, including differential
and algebraic geometry setting of the Toda systems coupled to
the matter fields.  For this questions, the formulation of the corresponding
problem in terms of Lax operators of a Generalized mkdV hierarchy \ct{mkdv}
should
be useful. Second, it is very believable that the
general solution for the matter fields can be presented in a compact
form; in other words, it seems to be possible to resolve the
recurrent formula \rf{rec}. Third, it would be useful to obtain,
in some suitable parametrisation for the mappings $B$ and $F^{\pm}$,
an explicit expression for the Lagrange function or the
effective action corresponding to our system \rf{em1}--\rf{em3}.
Having such an expression, one can directly get the formula for
the energy--momentum tensor, and hence calculate in a more
simple way the masses of solitons generated by the spontaneously
breakdown of the conformal symmetry. In cases
where this is not possible, one
can work with the WZNW fields, in terms of which the matter fields
can be written locally.

The construction of the multisoliton solutions can be made in a
straightforward way using the methods presented in this paper. That would be
important in the study of the classical scaterring of the solitons, which
could give us valuable information to construct the corresponding S--matrix,
using a bootstrap programme \ct{zamo,marco,dorey}.
Here, the question of massless
particles is of special importance \ct{Timo}.

Another point to be explored is the construction of the local conserved
charges of these systems using for instance, the methods of refs.
\ct{chargesot,charges}, where the flat
connection is gauge transformed into an
abelian (Heisenberg) subalgebra of the affine Kac-Moody algebra $\cgh$.

\section{Acknowledgements}
One of the authors
 (M. S.) wishes to acknowledge the warm
hospitality of the Laboratoire de Physique
Th\'eorique de l'\'Ecole Normale Sup\'erieure de Paris, and Facultad
de F\'\i sica of the  Universidade de Santiago
de Compostela. The authors are indebted  to J. L. Miramontes for his
careful and critical reading of the manuscript and for his very relevant and
enlightning comments. They are also very grateful for discussions with
E. Br\'ezin, A. Georges,
 J. F. Gomes, S. P. Novikov, A. V. Razumov,
P. Wiegmann, A. Zee,
 and A. H. Zimerman.
This work was partially supported by the Russian Fund for Fundamental
Research and International Science Foundation.
L.A.F. was partially supported
by Ministerio de Educaci\'on y Ciencia (Spain).

\end{document}